\PassOptionsToPackage{square,comma,numbers,sort&compress,super}{natbib}
\documentclass[reprint,superscriptaddress,amsmath,amssymb,aps,prx,twocolumn,nofootinbib,notitlepage,noeprint]{revtex4-2}
\usepackage[version=3]{mhchem}
\usepackage[english]{babel}
\usepackage{placeins}
\usepackage{dsfont}

\usepackage{amssymb,amsmath,amsthm,mathrsfs,mathtools}
\usepackage[usenames,dvipsnames]{xcolor}
\usepackage{comment,afterpage,accents,subcaption,braket,booktabs,gensymb,graphicx,epstopdf,float,multirow,algorithm,algpseudocode}
\usepackage{dcolumn}
\usepackage{bm}
\usepackage[colorlinks]{hyperref}
\hypersetup{colorlinks=blue,
            breaklinks,
            urlcolor=black,
            linkcolor=red,
            citecolor=blue}
\usepackage[normalem]{ulem}
\usepackage[font=small]{caption}
\usepackage[labelsep=period]{caption}
\makeatletter
\renewcommand{\fnum@figure}{FIG. \thefigure}
\makeatother
\makeatletter
\renewcommand{\fnum@table}{TABLE \thetable}
\makeatother

\begin{document}

\title{A quantum eigenvalue solver based on tensor networks}

\author{Oskar Leimkuhler}
\email{ol22@berkeley.edu}
\affiliation{Department of Chemistry, University of California, Berkeley}
\affiliation{Berkeley Quantum Information and Computation Center, University of California, Berkeley, CA 94720, USA}
\author{K. Birgitta Whaley}
\email{whaley@berkeley.edu}
\affiliation{Department of Chemistry, University of California, Berkeley}
\email{whaley@berkeley.edu}
\affiliation{Berkeley Quantum Information and Computation Center, University of California, Berkeley, CA 94720, USA}

\date{\today}

\begin{abstract}
Electronic ground states are of central importance in chemical simulations, but have remained beyond the reach of efficient classical algorithms except in cases of weak electron correlation or one-dimensional spatial geometry. We introduce a hybrid quantum-classical eigenvalue solver that constructs a wavefunction ansatz from a linear combination of matrix product states in rotated orbital bases, enabling the characterization of strongly correlated ground states with arbitrary spatial geometry. The energy is converged via a gradient-free generalized sweep algorithm based on quantum subspace diagonalization, with a potentially exponential speedup in the off-diagonal matrix element contractions upon translation into compact quantum circuits of linear depth in the number of qubits. Chemical accuracy is attained in numerical experiments for both a stretched water molecule and an octahedral arrangement of hydrogen atoms, achieving substantially better correlation energies compared to a unitary coupled-cluster benchmark, with orders of magnitude reductions in quantum resource estimates and a surprisingly high tolerance to shot noise. This proof-of-concept study suggests a promising new avenue for scaling up simulations of strongly correlated chemical systems on near-term quantum hardware.
\end{abstract}

\maketitle

\section{Introduction}

Following a century of developments in \textit{ab initio} electronic structure theory since Hartree's self-consistent field \cite{Hartree_1928}, the chemical ground state problem remains intractable for large systems exhibiting strong electron correlation. In these cases a superposition of many Slater determinants may be necessary to characterize the ground state accurately, and the computation time and memory requirements on classical hardware can then scale exponentially with the system size, under a combinatorial explosion of the configuration space. In recent decades tensor network \cite{Penrose_1971,biamonte_tensor_2017,Bridgeman_2017} methods such as the density matrix renormalization group (DMRG) \cite{PhysRevLett.69.2863,white_ab_1999,chan_density_2011,chan_matrix_2016,baiardi_density_2020} have emerged as powerful tools to partially tame, if not exactly break, this so-called `curse of dimensionality'. Tensor networks promise to efficiently represent some classes of many-body quantum states by exploiting the locality of correlations among the parts of the system. A necessary but not sufficient condition for a system to be efficiently described by a tensor network is that it obeys an area law of entanglement \cite{Eisert_2010}, meaning that entropy measures between certain partitions of the system scale only with the number of sites lying on the partition boundaries, as compared to volume law entanglement that scales with the total number of sites in the bulk. Tensor networks have proven very powerful in simulating one-dimensional lattice systems, for which the matrix product state (MPS) \cite{Fannes1992,Klumper_1993} ansatz offers attractive qualities such as flexibility, variationality, and systematic convergence with increasing bond dimension. While matrix product states have had some success in quantum chemical settings \cite{white_ab_1999,chan_density_2011,sharma_spin-adapted_2012,sharma_low-energy_2014,chan_matrix_2016,baiardi_density_2020}, systems without a one-dimensional area law may require exponentially large bond dimensions, and even states that follow the law are not guaranteed to be efficiently simulable \cite{Schuch_2008}. Extensions to two-dimensional connectivity known as projected-entangled pair states (PEPS) \cite{verstraete2004renormalization} have been proposed, but these are generically hard to contract even in terms of average-case complexity \cite{Haferkamp_2020}. This reveals a broader challenge within the field of tensor networks, which is that generalizations beyond one-dimensional connectivity tend to incur exponentially hard classical contraction bottlenecks unless approximations or restrictions are applied \cite{Zalatel_2020}.

Quantum computers \cite{Feynman1986,Nielsen_Chuang_2010} present an alternative route to breaking the curse of dimensionality by encoding highly correlated ground states in entangled registers of qubits. An ideal quantum computer could in principle represent a state with arbitrarily large entanglement with no increase in the space requirement, and useful properties such as ground state energies may then be extracted via quantum phase estimation \cite{kitaev1995quantum,Aspuru_Guzik_2005,Su_2021,Ding_2023}. Given the significant technical challenges associated with preserving and manipulating an array of entangled qubits, algorithms designed for the current generation of noisy intermediate-scale quantum (NISQ) processors must utilize as little quantum resource as possible. Hybrid quantum-classical algorithms, such as the variational quantum eigensolver (VQE) and its contemporary variants \cite{Peruzzo2014,Tilly_2022,Grimsley2019,Ryabinkin2018,Mizukami_2020}, aim to achieve practical ground state energy estimates with extremely low depth quantum circuits at the expense of a greater measurement cost. While VQE algorithms have gathered interest due to their ease of quantum circuit implementation, the most widely studied ans\"atze based on single-reference unitary coupled cluster with single and double excitations (UCCSD -- we shall refer to related approaches as UCC-type) suffer from poor treatment of strong correlation, expensive gradient calculations, and barren plateaus in the training landscape \cite{Tilly_2022,McClean_2018}. The lack of strong correlation can be ameliorated by multi-reference methods such as the recently developed non-orthogonal quantum eigensolver (NOQE) \cite{PRXQuantum.4.030307}, which promises a quantum advantage for systems with finite numbers of radical sites exhibiting both strong and weak electron correlation. Furthermore, optimizing the UCCSD ansatz requires a two-qubit gate count per QPU call of $O(N^4)$, where $N$ is the number of molecular orbitals, which may be prohibitively expensive for medium-sized or large systems. While tensorial decompositions offer lower order polynomial scaling in practice \cite{motta_low_2021,PRXQuantum.4.030307}, and there is numerical evidence that adaptive methods such as ADAPT-VQE \cite{Grimsley2019} can reduce gate counts considerably, these cost reductions are system dependent and the worst-case scaling for the most challenging systems may still be quartic in $N$.

Given that quantum circuits are tensor networks with unitary constraints, several recent works \cite{bauer_quantum_2020,haghshenas_variational_2022,kim_robust_2017,huggins_towards_2019,borregaard_noise-robust_2021,jamet2023anderson,Lubasch_2020,Keever_2023,Huang_2023,Watanabe_2024,Miao_2023,miao2023convergence} have suggested that classical tensor network methods may be adapted to generate more compact and robustly trainable quantum circuit ans\"atze. In return, quantum computers may help to resolve the computational bottlenecks in the contraction of higher-dimensional tensor networks. Here we present a new hybrid quantum-classical algorithm, which we call a tensor network quantum eigensolver (TNQE), that solves for chemical ground state energies by constructing a highly compact wavefunction from a superposition of $M$ matrix product states of fixed bond dimension, $\chi$, expressed in rotated orbital bases. We show that computing expectation values with such a construction by classical tensor network contraction is exponentially costly in the general case, but is however easily translated into quantum circuits of depth $O(N)$. We emphasize that although the TNQE method employs a hybrid quantum-classical parameter optimization routine which satisfies the Rayleigh-Ritz variational principle, it is not a `VQE method' in the typical sense, in which a parameterized quantum circuit is optimized using energy gradients. In contrast, it is a gradient-free approach that iteratively sets up and solves a generalized eigenvalue problem in a basis of non-orthogonal matrix product states. It can therefore be regarded as a member of the emerging family of quantum subspace diagonalization (QSD) methods \cite{McClean_2017,PRXQuantum.4.030307,Huggins_2020,PhysRevA.105.022417,francis2022subspace,motta2023subspace}, which use a QPU to evaluate the Hamiltonian and overlap matrix elements within a subspace of low energy state vectors, and as such may also be extendable to the calculation of low-lying excited states.

We know of several existing proposals for MPS-based ans\"atze within the VQE paradigm \cite{Fan2023,Rudolph2023,haghshenas_variational_2022,Meth_2022,khan2023preoptimizing}, which fall into two main categories. The first of these uses a single parameterized quantum circuit ansatz that encodes a matrix product state of bond dimension $\chi$, which can be efficiently prepared by quantum circuits with gate count at most $O(N\chi^2)$ \cite{Fan2023,haghshenas_variational_2022,Meth_2022}. In general, this requires the quantum many-body system under study to obey a one-dimensional area law of entanglement, unless some additional restriction on the form of the MPS tensors is enforced. It has therefore been argued that these single-reference MPS methods can at best achieve a polynomial advantage over classical optimizers \cite{haghshenas_variational_2022}. The second category uses an MPS as a classically pre-optimized initial state which is then augmented by a conventional parameterized quantum circuit, in an attempt to `short-cut' the non-linear VQE optimization past barren plateaus and local minima \cite{Rudolph2023,khan2023preoptimizing}. There is no guarantee of successfully side-stepping local minima by this method, although it has been shown to lead to an improvement in the cases studied. The circuit depth required by a single parameterized circuit to characterize generic chemical ground states is not well established, but is expected to be at least super-linear if not super-polynomial in $N$ in the general case.

Our method advances on these prior works in two important respects. First, the superposition of single-particle bases extends the reach of the TNQE ansatz beyond states obeying a one-dimensional area law of entanglement in a systematic manner. This has not been attempted previously and may now allow for circumvention of the simulability constraints of single MPS-based ans\"atze \cite{Schuch_2008}. Second, our optimization routine is entirely gradient-free and can instead be considered as a natural generalization of the DMRG sweep algorithm to linear combinations of non-orthogonal matrix product states. Recent results have shown that the canonical form of the MPS, which is utilized by the sweep algorithm, can provide rigorous guarantees of the absence of barren plateaus in the training landscape \cite{miao2024isometric,barthel2023absence}. This may significantly reduce the overall measurement cost of ground-state energy estimation compared to conventional gradient-based variational optimizers \cite{CerveroMartin2023barrenplateausin}. The results presented in this first study are consistent with this prediction. We show that for ground states the energy can be reliably converged via iterative quantum subspace diagonalizations, using $O(NM^2\chi^4)$ calls to a QPU per sweep, with a gate count for each QPU call that scales at worst as $O(N^2 + N\chi^2)$ and with circuit depths linear in $N$. We achieve chemical accuracy in numerical tests on two small chemical systems -- a stretched water molecule and an octahedral arrangement of six hydrogen atoms -- empirically demonstrating both significantly better converged energy estimates and a far lower sensitivity to shot noise than VQE with a single UCCSD ansatz circuit, resulting in orders of magnitude reductions in the estimated quantum resources for the H$_6$ cluster, with even greater reductions expected for larger systems.

\section{Theory}

\subsection{Preliminaries}

Expressed in a basis of $N$ molecular orbitals, the second-quantized electronic structure Hamiltonian takes the form
\begin{align}
\hat{H} = \sum_{pq=1}^N h_{pq}\,\hat{a}^\dag_p \hat{a}_q + \sum_{pqrs=1}^N h_{pqrs}\,\hat{a}^\dag_p \hat{a}^\dag_q\hat{a}_r\hat{a}_s,
\label{eq:ham_def}
\end{align}
where the one- and two-body coefficients $h_{pq}$ and $h_{pqrs}$ encode one- and two-electron integrals over the single-particle molecular orbital basis, and the creation and annihilation operators act on the Fock space of $\eta$-electron Slater determinants $\ket{\vec{k}} = \ket{k_1,\ldots,k_N}$ \cite{Helgaker_2000}. Each orbital occupancy index $k_p$ has $d$ possible values, where $d=2$ with occupancy values 0 or 1 when expressed in terms of spin-orbital sites, or $d=4$ with binary occupancy values 00, 01, 10, or 11 for spatial orbital sites, corresponding to the local basis states $\ket{\text{vac}}$, $\ket{\uparrow}$, $\ket{\downarrow}$, or $\ket{\downarrow\uparrow}$, respectively. The ground state problem consists in finding a superposition of Slater determinants $\ket{\Psi}$ that minimizes the expectation value $E_0=\braket{\Psi|\hat{H}|\Psi}$. The full configuration interaction (FCI) coefficient tensor $\Psi^{k_1...k_N}$ encodes the coefficients exactly describing the ground state of the system,
\begin{align}
    \ket{\Psi_\text{FCI}} = \sum_{\vec{k}}^{d^N}\Psi^{k_1...k_N}\ket{\vec{k}}.
    \label{eq:fci}
\end{align}
The FCI coefficient tensor can be decomposed via tensor-train factorization \cite{doi:10.1137/090752286} -- a sequence of index groupings and singular value decompositions (SVDs) -- which is equivalent to representing the wavefunction in a tensor network format known as a matrix product state (MPS) or a tensor train,
\begin{align}
    \ket{\Phi_\text{MPS}} = \sum_{\vec{k}}^{d^N}\Big[\sum_{l_1...l_{N-1}}^{\chi} \Phi_{l_1}^{k_1}\Phi_{l_1l_2}^{k_2}\cdots\Phi_{l_{N-2}l_{N-1}}^{k_{N-1}}\Phi_{l_{N-1}}^{k_N}\Big]\ket{\vec{k}},
    \label{eq:mps}
\end{align}
\begin{figure}
\centering
\includegraphics[width=0.41\textwidth]{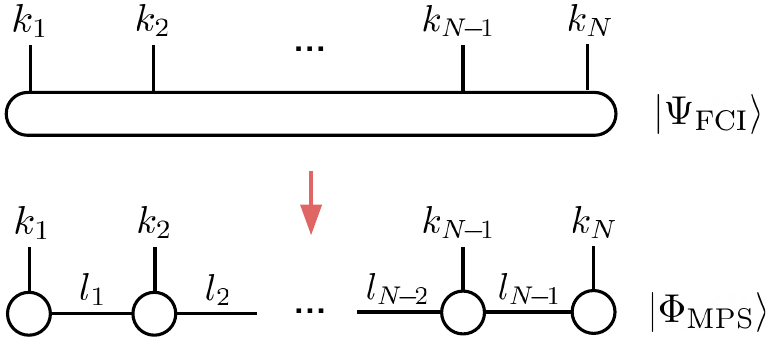}
\caption{A tensor network diagram depicting the tensor-train (TT) factorization of the FCI coefficient tensor in Eq. \ref{eq:fci}, resulting in a matrix product state defined in Eq. \ref{eq:mps}.}
\label{fig:tt_factor}
\end{figure}
where the $\Phi^{k_p}$ are the site tensors, each with $\chi^2d$ free parameters, so that the total number of tensor parameters is $O(N\chi^2d)$ (see Figure \ref{fig:tt_factor}). The generality of the SVD ensures that, before truncation of singular values, the MPS preserves all of the information in the FCI wavefunction and hence is an equivalent representation of the true ground state. The Schmidt decomposition allows quantification of the entanglement between any left block $A=\{1,\ldots,p\}$ and right block $B=\{p+1,\ldots,N\}$ of the MPS \cite{Vidal_2003}, which is written as
\begin{align}
\ket{\Phi_\text{MPS}} = \sum_{l=1}^{\chi}\sigma_{l}\ket{\Phi_{l}^A}\otimes \ket{\Phi_{l}^B},
\label{eq:schmidt}
\end{align}
where $\sigma_1\geq...\geq\sigma_\chi$ are the singular values, and the left and right Schmidt vectors have the important property $\braket{\Phi_l^A|\Phi_{l'}^A}=\braket{\Phi_l^B|\Phi_{l'}^B}=\delta_{ll'}$. The Schmidt rank is upper bounded by the bond dimension $\chi\leq \,\min{(d^{N_A},d^{N_B})}=d^{N/2}$ at the central bond. The MPS can be placed into this so-called canonical form by gauge transformations of the tensors, such that the canonical center is the diagonal matrix of singular values, as depicted in Fig. \ref{fig:isometric_form}a (see Appendix \ref{app:A}). The von Neumann entropy across the partition is then obtained as
\begin{align}
\mathcal{S}_{AB} = -\sum_{l=1}^{\chi}\sigma_l^2\ln\sigma_l^2.
\label{eq:vn_ent}
\end{align}
The memory cost can be reduced at the expense of accuracy by truncating $\chi$ at each bond, equivalent to projecting onto the first $\chi$ left and right Schmidt vector pairs. This truncation necessarily enforces a notion of locality to the ansatz; an MPS with fixed bond dimension $\chi\ll d^{N/2}$ obeys a one-dimensional area law of entanglement between the molecular orbitals, with maximal entropy between the left and right blocks equal to $\ln(\chi)$, seen by setting all singular values to be equal in Equations \ref{eq:schmidt} and \ref{eq:vn_ent}. In the MPS format expectation values can be extracted by efficient tensor network contraction with an analogous tensor-train factorization of the operator, known as a matrix product operator (MPO) \cite{chan_matrix_2016}, so that the sum over $d^N$ Slater determinants need not be explicitly computed. This is the basis for efficient classical ground state solvers, whereby an MPS is initialized with random parameters and optimized variationally to minimize the expected energy. This is typically by the one- or two-site DMRG sweep algorithm, wherein the canonical center is contracted over a single site or a pair of sites (Figs. \ref{fig:isometric_form}b and \ref{fig:isometric_form}c) which are then locally optimized \cite{chan_matrix_2016}.

\begin{figure}
\centering
\includegraphics[width=0.41\textwidth]{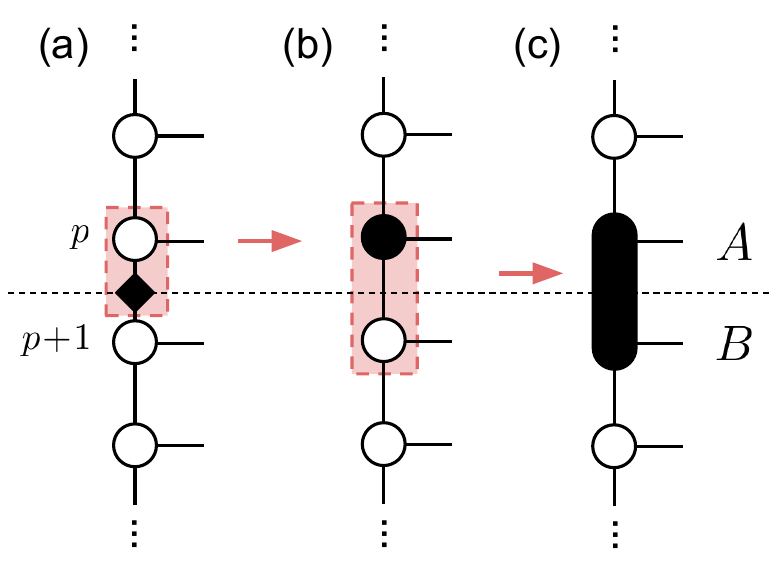}
\caption{Tensor network diagrams depicting a matrix product state in canonical form. The black tensor denotes the canonical center in each case. The arrows indicate contraction (a.k.a. `bubbling') of the highlighted tensors. (a) The canonical center (diamond-shaped tensor) is a diagonal matrix of singular values corresponding to the Schmidt decomposition over partitions $A$ (top) and $B$ (bottom) as in Eq. \ref{eq:schmidt}. (b) The canonical center is contracted into the single-site tensor at site $p$ (in $A$ for this example). (c) The canonical center is contracted into the two-site tensor at sites $p$ and $p+1$.}
\label{fig:isometric_form}
\end{figure}
\begin{figure}
\centering
\includegraphics[width=0.46\textwidth]{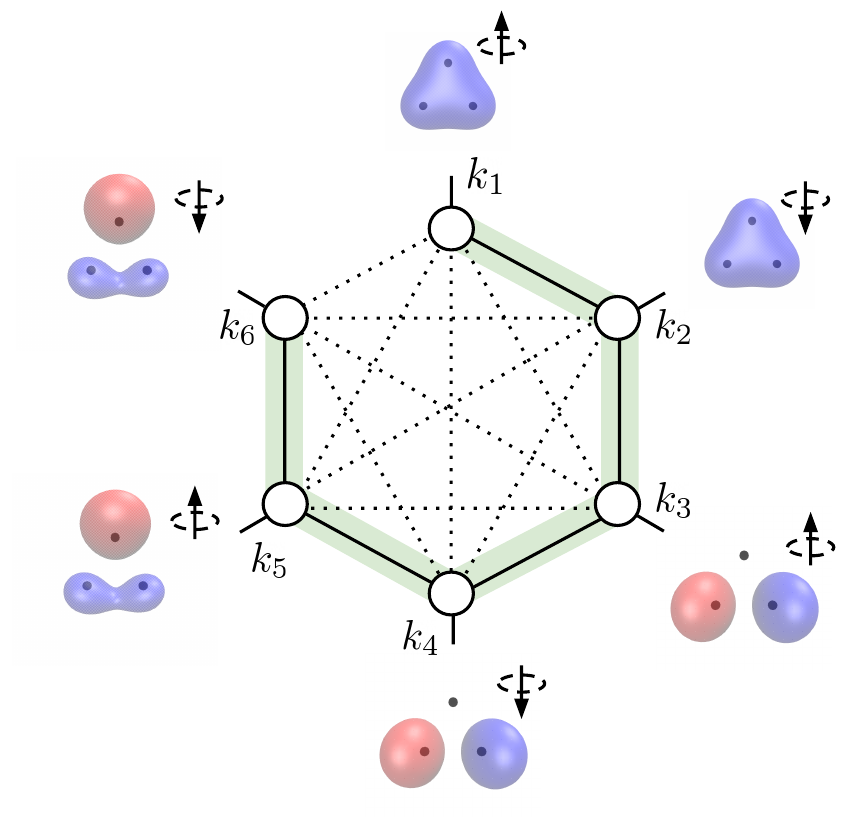}
\caption{A molecular orbital graph indicating all possible tensor network bond indices between $N=6$ spin-orbital sites. The solid lines denote the bond indices of a matrix product state as defined in Eq. \ref{eq:mps}. The restricted Hartree-Fock orbitals of the H$_3^+$ cation in the STO-3G basis set are provided as a minimal example of a closed-shell molecule with six spin-orbital sites.}
\label{fig:mo_graph}
\end{figure}
A more general tensor network state can be depicted by a graph with $N$ nodes representing the sites, in this case the molecular orbitals, and up to $N\choose 2$ edges representing the bonds, as in Fig. \ref{fig:mo_graph}. An MPS is then described by a path visiting each node via $N-1$ unique edges, a PEPS corresponds to a graph with $2N-2\sqrt{N}$ edges, and so on, up to a tensor network with full connectivity between all pairs of nodes. A fully connected tensor network is not a classically efficient description of a quantum state, requiring $O(N\chi^{N-1})$ tensor parameters. The set of all fully connected states also contains that of all PEPS, the exact contraction of which is \#P-hard \cite{Schuch_2007}, and is thus believed to be exponentially hard even for a quantum computer up to arbitrary precision. In practice, however, we are often more interested in approximate contraction up to an additive error, which is BQP-complete for tensor networks defined on graphs of bounded degree \cite{Arad_2010}.

Operators in the fermionic Fock space can be mapped onto the state space of a quantum register under the Jordan-Wigner transformation \cite{Jordan1928}, establishing a one-to-one correspondence between the spin-orbitals and the qubits, both of which are local Hilbert spaces of dimension $d=2$. The fermion operators are then represented as a linear combination of unitary Pauli strings,
\begin{align}
\hat{a}_p \mapsto (\hat{X}_p - i\hat{Y}_p)\hat{Z}_{p-1}\cdots\hat{Z}_1,
\end{align}
where $\hat{X}_p$ is short-hand for $\hat{\mathds{1}}^{\otimes p-1}\otimes\hat{X}\otimes\hat{\mathds{1}}^{\otimes N-p}$. The string of Pauli $\hat{Z}$ gates on the qubits $1,\ldots,p-1$ enforces the fermionic anticommutation relations. The electronic structure Hamiltonian in Eq. \ref{eq:ham_def} can then be fully specified in any orbital basis with $O(N^4)$ Pauli strings ($\hat{P}_\alpha$), written as
\begin{align}
\hat{H} = \sum_\alpha h_\alpha\hat{P}_\alpha.
\label{eq:pauli_decomp}
\end{align}
Under this representation the Slater determinants $\ket{\vec{k}}$ are equivalent to the computational basis vectors of the qubit register (also known as `bitstring' states). Near-term variational quantum eigenvalue solvers (VQEs) involve optimizing the parameters of a polynomial-size quantum circuit $\hat{U}$, which prepares some ansatz state $\ket{\phi} = \hat{U}\ket{0}$, to minimize the expectation value $\braket{\phi|\hat{H}|\phi}$ as computed from the Pauli decomposition in Eq. \ref{eq:pauli_decomp}. A common choice of ansatz circuit for quantum chemistry is fermionic unitary coupled cluster with single and double excitations (UCCSD) \cite{Tilly_2022}.

Alternatively, quantum multi-reference methods use a wavefunction ansatz constructed from a linear combination of efficiently preparable reference states,
\begin{align}
    \ket{\psi} = \sum_{j=1}^{M}c_j\ket{\phi_{j}},
    \label{eq:tnqe}
\end{align}
subject to the constraint that $\ket{\psi}$ is normalized and thus strictly variational. For example, the NOVQE method \cite{Huggins_2020} uses UCC-type reference states, where the gate rotation angles are computed using a hybrid quantum-classical variational gradient-based optimizer, while the NOQE method \cite{Baek_2022} also uses UCC-type reference states, but instead selects the circuit parameters using classical chemistry heuristics, avoiding all hybrid variational parameter optimization. Given a set of $M$ arbitrary reference states $\{\ket{\phi_j}\}_{j=1}^M$, the optimal coefficients $\{c_j\}$ to approximate the ground state of the Hamiltonian by their linear combination are obtained by solving the generalized eigenvalue problem
\begin{align} \label{eq:GEE}
    \textbf{HC}=\textbf{SCE},
\end{align}
where the elements of the matrix pencil ($\textbf{H}$,$\textbf{S}$) are given by 
\begin{align}
    H_{ij} = \braket{\phi_i|\hat{H}|\phi_j}, \quad S_{ij} = \braket{\phi_i|\phi_j}.
    \label{eq:mat_els}
\end{align}
The matrix $\textbf{E}$ is a diagonal matrix of generalized eigenvalues $E_1\leq\cdots\leq E_M$. The optimal coefficients corresponding to the ground state estimate $E_1$ are then given by the first column of $\textbf{C}$, i.e., $c_j=C_{j1}$. This technique of solving Eq. \ref{eq:GEE} on a classical computer with matrix elements obtained by a quantum computer has come to be known more broadly as quantum susbpace diagonalization \cite{motta2023subspace}.

In the following sections we will make frequent use of the direct equivalence between tensors subject to unitary constraints and quantum gates. In the figures we will use circles and rounded shapes to denote classical tensors in tensor network diagrams, and we will use squares and rectangles to denote quantum gates in circuit diagrams. For an in-depth introduction to tensor network notation and the relation to quantum circuits, see e.g. \cite{biamonte_tensor_2017,Bridgeman_2017}.
\subsection{TNQE ansatz}
\begin{figure}
\centering
\includegraphics[width=0.48\textwidth]{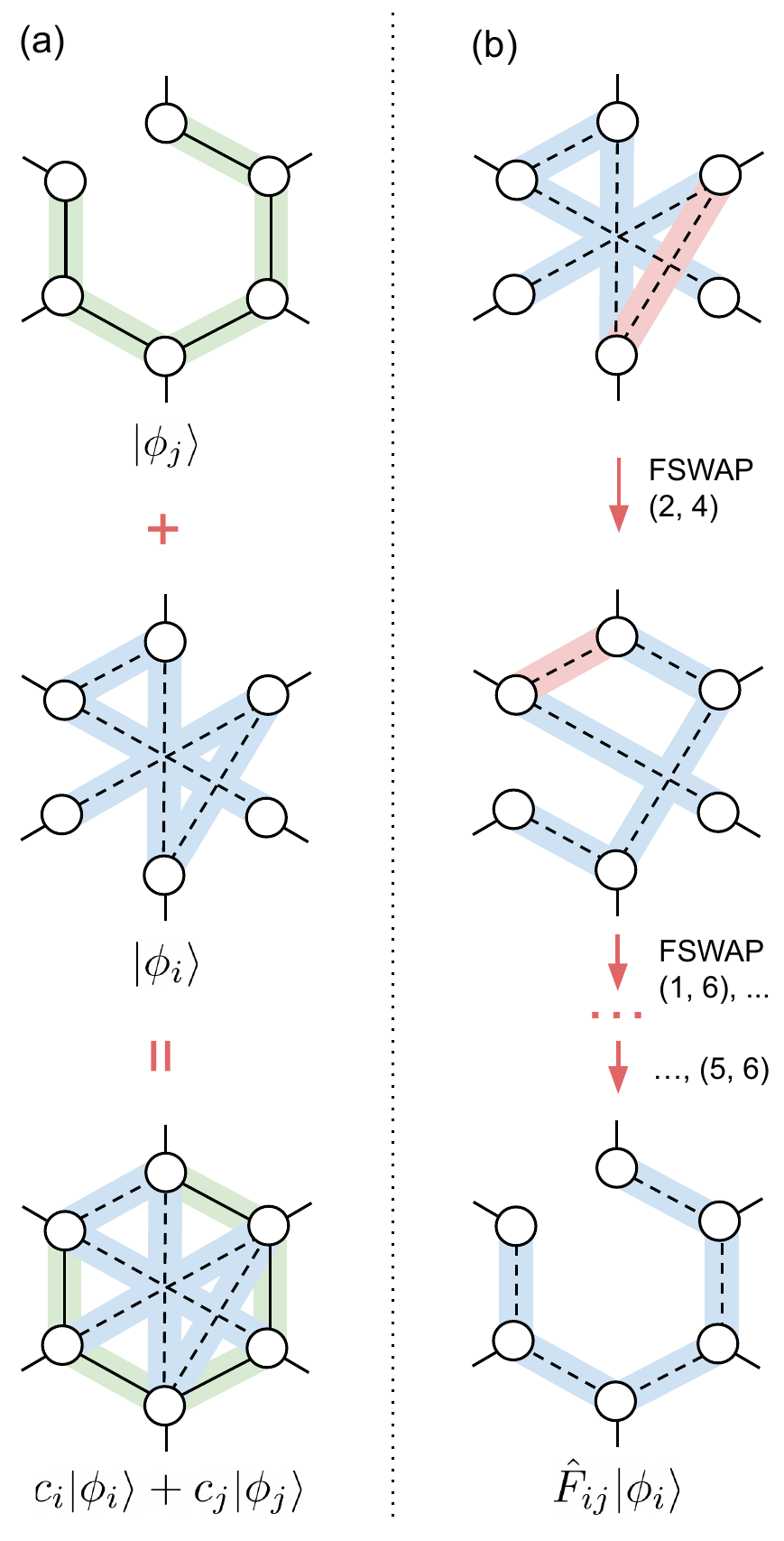}
\caption{Graph representations of matrix product states. (a) MPSs $\ket{\phi_j}$ and $\ket{\phi_i}$ with different site orderings over the same set of $N=6$ molecular orbitals, and the superposition state $c_i\ket{\phi_i}+c_j\ket{\phi_j}$. (b) A sequence of FSWAP operations $\hat{F}_{ij}$ transforms the MPS $\ket{\phi_i}$ into the orbital ordering of $\ket{\phi_j}$, giving rise to the tensor network in Fig. \ref{fig:mat_el}a.}
\label{fig:mo_graphs}
\end{figure}
The TNQE ansatz is a quantum multi-reference state, with the form of $\ket{\psi}$ in Equation \ref{eq:tnqe}, which is a linear combination of $M$ non-orthogonal matrix product states $\{\ket{\phi_j}\}_{j=1}^M$ of fixed bond dimension $\chi$, as defined in Eq. \ref{eq:mps}, expressed in \textit{different} orbital bases. Thus the TNQE wavefunction is specified by the $O(MN\chi^2)$ MPS tensor parameters, the choice of single-particle basis for each MPS, and the coefficients $\{c_j\}$, which are subject to the constraint that $\ket{\psi}$ is normalized. This wavefunction form is strictly variational, and can be seen to inherit size consistency from its closest classical analogues, namely the non-orthogonal configuration interaction \cite{Sundstrom_2014,Thom_2009} and orbital-optimized DMRG \cite{Krumnow_2016,Zgid_2008,Ghosh_2008}. While each MPS obeys a one-dimensional area law within its own single-particle basis, their superposition allows for the efficient description of a broader class of wavefunctions. This additional flexibility also incurs an exponential separation in complexity between the classical and quantum evaluations of the off-diagonal matrix elements in Eqs. \ref{eq:mat_els}, as we shall now demonstrate.

\begin{figure}
\centering
\includegraphics[width=0.47\textwidth]{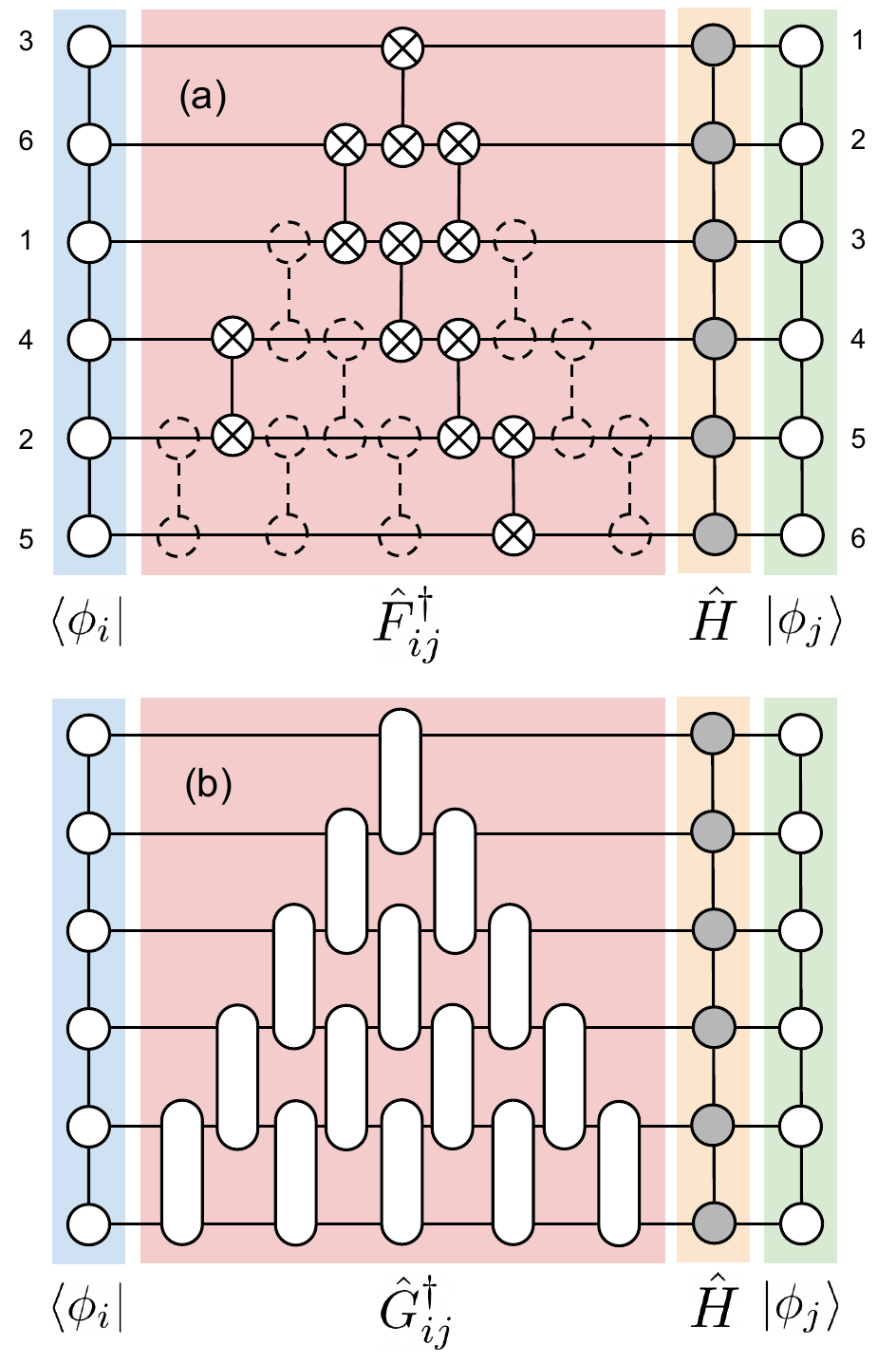}
\caption{Tensor networks to compute the off-diagonal matrix element $\braket{\phi_i|\hat{H}|\phi_j}$ between matrix product states with permuted or rotated orbitals, given an efficient MPO representation of $\hat{H}$. (a) The MPSs are related by a permutation over the same set of orbitals. The FSWAPs, denoted by pairs of crossed tensors, are selected by a bubble sort comparator network. Each comparator applies an FSWAP to rearrange neighboring orbitals if their order does not match the order in which they appear on the right. The dashed outlines indicate the positions of non-swapping comparators. (b) The MPSs are related by an arbitrary orbital rotation $\hat{G}_{ij}$, which is decomposed into a pyramidal structure of Givens rotation tensors  (see Eq. \ref{eq:givens_decomp}).}
\label{fig:mat_el}
\end{figure}

To gain some intuition we shall first introduce the restricted case of orbital permutations, i.e., the same set of orbitals but with different site orderings, before generalizing to arbitrary rotations of the single-particle basis. In this first case, the superposition in Eq. \ref{eq:tnqe} allows for the combination of matrix product states with different site connectivity, as shown in Figure \ref{fig:mo_graphs}a. The set of all states that can be described by Eq. \ref{eq:tnqe} using both $M$ and $\chi$ of size polynomial in $N$ may then be expected to lie somewhere in-between that of a single matrix product state and that of a fully connected tensor network. If the orbital sets were identical, then the off-diagonal matrix element computations in Equations \ref{eq:mat_els} would be tractable by classical tensor network contraction, assuming an efficient MPO representation of $\hat{H}$. However, when the orbital sets differ, additional tensors must be inserted to transform between the representations. An arbitrary permutation of the orbitals is achieved with the correct treatment of fermionic antisymmetry via a nearest-neighbor fermionic-SWAP (FSWAP) network \cite{Verstraete_2009}. The sequence of FSWAP operations $\hat{F}_{ij}$ to rearrange the orbital ordering of $\ket{\phi_i}$ to that of $\ket{\phi_j}$ is illustrated in Fig. \ref{fig:mo_graphs}b. Representing these operations with classical tensors results in the tensor network in Fig. \ref{fig:mat_el}a to compute the matrix element $\braket{\phi_i|\hat{H}|\phi_j}$, which has a cost of contraction that rises steeply with the depth of the FSWAP network. Any pair of orderings can be connected by a path-restricted sorting network of depth lower bounded by $O(N)$ \cite{banerjee_sorting_2019,Knuth98}, e.g., a bubble sort \cite{10.1145/320831.320833}. The application of each FSWAP gate to the MPS on the left can grow the bond dimension by up to a factor of $d$, so $\chi$ will rise exponentially with $N$ unless severe truncations are applied. This implies that a large amount of entanglement can be generated in the ansatz through the site rearrangement, even though each matrix product state is individually of low bond dimension. We note that while at first glance these tensor network contractions appear to be classically hard, the matrix elements can in principle be efficiently approximated by randomly sampling computational basis vectors from the matrix product state (see Appendix \ref{app:C}). This suggests the potential for a powerful, although limited, new class of quantum-inspired classical tensor network methods via random sampling. Because FSWAP networks belong to the Clifford subgroup, this is somewhat comparable to a multi-reference generalization of the recently proposed Clifford-augmented matrix product states \cite{Qian_2024,lami2024quantumstatedesignsclifford}. We may say that this permutation-only variant conveys the basic intuition behind TNQE but does not fulfill its potential as a hybrid quantum-classical method. 
\begin{figure}
\centering
\includegraphics[width=0.29\textwidth]{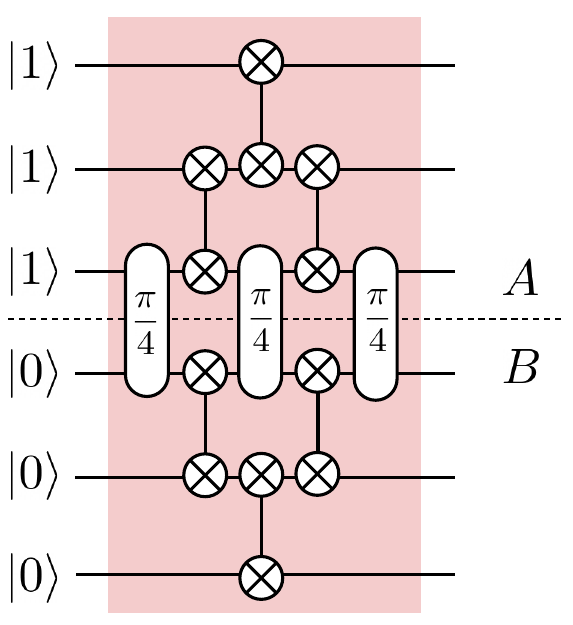}
\caption{A sequence of $\frac{\pi}{2}$ and $\frac{\pi}{4}$ orbital rotations transforms the unentangled state ${\ket{1}^{\otimes N_A}\otimes\ket{0}^{\otimes N_B}}$ to a state with maximal Schmidt rank and von Neumann entropy across the $A/B$ partition, where $N_A=N_B=N/2$, shown above for $N=6$ and explained in Appendix \ref{app:D}.}
\label{fig:mat_el2}
\end{figure}

We now show that the restriction to orbital permutations can be lifted to create a far more expressive ansatz under arbitrary rotations of the single-particle basis. Kivlichan \textit{et al.} \cite{Kivlichan_2018} have shown how to decompose an arbitrary rotation of the molecular orbitals $\hat{G}_{ij}$ into a sequence of nearest-neighbor Givens rotation gates of linear depth in $N$,
\begin{align}
\hat{G}_{ij} = \prod_{(p,\theta)\in \Theta_{ij}} \hat{g}_{p,p+1}(\theta),
\label{eq:givens_decomp}
\end{align}
where $\Theta_{ij}$ is a set of $N\choose 2$ pairs $(p,\theta)$ of site positions and rotation angles, and a Givens rotation between orbitals $p$ and $q$ with rotation angle $\theta$ is written as
\begin{align}
\hat{g}_{pq}(\theta) = \exp\left[ \theta (\hat{a}_p^\dag \hat{a}_q - \hat{a}_q^\dag \hat{a}_p) \right].
\end{align}
 This decomposition follows from a QR factorization of the $N\times N$ orbital rotation matrix, and is completely general as a consequence of the Thouless theorem \cite{THOULESS1960225}. As shown in Fig. \ref{fig:mat_el}, there is an obvious resemblance between the sequence of Givens rotations to perform an arbitrary orbital rotation and the comparator network that implements an arbitrary orbital permutation. The connection becomes clear when one considers that the FSWAP gate is a special case of a Givens rotation gate with rotation angle $\theta=\frac{\pi}{2}$, up to a change of phase on one of the orbitals (see Appendix \ref{app:A3}), so that in some sense the transformation $\hat{G}_{ij}$ can be thought of as a `quantum sorting network'. Although there is no longer an adequate orbital graph representation for the transformed state $\hat{G}_{ij}\ket{\phi_i}$, it is clear that any path rearrangement can be recovered with the appropriate set of discrete angles $\theta\in\{0,\frac{\pi}{2}\}$. The generalization to arbitrary continuous angles then implies a kind of smooth distribution over exponentially many possible paths.
 
In quantum mechanics, the entanglement of a system is dependent on the global basis in which it is measured; a system of low entanglement in one basis may have very high entanglement when expressed in a different basis. For instance, the tensor network in Fig. \ref{fig:mat_el2} depicts a linear-depth sequence of orbital rotations that transforms a completely unentangled bitstring state, which corresponds to a $d=2$, $\chi=1$ MPS, to a state with maximal Schmidt rank at the central partition, i.e. $\chi=2^{N/2}$ (this transformation is explained in Appendix \ref{app:D}). This appears to suggest that one may write down an MPS superposition ansatz with no classically efficient representation in any common orbital basis. It is argued in Appendix \ref{app:C} that in contrast to the restricted case of orbital permutations, the matrix elements between matrix product states differing by arbitrary orbital rotations cannot be efficiently evaluated via random sampling, so there is no known classical algorithm to efficiently compute Equations \ref{eq:mat_els} in this case.
\begin{figure*}
\centering
\includegraphics[width=\textwidth]{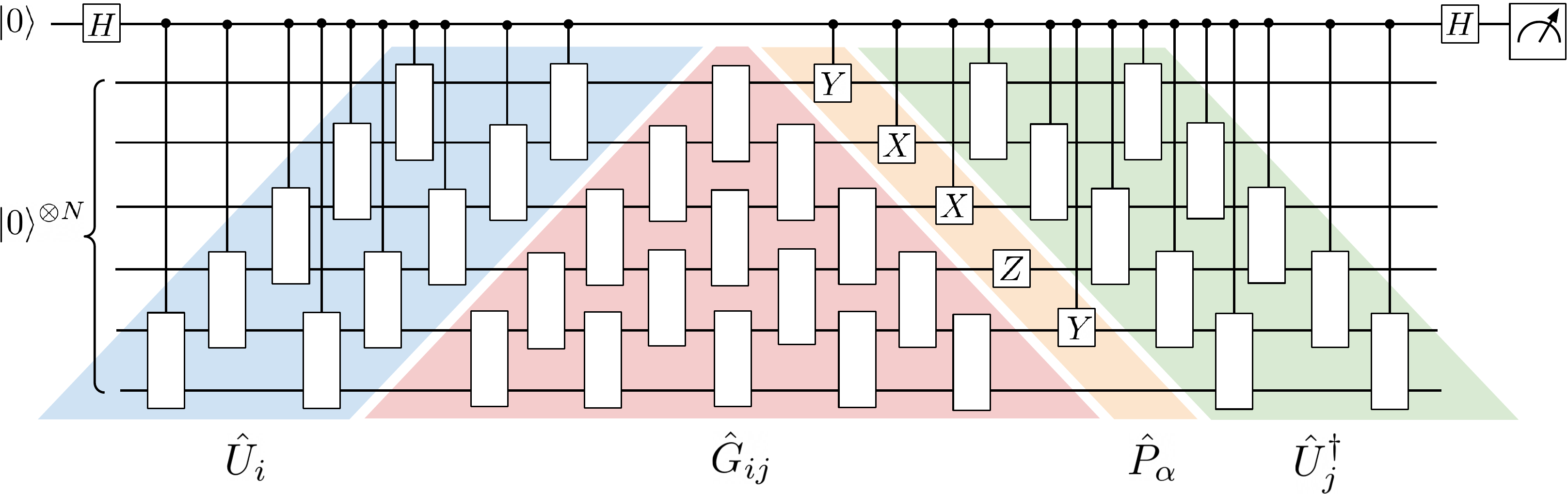}
\caption{A quantum circuit used to compute the matrix element $\braket{\phi_i|\hat{H}|\phi_j}$. This circuit requires controlled unitaries $\hat{U}_i$ and $\hat{U}_j$ to prepare $\ket{\phi_i}$ and $\ket{\phi_j}$ from the all-zero state in their respective orbital bases, an efficient Pauli string decomposition of $\hat{H}$ as in Eq. \ref{eq:pauli_decomp}, and an orbital rotation $\hat{G}_{ij}$ decomposed into Givens rotation gates as defined in Eq. \ref{eq:givens_decomp}. The details of the Hadamard test evaluation and the circuit construction are provided in Sections \ref{sec:meth_had} and \ref{sec:meth_circ}.}
\label{fig:qu_circ}
\end{figure*}

\subsection{TNQE algorithm}

Converging the expected energy of the TNQE asnatz requires an efficient optimization routine for both the tensor parameters and the orbital rotation angles. We introduce a generalized sweep algorithm (Algorithm \ref{gen_sweep_alg}) to co-optimize both of these quantities. The intuition behind this approach is that each MPS should `learn' those specific entanglement features which are most efficiently represented within its own single-particle basis. This means that the tensor parameters should be informed by the choice of orbital bases, and vice-versa.

First, we show how the DMRG sweep algorithm can be naturally extended to optimize the TNQE ansatz parameters via quantum subspace diagonalization (lines 4-7 of Algorithm \ref{gen_sweep_alg}), resulting in a gradient-free optimization strategy that converges reliably even in the presence of relatively large noise perturbations to the matrix elements. A key step in this procedure is the decomposition of the local two-site tensor ($T$) into a basis of `one-hot' tensors (line 4), which are tensors of the same dimension as $T$ that have
a single entry equal to 1 and all other entries 0 (see Section \ref{sec:parameter_opt} and
Appendix \ref{app:A1}). These span the space of all states that could be obtained by locally updating the MPS parameters. A QPU is needed to compute the expanded subspace matrix elements in line 5, with all of the additional co-processing steps being executed entirely on classical hardware. Second, it follows from Equation \ref{eq:givens_decomp} that any orbital rotation can be built up from pairwise orbital rotations, and in this work we explore this `bottom-up' perspective to manipulate the orbital entanglement during the MPS optimization. We show how the rotation angles can be classically co-optimized (line 8), with no additional QPU calls, to reduce the bond entanglement during the parameter sweep, enacting a `transfer' of entanglement between the bonds of the MPS and the orbital rotations. We note that similar strategies have been proposed to update the site ordering \cite{Li_2022} or the orbital basis \cite{Krumnow_2016} of a single MPS ansatz for use in classical methods.

Each sweep of Algorithm \ref{gen_sweep_alg} requires $O(NM^2\chi^4)$ matrix element evaluations to optimize all the states simultaneously, which scales favorably provided that $\chi$ is kept small and the resources of the ansatz are increased through $M$. In numerical experiments we have found it to be an effective strategy to increase $M$ in stages until convergence in the energy estimate is achieved. The first MPS ($M=1$) is constructed in orbitals obtained by a restricted Hartree-Fock calculation, and initialized with classical DMRG. At the next stage $(M=2)$, a new MPS is added in the same set of orbitals, with random tensor parameters. The entire subspace is then optimized by Algorithm \ref{gen_sweep_alg} for some number of sweeps. The same procedure is then repeated at each stage ($M=3,4,\ldots$), with each new MPS initialized with random tensor parameters, starting in the same set of orbitals as the previous MPS. With this iterative subspace construction, the TNQE algorithm is essentially a `black-box' routine, where in principle the system information is entirely supplied by the Hamiltonian coefficients, and the only free parameter is the fixed bond dimension cutoff $\chi$. In practice, however, some additional fine-tuning of the optimizer and system dependent parameter initializations may improve convergence, the exploration of which we leave to future work. A more detailed explanation of the TNQE algorithm with further derivations, intermediate steps, and lower-level pseudocode, is provided in Appendix \ref{app:A}.

\begin{algorithm}[H]
\caption{ \raggedright Generalized sweep algorithm (high level)}\label{alg:gen_sweep}
\begin{algorithmic}[1]
\For{$p=1,\ldots,N-1$}
\item[]
\State Put each MPS in canonical form centered at site $p$
\State Contract MPS two-site tensors ($T$) over sites $p,p+1$
\item[]
\State Decompose each $T$ into `one-hot' tensors
\State Query QPU for expanded subspace matrices $\textbf{H}'$, $\textbf{S}'$
\State Solve generalized eigenvalue problem on CPU
\State Update $T$ parameters from solution matrix $\textbf{C}'$
\item[]
\State Apply local two-site FSWAPs or orbital rotations
\State SVD of $T$ and truncate to $\chi$ singular values
\item[]
\EndFor
\end{algorithmic}
\label{gen_sweep_alg}
\end{algorithm}

\subsection{Matrix elements}

Figure \ref{fig:qu_circ} shows a quantum circuit used to efficiently evaluate the off-diagonal matrix elements in Equation \ref{eq:mat_els}, equivalent to the tensor network contraction in Fig. \ref{fig:mat_el}b, given linear qubit connectivity on the system register and all-to-one connectivity to a single ancilla qubit. By measuring this ancilla in the $z$-basis, one obtains the matrix elements required in line 5 of Algorithm \ref{gen_sweep_alg} up to standard error $\delta$ with $O(\delta^{-2})$ repetitions (see Section \ref{sec:meth_had}). This circuit requires a controlled state preparation unitary for each MPS. There are several proposals for MPS quantum circuit encodings \cite{Schon_2005,fomichev2024initial,ran_encoding_2020,Malz_2024}, the simplest exact approach being the direct synthesis of unitary encodings of each MPS tensor \cite{Schon_2005,fomichev2024initial}, with a total gate count that scales at worst as $O(N\chi^2)$, following standard asymptotic results for generic unitary synthesis \cite{Shende_2006,low2018trading}. For near-term hardware we present here a cost analysis based on the approximate disentangler technique pioneered by Ran \cite{ran_encoding_2020} and developed by others \cite{dov2022approximate,Rudolph_2024}, but in principle any encoding scheme may be substituted. This technique is briefly explained in Section \ref{sec:meth_circ}, wherein we show how to cheaply add controls to the disentanglers, then decompose the entire circuit into CNOT gates and single-qubit rotations to derive an exact two-qubit gate count for each circuit in terms of the number of qubits $N$ and the disentangler depth $D$, resulting in
\begin{align}
    n_\text{CNOT} = N^2 + (16D-1)N-16D.
    \label{eq:CNOT}
\end{align}
We have omitted the cost of the controlled Pauli string, which will be constant in $N$ for any $k-$local fermionic observable since the Pauli $\hat{Z}$ strings of the Jordan-Wigner mapping act trivially on the vacuum state. The scaling of $D$ with $\chi$ will vary depending on the MPS parameters, but is not expected to be more expensive than $O(\chi^2)$. The intuition behind the TNQE ansatz, which may be tested empirically, is that $\chi$ can be made constant in $N$, provided that the non-local correlations are recovered through the inclusion of multiple references in different orbital bases. This comes at the cost of a greater number of MPS reference states, $M$, which appears as a quadratic factor in the number of matrix elements, and hence in the overall measurement cost. With this choice of fixed-bond ansatz, Eq. \ref{eq:CNOT} presents a substantive advantage over existing ans\"atze, such as the UCC-type ans\"atze, that are characterized by a two-qubit gate count per circuit that scales as $O(N^4)$ \cite{Tilly_2022}.

With a single ancilla qubit, the layer depth in terms of CNOT gates and single-qubit rotations is given by
\begin{align}
L = 28ND + 17
\label{eq:layerd}
\end{align}
(see Section \ref{sec:meth_circ}). This results from having to apply each controlled disentangler in sequence, introducing an unwanted scaling of $O(ND)$. If, however, the hardware is of sufficient fidelity to prepare and maintain a coherent superposition of the all-zero and all-one states over $N/2$ qubits, known as a Greenberger-Horne-Zeilinger (GHZ) state, then this limitation can be circumvented at the cost of using $N/2$ ancillas. With this modification up to $N/2$ controlled rotations can be applied in parallel with control from different ancilla qubits. The layer depth is then given by 
\begin{align}
L_\text{GHZ} = 28N + 56D - 13, 
\label{eq:ghz_layerd}
\end{align}
which may be worth the additional overhead of $N/2$ ancillas and preparation of the GHZ state. In either case the depth of the circuit grows only linearly in the number of qubits for a fixed bond dimension $\chi$.

\subsection{Parameter optimization}
\label{sec:parameter_opt}

\begin{figure*}
\centering
\includegraphics[width=0.96\textwidth]{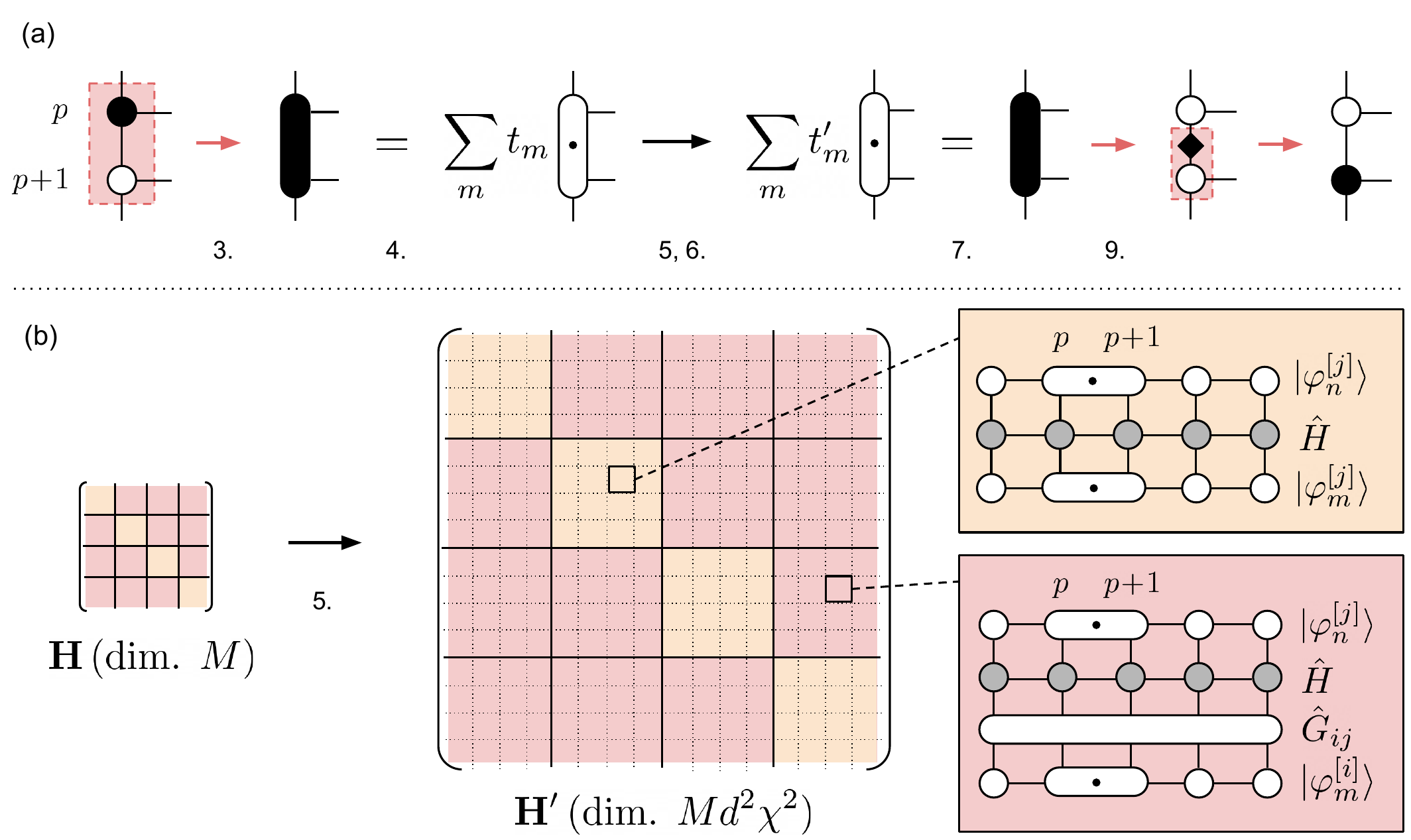}
\caption{Tensor network schematics for the two-site tensor parameter updates in the generalized sweep algorithm. (a) Local tensor operations corresponding to the lines of Algorithm \ref{gen_sweep_alg}, omitting the orbital update step. The canonical center at site $p$ (in black) is contracted into the two-site tensor $T$ (line 3), which is decomposed over $d^2\chi^2$ one-hot (dotted) tensors as in Equation \ref{eq:oht_exp} (line 4). After solving the generalized eigenvalue problem in the expanded subspace (lines 5, 6) the two-site tensor parameters are updated according to Eq. \ref{eq:param_upd} (line 7). The two-site tensor is then decomposed back into single-site tensors by SVD (line 9). (b) The Hamiltonian subspace matrix $\textbf{H}$ is expanded over one-hot tensor decompositions of the two-site tensors at sites $p$ and $p+1$, as in line 5 of Algorithm \ref{gen_sweep_alg} and Eq. \ref{eq:subs_exp} (for legibility, the number of elements depicted in each matrix block is not to scale). The tensor networks on the right compute the matrix elements in the diagonal ($i=j$) and off-diagonal ($i\neq j$) blocks. The expanded overlap matrix $\textbf{S}'$ is computed by similar tensor networks, but removing the Hamiltonian MPO. Each off-diagonal matrix element can be efficiently obtained using a quantum circuit encoding of the tensor network as in Fig. \ref{fig:qu_circ}.}
\label{fig:mat_exp}
\end{figure*}

Matrix product states are commonly optimized to lower the expected energy by a two-site sweep algorithm, whereby the MPS is put into canonical form centered on a two-site tensor over the sites $p$ and $p+1$ (Figure \ref{fig:isometric_form}c), which are then contracted, optimized, and decomposed back into site tensors. Here we explain how to generalize the conventional two-site algorithm to a linear combination of non-orthogonal matrix product states.  The key idea behind the generalized sweep algorithm is to expand each MPS into an orthonormal basis that represents the local degrees of freedom at sites $p$ and $p+1$. This follows from an expansion of the local two-site tensor as a sum over a set of orthonormal tensors, together with the canonical form of the MPS. By solving a generalized eigenvalue problem, like that in Eq. \ref{eq:GEE}, in the expanded subspace, a new set of optimal parameters are obtained for each two-site tensor to minimize the new ground state energy estimate $E_1'\leq E_1$.

More concretely, each two-site tensor $T$ is decomposed as a sum over the set of all unique tensors with the same indices as $T$ and a single element equal to 1, with all other elements equal to 0. We will refer to these as `one-hot' tensors, analogous to one-hot vectors that are common in computer science \cite{harris_digital_2010}. This sum over the one-hot tensors, indexed by $m$, is initially weighted by the two-site tensor parameters $\{t_m\}$ (see Fig. \ref{fig:mat_exp}a). We will further define the one-hot state, $\ket{\varphi^{[j]}_m}$, obtained by replacing $T$ in $\ket{\phi_j}$ with the unique one-hot tensor indexed by $m$. Because the MPS is in canonical form  centered at the sites $p$ and $p+1$, the one-hot states represent a decomposition of $\ket{\phi_j}$ into an orthonormal basis,
\begin{align}
    \ket{\phi_j} = \sum_{m=1}^{d^2\chi^2} t^{[j]}_m\ket{\varphi^{[j]}_m}, \quad
    \braket{\varphi^{[j]}_m|\varphi^{[j]}_n} = \delta_{mn}, \quad \sum_{m}\big|t^{[j]}_m\big|^2=1,
    \label{eq:oht_exp}
\end{align}
which spans the space of all possible matrix product states that could be obtained by locally updating the tensor parameters at sites $p$ and $p+1$ with any values satisfying the normalization condition. The optimal values of the tensor parameters can then be found by diagonalizing in this expanded basis. Each MPS can be decomposed simultaneously from two-site tensors located at the same or different sites, resulting in a new matrix pencil $(\textbf{H}',\textbf{S}')$ in the expanded subspace of dimension $Md^2\chi^2$, with matrix elements given by
\begin{align}
H'_{im,jn} = \braket{\varphi^{[i]}_m|\hat{H}|\varphi^{[j]}_n}, \quad S'_{im,jn} = \braket{\varphi^{[i]}_m|\varphi^{[j]}_n}.
\label{eq:subs_exp}
\end{align}
Equation \ref{eq:subs_exp} describes the simultaneous expansion of all reference states, such that $im$ denotes the combined row or column index $(i-1)d^2\chi^2 + m$, as in Fig. \ref{fig:mat_exp}b. One may also choose to optimize only a single MPS, or any subset thereof. After solving the generalized eigenvalue problem in the expanded subspace, 
\begin{align}
\textbf{H}'\textbf{C}'=\textbf{S}'\textbf{C}'\textbf{E}',
\label{eq:exp_gen_eig}
\end{align}
the new optimal coefficients $\{t'^{[j]}_m\}$ for each two-site tensor are obtained by slicing and normalizing from the first column of the expanded solution matrix $\textbf{C}'$,
\begin{align}
t'^{[j]}_m = \frac{C'_{jm,1}}{\sqrt{\sum_n|C'_{jn,1}|^2}}.
\label{eq:param_upd}
\end{align}
Further details and derivations of Eqs. \ref{eq:oht_exp}, \ref{eq:subs_exp}, and \ref{eq:param_upd} are provided in Appendix \ref{app:A1}. Optimizing all states simultaneously requires 
$O(M^2\chi^4)$ matrix element evaluations per site decomposition, so a full sweep over all sites requires $O(NM^2\chi^4)$ matrix elements, each of which may be efficiently obtained by a quantum circuit as in Fig. \ref{fig:qu_circ}. In practice, the number of one-hot tensors can be reduced by enforcing particle number and $z$-spin conservation through an internal block-sparse structure of the tensors \cite{Singh_2010,Singh_2011}. The effect of ill-conditioning of the generalized eigenvalue problem has been rigorously studied in Ref. \cite{epperly_2021}. In practice the condition number of $\textbf{S}'$ can be effectively regulated by discarding subspace vectors with a high degree of linear dependence (see Section \ref{sec:meth_reg}).

\subsection{Orbital rotations}

\begin{figure}
    \centering
    \includegraphics[width=0.42\textwidth]{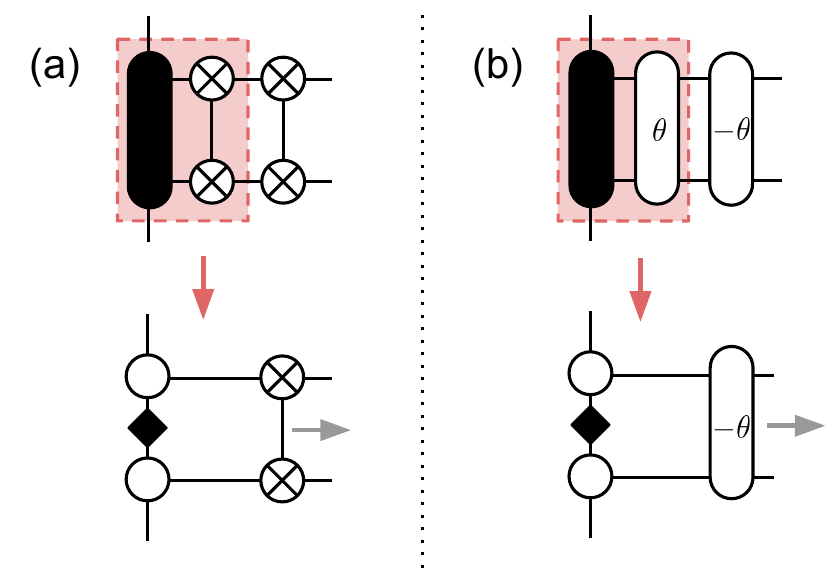}
    \caption{Tensor network schematics for the orbital rotation heurisitics. (a) A schematic of the FSWAP insertion. The vertical arrow denotes the combined operations of contraction into the two-site tensor followed by SVD in tensor network diagrammatic notation \cite{biamonte_tensor_2017,Bridgeman_2017}. The horizontal arrow indicates merging into $\hat{G}_{ij}$ if the update is accepted. (b) A schematic of the steps to compute the truncation error cost function $\xi(\theta)$ in Eq. \ref{eq:theta_opt} for the classical univariate optimization to determine $\theta_\text{opt}$. The adjoint rotation with angle $-\theta_\text{opt}$ is then merged into $\hat{G}_{ij}$.}
    \label{fig:orb_rot}
\end{figure}

After each parameter update the optimized two-site tensors are decomposed into single-site tensors via SVD with a maximum Schmidt rank of $d\chi$, which is then truncated to the $\chi$ largest singular values. At this step some information is discarded by projecting onto the space of the first $\chi$ left and right singular vector pairs of the Schmidt decomposition (see Equation \ref{eq:schmidt}). Given that entanglement is a basis-dependent quantity, it can be reduced before the truncation step by a local rotation of the orbitals $p$ and $p+1$ without loss of information. Our pairwise orbital co-optimization strategy is similar to previous approaches that have been developed to reduce the bond truncation error of a single MPS reference state, either by swapping neighboring sites \cite{Li_2022}, or by applying a local pairwise orbital rotation \cite{Krumnow_2016}. The procedure is outlined with tensor network diagrams in Figure \ref{fig:orb_rot}. Consider first the restricted case of a pairwise orbital permutation. The double application of an FSWAP gate is equivalent to the identity and so does not alter the state encoded by the two-site tensor. One of the FSWAPs is then contracted into the two-site tensor on the left prior to the SVD and the truncation error is computed as 
\begin{align}
\xi = 1 -\sum_{l=1}^{\chi}\sigma^2_l.
\end{align}
The orbital rearrangement is accepted if the truncation error at the bond is reduced by the FSWAP contraction, in which case the second FSWAP is then merged into $\hat{G}_{ij}$ by updating the $N\times N$ orbital rotation matrix and re-computing the rotation angles. Otherwise, the FSWAP insertion is rejected, and the two-site tensor parameters are reverted to their original values. A similar method to reduce the bond entanglement by swapping neighboring site indices was studied in Ref. \cite{Li_2022} for dynamical simulations in a non-fermionic context, which is equivalent to substituting the generic SWAP gate in the procedure above.

To apply an arbitrary pairwise orbital rotation, the unitary $\hat{g}(\theta)$ is classically optimized to find the value of $\theta_\text{opt}$ that minimizes the truncation error,
\begin{align}
\theta_\text{opt} = \text{arg}\min_{\theta}[\xi(\theta)].
\label{eq:theta_opt}
\end{align}
This is a purely classical co-optimization step with a cost $O(\chi^3)$ that derives solely from performing the SVD on the contracted two-site tensor, thus requiring no additional QPU calls or energy gradient calculations. The adjoint rotation $\hat{g}^\dag(\theta_\text{opt}) = \hat{g}(-\theta_\text{opt})$ is then merged into $\hat{G}_{ij}$. In a sense this unitary `encodes' a part of the information from the optimized two-site tensor, which is then merged into the basis rotation instead of being discarded by the singular value truncation.The key insight behind this approach is that switching the focus of the orbital optimization from directly minimizing the energy estimate to instead maximizing the fidelity of the wavefunction by reducing the truncation error, in combination with a DMRG-like optimization procedure for the MPS, can allow for more rapid and robust convergence of the orbital rotation angles. Essentially the same idea was proposed for orbital optimization in a fully classical context with a single MPS reference in Ref. \cite{Krumnow_2016}. In this classical setting only a single orbital basis must be updated, which amounts to a transformation of the fermionic Hamiltonian coefficients. By contrast, our multi-reference ansatz requires a set of distinct orbital rotation operators, $\hat{G}_{ij}$, to transform between every pair of orbital bases. Note that in line 8 of Algorithm \ref{gen_sweep_alg} only a single Givens rotation angle is optimized for each MPS in order to minimize its local bond truncation error. However, when this new local orbital rotation is `merged' into each of the $\hat{G}_{ij}$, every $N\times N$ orbital transformation matrix must be updated separately, after which, in order to factorize them back into Givens rotation circuits as in Eq. \ref{eq:givens_decomp}, all of the rotation angles need to be re-computed, using the technique of Ref. \cite{Kivlichan_2018}.

In practice it is observed that interleaving orbital permutation sweeps with orbital rotation sweeps is highly effective at breaking the optimizer out of local minima. This corresponds to alternating between rearranging the orbitals to different site positions via nearest neighbor swapping, and mixing of the orbitals through nearest neighbor quantum superpositions. After the rotation step it is found to be beneficial to further mitigate the penalty in the expected energy incurred by singular value truncation using a sequence of single-site decompositions, with no additional matrix element computations required, as explained in Section \ref{sec:meth_mit} and Appendix \ref{app:A4}.

\section{Results}

\subsection{Numerical benchmarking}
Here we evaluate the performance of the TNQE method on a stretched water molecule and on an octahedral arrangement of six hydrogen atoms. All calculations are performed in the STO-3G minimal basis set, corresponding to seven spatial molecular orbitals for H$_2$O and to six spatial orbitals for octahedral H$_6$. This would be equivalent to running quantum circuits on system registers of 14 and 12 qubits respectively under the Jordan-Wigner mapping. We have chosen to focus on spin-restricted rotations over the spatial orbitals; we suggest experimentation with spin-symmetry breaking as a topic of future work. We benchmark the method against classical DMRG calculations and against a simple sparse matrix implementation of VQE with a single UCCSD ansatz circuit (VQE-UCCSD) under a first-order Trotter decomposition. For details of the numerical calculations see Section \ref{sec:meth_num}.
\begin{figure*}
\centering
\includegraphics[width=\textwidth]{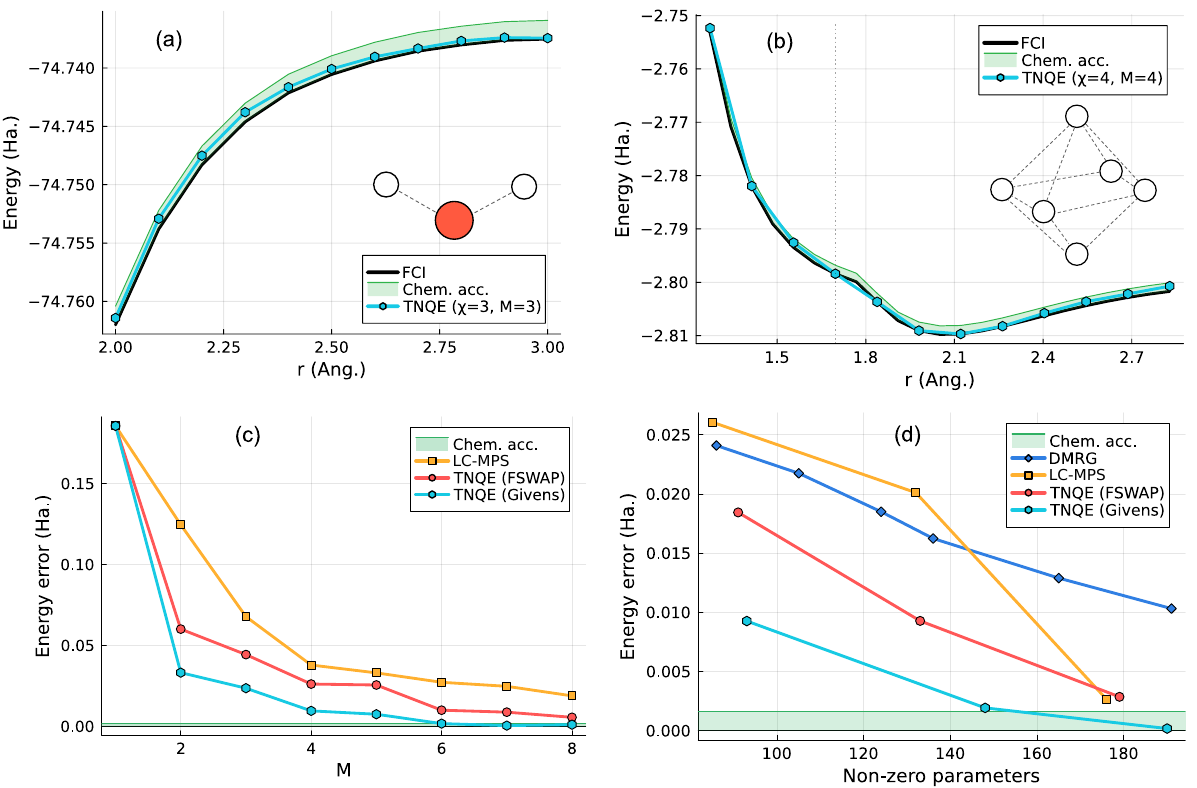}
\caption{Performance results from numerical calculations in the STO-3G basis set. (a) TNQE ($\chi=3$, $M=3$) energies for a stretched H$_2$O molecule from $r=2\text{\AA}$ up to $r=3\text{\AA}$. (b) TNQE ($\chi=4$, $M=4$) energies for octahedral H$_6$ from $r=1.13\text{\AA}$ up to $r=2.83\text{\AA}$. The `kink' in the FCI curve at $\sim1.75\text{\AA}$ is an artefact of the minimal basis set. (c) Comparison of LC-MPS, TNQE-F (FSWAP), and TNQE-G (Givens) energies at $\chi=3$ with increasing $M$ for octahedral H$_6$ at $r=1.70\text{\AA}$. (d) Comparison of DMRG, LC-MPS, TNQE-F (FSWAP), and TNQE-G (Givens) energies with increasing non-zero parameter count for octahedral H$_6$ at $r=1.70\text{\AA}$. The DMRG bond dimension is increased from $\chi=8$ to $\chi=13$, while the TNQE bond dimension is fixed at $\chi=4$ and $M$ is increased from 2 to 4.}
\label{fig:performance_plots}
\end{figure*}

\begin{figure}
\centering
\includegraphics[width=0.49\textwidth]{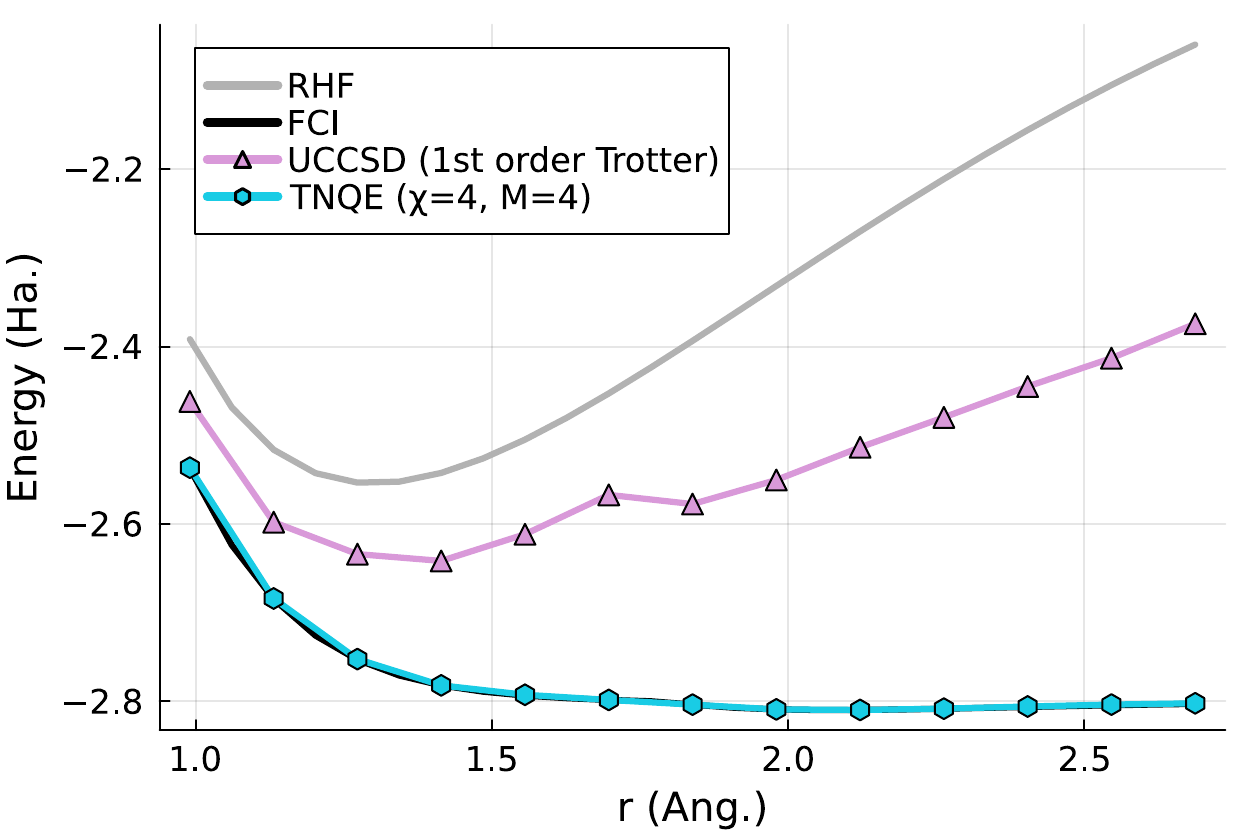}
\caption{TNQE ($\chi=4$, $M=4$) energies for octahedral H$_6$ in the STO-3G basis compared against RHF and VQE-UCCSD (which computes the single-reference UCCSD energy)} from $r=0.99\text{\AA}$ up to $r=2.69\text{\AA}$. The FCI curve appears underneath the TNQE curve at this energy scale. The `kink' in the UCCSD energy curve at $\sim1.75\text{\AA}$ is an artefact arising from the minimal basis set.
\label{fig:pe_curve_1}
\end{figure}

In Figure \ref{fig:performance_plots}a we demonstrate that TNQE can be reliably converged to chemical accuracy for the stretched water molecule with O---H bond length between $2$-$3\text{\AA}$, with bond dimension $\chi=3$ and subspace dimension $M=3$. This dissociation region is known to be challenging for UCC-type ans\"atze \cite{Mizukami_2020}. The ease of convergence with low computational resources suggests that systems with weak to moderate amounts of electron correlation are not challenging for TNQE. In Fig. \ref{fig:performance_plots}b we show that TNQE also reliably attains chemical accuracy for the more strongly correlated octahedral H$_6$ system with $\chi=4$ and $M=4$ across a wide range of internuclear separations, demonstrating that TNQE can also handle cases of much stronger electron correlation with a very manageable increase in the computational resources. Fig. \ref{fig:pe_curve_1} shows that the TNQE energy is drastically more accurate than the VQE-UCCSD energy estimate for this system across all bond lengths.

To assess the prospects for quantum advantage, the convergence of TNQE with increasing subspace dimension $M$ is compared against two classically tractable variants of the method, the first of which removes all of the orbital rotations and permutations. We refer to this variant, which is tractable by classical tensor network contraction, as linear combinations of matrix product states (LC-MPS). The second variant restricts the orbital rotations to orbital permutations with FSWAP networks. As explained in Appendix \ref{app:C}, this variant is in principle dequantizable by classical random sampling. We will refer to this variant in the text as TNQE-F. The TNQE ansatz using Givens rotations with arbitrary angles is not tractable using any known classical algorithm, and we refer to this as TNQE-G. Figure \ref{fig:performance_plots}c shows a clear advantage for TNQE-G over both LC-MPS and TNQE-F in the octahedral H$_6$ system at $1.70\text{\AA}$ with $\chi=3$. While TNQE-G touches the threshold for chemical accuracy at $M=6$, the dequantizable TNQE-F and classical LC-MPS sit at $\sim10$ mHa and $\sim30$ mHa respectively for $M=6$, and have still not converged to chemical accuracy at $M=8$.

In Fig. \ref{fig:performance_plots}d TNQE is benchmarked against classical DMRG by directly comparing the total number of tensor parameters in the non-zero blocks: this number scales as $O(N\chi^2)$ for DMRG and as $O(MN\chi^2)$ for TNQE. For this example, we scale up the parameters in the TNQE variants at a fixed value of $\chi=4$ by increasing $M$ from 2 to 4. For the DMRG curve we have $M=1$ by definition and we increase $\chi$ from 8 up to 13. For the octahedral H$_6$ system at $r=1.70\text{\AA}$ we observe that TNQE achieves far better energies at low parameter counts. However we are careful not to claim this as proof that TNQE outperforms DMRG for this small system, and we mention several caveats to this comparison. First, at larger bond dimensions $(\chi\geq 15)$ the DMRG curve converges almost exactly to the FCI energy, with a slightly decreased parameter count of $\sim180$, roughly on par with the number of parameters for TNQE at $\chi=4$, $M=4$. This may be due to a rearrangement of the internal block-sparse structure at larger bond dimensions. It may also be possible to improve the convergence of the DMRG curve with localized orbitals or other classical techniques. There is also some ambiguity regarding the inclusion of the Givens rotation angles in the total parameter count of TNQE. These are not `free' parameters in the TNQE ansatz and are mutually interdependent, so it is not clear how or whether they should be counted. Nonetheless, the results at low parameter counts are encouraging for the prospect of observing a more decisive advantage in larger strongly correlated systems. Furthermore, these results have all been obtained with superpositions over spin-restricted orbital bases. This restriction can easily be lifted to include orbital bases of broken spin symmetry, which may reveal a yet stronger improvement in the energy compared to the standard spin-restricted variants of DMRG \cite{li_spin-projected_2017}.
\begin{table}
\centering
\begin{tabular}{llllll}
\toprule
\multicolumn{2}{l}{Cost metric} & \multicolumn{2}{l}{TNQE} & \multicolumn{2}{l}{VQE-UCCSD} \\
\midrule
1 & CNOTs per circuit & 1.2e3 & & 3.0e3 & \\
2 & Qubits (GHZ) & 13 & (18) & 12 &\\
3 & Layer depth (GHZ) & 2.0e3 & (6.6e2) & 3.9e3 & \\
4 & QPU calls (batches) & 5.6e5 & (2.4e2) & 2.5e4 & (1.6e4) \\
5 & Noise tol. (overlap) & 1.0e-4 & (1.0e-5) & 1.0e-8 &   \\
6 & Total shots & 6.4e16 & & 5.4e23 & \\
7 & Total CNOTs & 7.7e19 & & 1.6e27 & \\
\midrule
\multicolumn{2}{l}{Correlation energy} & $99.7$\% & & $33.2$\% & \\
\bottomrule
\end{tabular}
\caption{Quantum resource estimates for the VQE-UCCSD (first order Trotter) and TNQE ($\chi=4$, $M=4$) calculations for the octahedral H$_6$ system in the STO-3G basis set. The number of qubits (line 2) and the layer depth (line 3) of the TNQE circuits are given for both a single ancilla qubit and a GHZ ancilla register. In line 4 each QPU call corresponds to one VQE expectation value, or to one TNQE matrix element, which can be executed in parallelizable batches. line 6 shows the estimated sum total of circuit repetitions to resolve all of the QPU calls to within the noise tolerance (line 5). The total CNOTS (line 7) is then the product of the total shots (line 6) and the CNOT gates per circuit (line 1).}
\label{tab:res_est}
\end{table}
\begin{figure}
\centering
\includegraphics[width=0.49\textwidth]{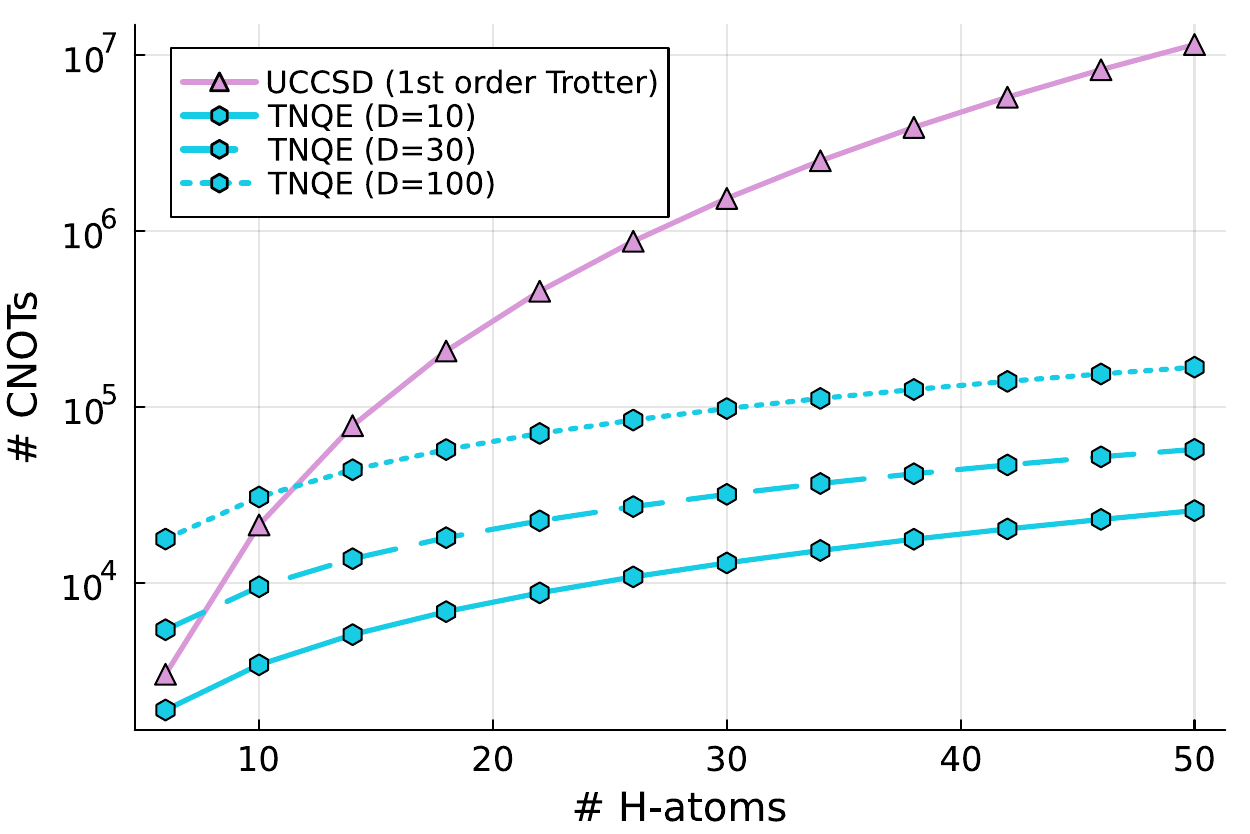}
\caption{Comparison of UCCSD and TNQE CNOT counts per circuit for systems with increasing numbers of hydrogen atoms, with arbitrary geometric structure, in the minimal basis set. The UCCSD gate counts are estimated using a CNOT-efficient encoding scheme \cite{Yordanov_2020}, and the TNQE gate counts are obtained from Eq. \ref{eq:CNOT} with disentangler depths $D$ of $10$, $30$, and $100$ (see Appendix \ref{app:B}).}
\label{fig:cnot_comp}
\end{figure}
\subsection{Cost estimates}
We provide detailed estimates of the exact CNOT gate count and circuit depth for TNQE (with both the one-ancilla and the GHZ-ancilla circuit variants) and VQE-UCCSD Trotterized to first order, using a CNOT-efficient circuit encoding of the UCC fermion operators \cite{Yordanov_2020}, for the octahedral H$_6$ system (see Appendix \ref{app:B}). We also estimate the total number of QPU calls to converge the energy estimate, where each QPU call corresponds to a single Hamiltonian or overlap evaluation, noting that the QPU calls in TNQE are highly parallelizable. We simulate shot noise in the matrix elements and expectation values with standard error $\delta$, and we find that while VQE-UCCSD fails to converge with noise levels of $\delta=10^{-7}$ and requires $\delta=10^{-8}$ to match the noiseless result to within 3 mHa after 100 iterations, TNQE reliably converges to within chemical accuracy after 48 sweeps with $\delta=10^{-4}$ in the Hamiltonian matrix elements and $\delta=10^{-5}$ in the overlaps. We hypothesize that this may be due in part to effective regulation of the subspace condition number, as detailed in Appendix \ref{app:A2}. From this we derive estimates for the number of shots to resolve each QPU call to these levels of precision, and hence the total number of CNOT gate executions required for the computation. This is the total number of two-qubit entangling gate operations ($n_\text{g}$), which gives a sense of the overall quantum compute requirement; with access to $N_\text{q}$ qubits (in total across all available QPUs), one can implement at most $N_\text{q}/2$ two-qubit gates in parallel, so the number of entangling gate layers that must be executed in sequence is bounded from below by $2n_\text{g}/N_\text{q}$, and from above by $n_\text{g}$. A greater overall two-qubit gate count necessitates either more sequential gate layers, raising the lower bound on the end-to-end execution time, or access to a greater number of qubits, thus increasing the hardware budget requirement. We summarize these findings in Table \ref{tab:res_est} for the octahedral H$_6$ system at $r=1.70\text{\AA}$ in the STO-3G basis. We comment on the high prefactors in the number of shots for both methods in Section \ref{sec:discussion}. While the primary driver behind our cost reductions in the octahedral H$_6$ system is the high noise tolerance of the TNQE optimizer, we show in Figure \ref{fig:cnot_comp} that the CNOT gate count per circuit also scales much more favorably for TNQE with increasing system size, observing a reduction in the CNOT count of nearly two orders of magnitude for 50 hydrogen atoms with disentangler depths as high as $D=100$ (see Section \ref{sec:meth_circ} for an explanation of the disentangler depth).
\section{Discussion} \label{sec:discussion}

We have presented a new class of hybrid quantum-classical algorithm, which we call a tensor network quantum eigensolver, to evaluate chemical ground state energies on near-term quantum computers. We have combined techniques from classical tensor networks, variational quantum algorithms, and quantum subspace diagonalization to eliminate the reliance on costly and unreliable gradient-based optimizers. In the TNQE method, as in classical DMRG, the efficient reduction of truncation error is the key to successfully optimizing for the molecular ground state wavefunction, which is now enabled through the superposition of matrix product states in different orbital bases. Although the TNQE ansatz is constructed using MPS of low bond dimension, the orbital rotations can mimic the entanglement structure of systems that do not follow a one dimensional area law, as is likely to be the case in many real-world instances of strong electron correlation such as cuprates \cite{Imada_1998} or the oxygen-evolving complex of photosystem II \cite{askerka_o2-evolving_2017}. We have demonstrated reliable convergence to chemical accuracy in small chemical systems, with low parameter counts that suggest a possible regime of practical quantum advantage over classical DMRG. Additionally we report a high tolerance to shot noise and efficient use of quantum resources, with per-circuit gate counts of $O(N^2)$ and circuit depths of $O(N)$, signalling the potential of the method to scale well to larger systems in the NISQ era.

The argument for realizing an exponential quantum advantage with TNQE rests on the degree to which the orbital rotations are of practical utility in the energy calculations. This should now be rigorously tested on larger and more complex chemical problems. We note that the orbital rotation heuristics we have developed appear to be improvable from both a quantum information and a quantum chemistry perspective, for example using the quantum mutual information \cite{Rissler_2006}, or initializing the orbitals from unrestricted Hartree-Fock calculations with broken spin symmetry \cite{Thom_2009}. Achieving practical results at scale will require further research into a number of open questions. First, the gate count, the number of measurements, and the classical cost of solving the generalized eigenvalue problem scale cheaply with the number of matrix product states, as $O(1)$, $O(M^2)$, and $O(M^3)$ respectively, but somewhat less favorably with the bond dimension, at worst as $O(\chi^2)$, $O(\chi^4)$, and $O(\chi^6)$. Establishing the optimal tradeoff between $M$ and $\chi$ will be necessary to fully characterize the algorithmic complexity, as well as the scaling of both of these quantities with $N$, which is expected to be strongly system-dependent. The method also requires the efficient quantum circuit encoding of a large number of matrix product states to high fidelity. It is not known how this cost will scale using the disentangler method \cite{ran_encoding_2020}, in terms of both the classical overhead and the circuit depth, and the optimal encoding of matrix product states is currently an ongoing topic of research. It is also unknown how the number of sweeps required for convergence will scale with the system size. Finally, a deeper understanding of the noise resilience and the identification of possible regimes where this breaks down will be important for implementations on real quantum hardware with both stochastic and coherent gate errors. As with any quantum algorithm, it is expected that the incorporation of some form of error mitigation or error correction will be vital for any successful practical implementation.

Compared with the closely related NOQE method \cite{PRXQuantum.4.030307}, TNQE offers more compact circuits, greater flexibility, and systematic convergence to the ground state. However, these advantages come at the cost of a greater number of measurements to converge the expected energy. While the estimate of $\sim10^{16}$ shots for a TNQE calculation on 12 system qubits is orders of magnitude cheaper than the estimate of $\sim10^{23}$ for VQE using UCCSD, the high measurement prefactor will need to be addressed in future work. Although the circuit repetitons are highly parallelizable, running calculations on $\sim10^{14}$ QPUs in parallel is prohibitive with current hardware availability. The shot noise limit, $n_\text{S}\propto \delta^{-2}$, is a generic problem facing near-term algorithms, and has been frequently discussed in the context of VQE \cite{Tilly_2022,Wecker_2015}. It may be possible to use classical shadows to reduce the measurement prefactor by exploiting the operator structure of the chemical Hamiltonian \cite{Huang_2020}, although the comparative measurement scaling with increasing system size is unknown. Alternatively, improvements in hardware fidelity may enable amplitude amplification teqhniques \cite{Brassard_2002} to provide a quadratic reduction in the measurement scaling, to $n_\text{S}\propto \delta^{-1}$, for a tradeoff in the circuit depth.

Tensor networks have played a central role in the `dequantization' of quantum algorithms and simulations \cite{Begusic_2024,shin2023dequantizing}, providing a richer understanding of the arguments for and against both near-term and fault-tolerant quantum advantage \cite{lee2023evaluating}. Here we have taken a complementary approach, identifying tensor network contractions that are thought to be classically hard and exploiting their efficient quantum evaluation as algorithmic components. This design philosophy yields quantum algorithms that are not limited by classical contraction constraints, and are by construction difficult to spoof with tensor networks. In the domain of quantum chemical simulation this provides new tools to extend tensor network descriptions of chemical systems to higher connective geometries and to study highly entangled states of chemical interest with low-depth quantum circuits. We argue that far from being antithetical to quantum computing, the tensor network paradigm is compatible with and actively beneficial for emerging quantum algorithms, enabling a virtuous cycle between the development of both quantum-inspired classical methods and classical-inspired quantum methods. We anticipate the techniques developed in this work finding applications beyond quantum chemistry, in fields such as condensed matter physics and quantum machine learning.

\section{Methods}
\label{sec:meth}

\subsection{Numerical details}
\label{sec:meth_num}

We have presented above the results of numerical simulations using the ITensor Julia library \cite{itensor} for the classical tensor network parts of the algorithm, with conversion to a sparse matrix representation to evaluate the off-diagonal matrix elements. The code is made publicly available (see Section \ref{sec:code_avail}). For numerical convenience we have used spatial orbital sites with $d=4$, which can be easily converted to the $d=2$ representation for quantum circuit encoding by use of SVDs. The parameter counts were reduced by enforcing particle number and $z$-spin symmetry using a standard block-sparse representation \cite{Singh_2010,Singh_2011}. An initial orbital basis was computed via restricted Hartree-Fock calculations in the PySCF Python package \cite{Sun_2020,Sun_2018,Sun_2015}, and an initial orbital ordering was selected using mutual information heuristics common in DMRG calculations \cite{Rissler_2006,ali_ordering_2021}. The effect of shot noise on the quantum processor was emulated by adding random Gaussian perturbations to the elements of $\textbf{H}'$ and $\textbf{S}'$ with standard error $\delta$. The algorithm was observed to reliably converge to chemical accuracy in the systems studied, almost always converging in the first run, and if not, then within two or three runs. We note that in rare cases, it was found to be beneficial to initialize the first RHF ordering at random instead of using the mutual information heuristic. The occurrence of these rare cases followed no discernible pattern, thus we attribute this to the inadequacy of the heuristic used, and we believe that a more robust starting guess for the RHF ordering could be developed. In these cases, when the random initial ordering was substituted, the algorithm would again typically converge in the first run, and if not, then within two or three runs.

The VQE-UCCSD benchmark was optimized via the L-BFGS algorithm \cite{Liu1989} with numerical gradients starting from zero amplitudes. While some reduction in the number of QPU calls for these benchmarking calculations may be possible by UCC parameter initialization with MP2 or projective coupled-cluster amplitudes \cite{hirsbrunner2023mp2}, and by use of analytic gradient calculations \cite{Izmaylov_2021}, since our goal was to assess the feasibility and scalability of TNQE relative to a standard method, we have left a more detailed comparison against state-of-the-art VQE methods to future work.

\subsection{Matrix element Hadamard test}\label{sec:meth_had}

Here we assume the $d=2$ representation so that $N$ is equal to the number of system qubits under the Jordan-Wigner transformation. For the quantum evaluation of the matrix elements we first require a unitary circuit to prepare each $\ket{\phi_j}$ from the vacuum state in their respective orbital bases,
\begin{align}
    \ket{\phi_j} = \hat{U}_j\ket{0},
\end{align}
where $\ket{0}$ is understood to refer to the multi-qubit state 
$\ket{0}^{\otimes N}$. We also require the Givens rotation circuits $\hat{G}_{ij}$ to rotate each state $\ket{\phi_i}$ into the orbital basis of another state $\ket{\phi_j}$. Finally we assume an efficient Pauli string decomposition for the observable $\hat{H}$ in each orbital basis as in Equation \ref{eq:pauli_decomp}. The matrix elements are expanded as a sum over expectation values of a set of unitary products with respect to the vacuum state,
\begin{align}
\braket{\phi_j|\hat{H}\hat{G}_{ij}|\phi_i} = \sum_{\alpha}h_\alpha\braket{0|\hat{U}_j^\dag\hat{P}_\alpha\hat{G}_{ij} \hat{U}_i|0},
\end{align}
which can be evaluated in parallel via standard Hadamard test \cite{Cleve_1969} circuits as shown in Figure \ref{fig:qu_circ}. This circuit makes a conditional application of $\hat{U}_j^\dag\hat{P}_\alpha \hat{G}_{ij} \hat{U}_i$ with control from a single ancilla qubit. The particle number conserving operator $\hat{G}_{ij}$ acts trivially on the vacuum state, so does not require control. The ancilla qubit is measured in the $z$-basis at the end of the computation, and the real part of the expectation value is calculated from the probabilities
\begin{align}
\text{Re}\braket{\hat{U}^\dag_j\hat{P}_\alpha\hat{G}_{ij}\hat{U}_i} = P(0) - P(1).
\end{align}
The imaginary component can also be obtained via a change of phase of the ancilla qubit, however we note that the tensor elements in the TNQE ansatz can be restricted to real values without loss of generality, hence for real-valued observables there is no imaginary component of the expectation value. An overlap matrix element $\braket{\phi_j|\hat{G}_{ij}|\phi_i}$ can be evaluated with a single Hadamard test circuit omitting the Pauli string. For a number of shots $n_\text{S}$ the output of each Hadamard test circuit has variance $\leq 1/n_\text{S}$ \cite{Polla_2023}. Consequently each overlap matrix element can be resolved to standard error $\delta$ with $n_\text{S}\approx\delta^{-2}$, and, following a standard optimal measurement allocation over the Pauli strings \cite{Wecker_2015}, each Hamiltonian matrix element can be resolved up to standard error $\delta$ with
\begin{align}
n_\text{S} \approx \left(\frac{\sum_{\alpha}|h_\alpha|}{\delta}\right)^2.
\label{eq:nshots}
\end{align}
There are several approaches in the literature for reducing the number of Pauli terms which will be  important in the presence of gate noise \cite{Huggins_2021,inoue2023optimal}. We leave the incorporation of these methods to future work.

\subsection{Circuit compilation}
\label{sec:meth_circ}

\begin{figure}
    \centering
    \includegraphics[width=0.49\textwidth]{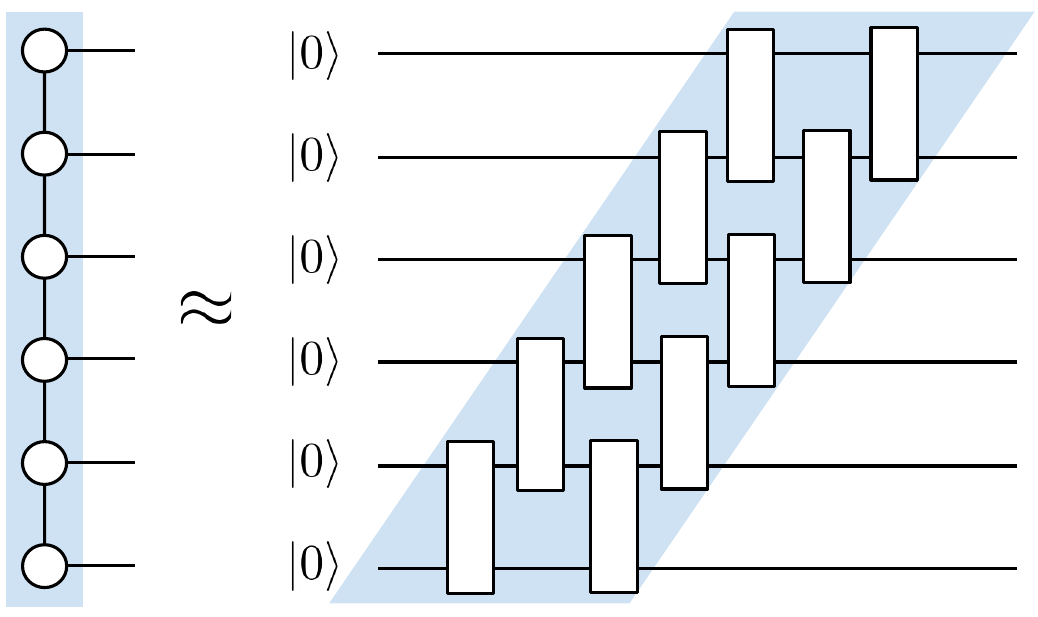}
    \caption{A circuit diagram illustrating the approximate preparation of a matrix product state via $D$ layers of disentangler gates, shown here with $D=2$.}
    \label{fig:disentanglers}
\end{figure}

Here we break down the gate cost and layer depth for the disentangler-based implementation of the matrix element quantum circuits. The disentangler method first introduced by Ran \cite{ran_encoding_2020} works by truncating the target MPS to $\chi=2$ and then computing a staircase-like circuit of one- and two-qubit unitaries that map the truncated MPS to the all-zero state (the one-qubit gate can then be then merged into the last two-qubit gate). Applying these gates to the untruncated MPS results in a new MPS that is closer in fidelity to the all-zero state. Repeating this procedure $D$ times results in a sequence of $D$ disentangler layers that, when applied in reverse to the all-zero state, approximately prepares the original MPS up to some desired fidelity, as shown in Figure \ref{fig:disentanglers}. The fidelity can be further improved by numerically re-optimizing each unitary via a QR factorization of a black-box optimized matrix, similar to the strategies advocated in Refs. \cite{dov2022approximate,Rudolph_2024}.

\begin{figure}
    \centering
    \includegraphics[width=0.49\textwidth]{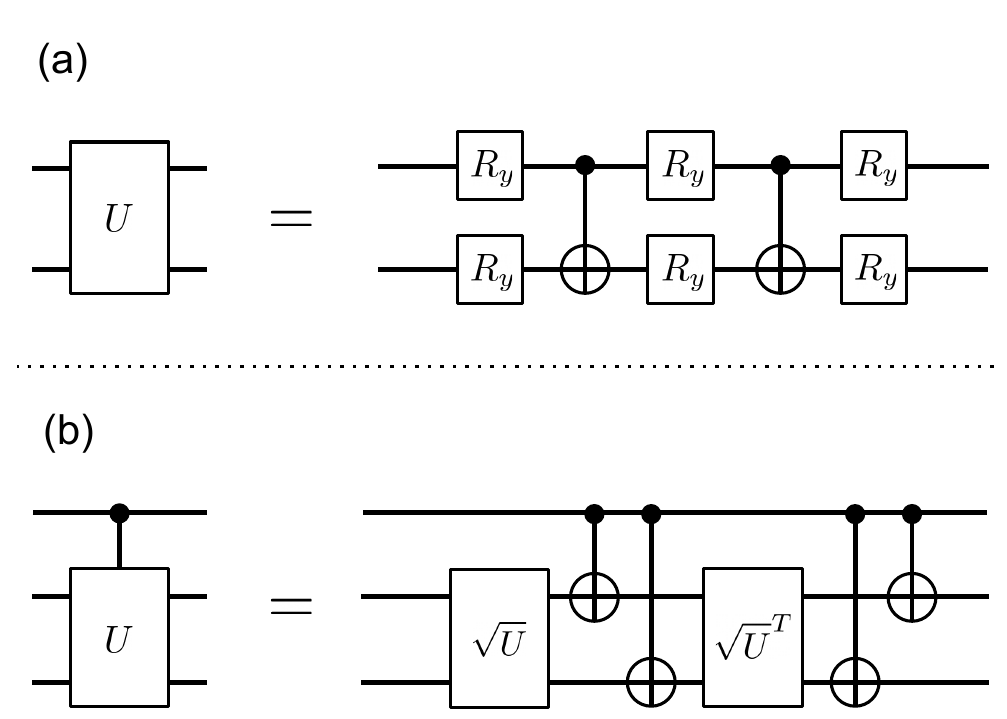}
    \caption{Orthogonal gate decompositions. (a) An orthogonal two-qubit gate $U$ with $\text{det}(U)=+1$ decomposes into two CNOT gates and six $R_y$ gates \cite{wei2012decomposition}. (b) Adding control to an orthogonal two-qubit gate $U$ with two orthogonal two-qubit gates and four CNOT gates \cite{zickert2021hands}.}
    \label{fig:ctrl_U}
\end{figure}

Without loss of generality, the matrix product states used in the TNQE ansatz can be made entirely real-valued, from which it follows that each two-qubit disentangling gate $U$ is a real-valued orthogonal transformation. We exploit this fact to reduce the prefactors in the CNOT gate count for each controlled disentangler. An orthogonal unitary $U$ satisfies $\text{det}(U)=\pm1$. Any orthogonal two-qubit gate with determinant $+1$ can be decomposed into two CNOT gates and six $R_y$ gates via the Cartan decomposition \cite{wei2012decomposition}, depicted in Figure \ref{fig:ctrl_U}a. Furthermore, for any orthogonal matrix with determinant $+1$ there exists an orthogonal matrix $\sqrt{U}$ such that $(\sqrt{U})^2=U$ and $\sqrt{U}\sqrt{U}^T=\sqrt{U}^T\sqrt{U}=\mathds{1}$. By computing this matrix, we may apply controlled-$U$ by a standard technique \cite{zickert2021hands}, with one application each of $\sqrt{U}$ and $\sqrt{U}^T$ and four additional CNOTs, as shown in Fig. \ref{fig:ctrl_U}b.
Thus adding control to an arbitrary two-qubit orthogonal gate introduces a factor of two to the synthesis cost plus an additional four CNOT gates. Note in the case that $\text{det}(U)=-1$, we may instead compute $\sqrt{-U}$, and after application of controlled-($-U$) flip the phase of the ancilla $\ket{1}$ state by applying a $Z$ gate to the ancilla qubit. We may then decompose each $\sqrt{U}$ into six single-qubit $R_y$ gates and two CNOT gates as before. We can do this regardless of the determinant of $\sqrt{U}$ because we may equivalently decompose $-\sqrt{U}$, as the global phase will cancel with that of $-\sqrt{U}^T$. Therefore each controlled two-qubit rotation requires a total of eight CNOT gates with a layer depth of 14. Note that if one were to simply replace the CNOT gates in Fig. \ref{fig:ctrl_U}a with Toffoli gates, each would decompose into six CNOT gates \cite{Shende_2009}, for twelve CNOT gates in total. Therefore the implementation in Fig. \ref{fig:ctrl_U}b reduces the overall CNOT count by a third.

We can evaluate a matrix element with two controlled MPS preparation unitaries as in Fig. \ref{fig:qu_circ}, each of which require $(N-1)D$ controlled two-qubit rotations, assuming one-all connectivity to the ancilla qubit. The $N\choose 2$ Givens rotation gates in the orbital rotation circuit are also orthogonal and have determinant equal to $+1$, and therefore can be implemented with two CNOT gates and five gate layers under the same two-qubit gate decomposition (Fig. \ref{fig:ctrl_U}a). Only three of the rotation unitaries contribute to the layer depth since the rest can be performed fully in parallel with the controlled disentanglers. We thus arrive at the Equations \ref{eq:CNOT}, \ref{eq:layerd}, and \ref{eq:ghz_layerd}.

\subsection{Regulating the subspace condition number}
\label{sec:meth_reg}

When solving the generalized eigenvalue problem in Equation \ref{eq:exp_gen_eig} (line 6 in Algorithm \ref{gen_sweep_alg}), the quantity of principal concern regarding numerical stability is the condition number of the expanded matrix $\textbf{S}'$. We can directly control this quantity by discarding one-hot states that have a high degree of linear dependence. Intuitively this representational flexibility is already present within the rest of the states in the subspace, so the discarded states are in some sense redundant. In fact the discarding step appears to be beneficial for convergence of the optimizer, possibly because it encourages each matrix product state to capture different features of the ground state. We formalize the linear dependence condition by measuring the squared norm of the projection of each of the one-hot states onto the subspace of the previous states. More concretely, let $\hat{P}^\ddagger$ be the projector onto the subset of the first $jn-1$ one-hot states (the upper-left block of the expanded state space). Let $\textbf{S}^\ddagger$ denote the reduced overlap matrix consisting of the upper-left block of $\textbf{S}'$ over the rows and columns $1,\ldots,jn-1$, and let $\vec{s}$ be the reduced vector taken from the first $1,\ldots,jn-1$ elements of the $jn$'th column of $\mathbf{S}'$, i.e., $s_{im}\equiv \braket{\varphi^{[i]}_m|\varphi^{[j]}_n}$, for $im=1,\ldots,jn-1$. Then the squared norm of the projection of $\ket{\varphi^{[j]}_n}$ onto the upper-left subspace is given by
\begin{align}
    \|\hat{P}^\ddagger\ket{\varphi^{[j]}_n}\|^2 = \vec{s}^{\,\dag}[\textbf{S}^\ddagger]^{-1}\vec{s}
\end{align}
(see Appendix \ref{app:A2}). We discard the one-hot state $\ket{\varphi^{[j]}_n}$ if one minus the squared norm is less than some set tolerance, e.g. $10^{-3}$. As a secondary check we also discard the one-hot state if the condition number of the new reduced overlap matrix including $\ket{\varphi^{[j]}_n}$ is above a certain threshold. Since we have discarded the linearly dependent states in the upper left block the matrix inverse should remain well-behaved (in practice we use the Moore-Penrose pseudoinverse with singular value tolerance $\sqrt{\varepsilon}\|\textbf{S}^{\ddagger}\|$ where $\varepsilon$ is the machine epsilon and $\|\cdot\|$ is the operator 2-norm).

We can further control the solution of the generalized eigenvalue problem by thresholding the singular values of $\textbf{S}'$ \cite{epperly_2021}. We have implemented two common thresholding strategies known as projection and inversion. For both methods, we begin by computing
\begin{align}
    \textbf{S}' = \textbf{U}\boldsymbol\Lambda\textbf{U}^\dag.
\end{align}
We filter the eigenvalues to keep only those greater than a chosen singular value tolerance, i.e. $\lambda>\epsilon$. Then by the projection method, we take the rectangular matrix of the remaining eigenvectors $\textbf{U}_{\epsilon}$ and solve
\begin{align}
    [\textbf{U}_{\epsilon}^\dag\textbf{H}'\textbf{U}_{\epsilon}]\textbf{C}'= [\textbf{U}_{\epsilon}^\dag\textbf{S}'\textbf{U}_{\epsilon}]\textbf{C}'\textbf{E}'.
\end{align}
Then the columns of the matrix product $\textbf{U}_{\epsilon}^\dag\textbf{C}'$ provide the approximate desired coefficients.
By the inversion method, we remove the eigenvalues $\lambda\leq\epsilon$ from $\boldsymbol\Lambda$, and then solve
\begin{align}
\textbf{U}\boldsymbol\Lambda_\epsilon^{-1}\textbf{U}^\dag\textbf{H}' = \textbf{C}'\textbf{E}' \textbf{C}'^\dag.
\end{align}
Letting $\vec{c}\,'$ be the first column of $\textbf{C}'$, the approximate desired coefficients are then provided by $\vec{c}\,'/\sqrt{\vec{c}\,'^{\dag}\textbf{S}'\vec{c}\,'}$.
Using a combination of these measures we find that in practice the optimizer explores the space surrounding the ground state in a highly restricted manner, such that the more ill-conditioned regions of the non-orthogonal state space are largely avoided. This is backed up empirically by our finding that we can reliably converge the energy estimate with Gaussian noise in the overlap matrix elements on the order of $10^{-5}$.

\subsection{Mitigating the energy penalty due to singular value truncation}
\label{sec:meth_mit}

\begin{figure}
    \centering
    \includegraphics[width=0.42\textwidth]{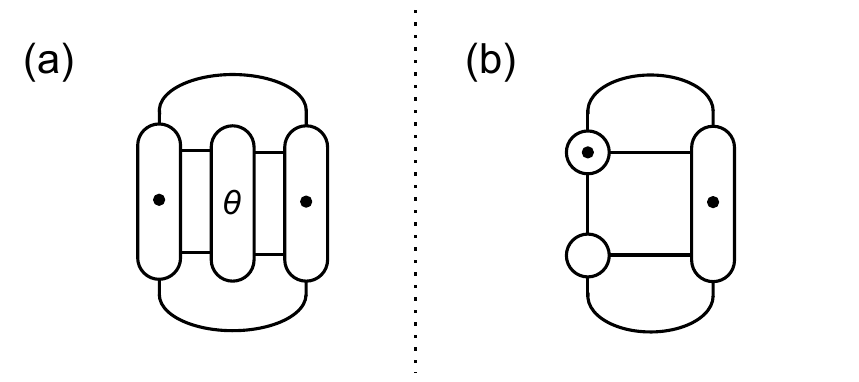}
    \caption{Tensor networks to compute the elements of the transformation matrices that are applied to the blocks of $\textbf{H}'$ and $\textbf{S}'$ during the single-site optimizations. (a) A tensor network to compute the elements of the $d^2\chi^2\times d^2\chi^2$ block transformation matrices $\{\textbf{G}^{[j]}\}$ corresponding to the local orbital rotation updates $\hat{g}(\theta)$ in the basis of the two-site one-hot states of reference state $j$. (b) A tensor network to compute the elements of the $d\chi^2\times d^2\chi^2$ isometries $\{\textbf{T}^{[j]}\}$ that, when applied to the blocks of the two-site expanded subspace matrices $\textbf{H}'$ and $\textbf{S}'$ (dim. $Md^2\chi^2$), yield the blocks of the single-site expanded subspace matrices (dim. $Md\chi^2$) (see Appendix \ref{app:A4} for a detailed explanation).}
    \label{fig:singlesite}
\end{figure}

In the final step of the generalized sweep algorithm (line 9 in Algorithm \ref{gen_sweep_alg}), after the orbital updates, the truncation error can still significantly and nontrivially affect the energy estimate of the ansatz. The optimal parameters of the two-site tensor after truncation are typically better than the previous parameters, however they may not be optimal for a pair of single-site tensors with a fixed singular value threshold. We mitigate this penalty due to truncation without growing the bond dimension by performing a sequence of alternating single-site decompositions at sites $p$ and $p+1$, starting from the two-site parameters following the initial truncation. This does not require any additional QPU calls, as all of the necessary information for this step is contained within the two-site expanded matrices $\textbf{H}'$ and $\textbf{S}'$. Instead, local classical tensor contractions are used to compute block transformation matrices (see Figure \ref{fig:singlesite}), which are then applied to the blocks of $\textbf{H}'$ and $\textbf{S}'$ to give the subspace matrices for the single-site subspace expansions. This technique is explained in detail in Appendix \ref{app:A4}.

Finally, the parameter update is rejected if the new ground state energy estimate is higher than the previous estimate by a fixed tolerance, on the order of $\sim 1$ mHa. We find that in combination these measures can effectively mitigate any remaining convergence issues due to truncation.

\section{Data availability}
The numerical data presented in this study are available from the corresponding author upon reasonable request.
\section{Code availability}
\label{sec:code_avail}
The code used to generate all of the numerical data presented in this study is publicly available at \url{https://github.com/oskar-leimkuhler/TNQE-Julia/}.
\section{Author contributions}
O.L. proposed the project, developed the algorithm, implemented and performed the classical numerical simulations, worked out the resource estimates, and wrote the initial draft of the paper. K.B.W. participated in discussions, provided guidance throughout the project, and contributed to the writing of the paper.
\section{Competing interests}
The authors declare no competing financial or non-financial interests.

\section{Acknowledgements}

We thank Unpil Baek, William J. Huggins, and Torin F. Stetina for many invaluable discussions.  This work was supported in part by the NSF QLCI program through grant number QMA-2016345, in part by the U.S. Department of Energy, Office of Science, Office of Advanced Scientific Computing Research under Award Number DE-SC0025526, and as part of a joint development agreement between UC Berkeley and Dow.

\appendix

\renewcommand\thefigure{\thesection.\arabic{figure}}  
\renewcommand\thealgorithm{\thesection.\arabic{algorithm}} 

\section{Details of the generalized sweep algorithm}\label{app:A}

In this appendix we explain the steps of the generalized sweep algorithm, as well as a procedure for iterative construction of the MPS subspace, which are then summarized with pseudocode in Algorithms \ref{alg:gen_sweep} and \ref{alg:sub_exp}.

\setcounter{figure}{0}

\subsection{Parameter optimization}\label{app:A1}

In classical \textit{ab initio} quantum chemistry, matrix product states are commonly optimized variationally using the two-site DMRG algorithm. Summarized at a high level, the MPS is orthogonalized to site $p$, after which the site tensors between sites $p$ and $p+1$ are contracted, optimized, and decomposed by SVD into new site tensors. The procedure is repeated over all the bonds of the MPS in sequence from left to right. In practical calculations the right blocks of the MPS are pre-cached while the left block is iteratively updated. In equation form, we write the orthogonalization to site $p$ as
\begin{align}
\ket{\phi} &= \sum_{\vec{k}}\Big[\sum_{l_1...l_{N-1}} \Phi_{l_1}^{k_1}\Phi_{l_1l_2}^{k_2}\cdots\Phi_{l_{N-1}}^{k_N}\Big]\ket{\vec{k}} \\
&= \sum_{\vec{k}}\Big[\sum_{l_1...l_{N-1}} U_{l_1}^{k_1}U_{l_1l_2}^{k_2}\cdots T_{l_{p-1}l_{p+1}}^{k_pk_{p+1}}\cdots V_{l_{N-1}}^{k_N}\Big]\ket{\vec{k}},
\end{align}
where $T_{l_{p-1}l_{p+1}}^{k_pk_{p+1}}$ is the two-site tensor formed from contracting the tensors at sites $p$ and $p+1$. In this form the tensors in the left and right blocks have the property that, for any sites $q<p$ and $r>p+1$,
\begin{align}
\sum_{k_1...k_q}\sum_{\substack{l_1...l_{q-1} \\ l'_1...l'_{q-1}}}(U^{k_1}_{l_1}U^{\dag k_1}_{l'_1})\cdots (U^{k_q}_{l_{q-1}l_q}U^{\dag k_q}_{l'_{q-1}l'_{q}}) &= \mathds{1}_{l_q l'_q}, \label{eq:orth_l} \\ 
\sum_{k_r...k_N}\sum_{\substack{l_{r-1}...l_{N-1} \\ l'_{r-1}...l'_{N-1}}}(V^{k_r}_{l_{r-1}l_r}V^{\dag k_r}_{l'_{r-1}l'_r})\cdots (V^{k_N}_{l_{N-1}}V^{\dag k_N}_{l'_{N-1}}) &= \mathds{1}_{l_{r-1} l'_{r-1}}. \label{eq:orth_r}
\end{align}

The two-site tensor $T$ is optimized to lower the expectation value $\braket{\phi|\hat{H}|\phi}$ and then decomposed back into site tensors by SVD with truncation from $d\chi$ to $\chi$ singular values,
\begin{align}
T_{l_{p-1}l_{p+1}}^{k_pk_{p+1}} \approx \sum_{l_p,l_p'}^\chi U_{l_{p-1}l_{p}}^{k_p}\Sigma_{l_{p}l'_{p}}V_{l'_{p}l_{p+1}}^{k_p+1} = \sum_{l_p}^\chi U_{l_{p-1}l_{p}}^{k_p}T_{l_{p}l_{p+1}}^{k_p+1},
\label{eq:svd_longform}
\end{align}
where the tensor $\Sigma$ is the diagonal matrix of singular values, which is contracted into $V$ in order to move the orthogonality center to site $p+1$. Note that we must renormalize the state since we have thrown away some information. 

We now demonstrate that finding the optimal coefficients for the two-site tensor is equivalent to diagonalizing in an expanded subspace spanned by an orthonormal set of states. We can decompose the tensor $T$ as a sum over ``one-hot'' tensors, which are tensors with the same indices as $T$ and a single nonzero element equal to one,
\begin{align}
T_{l_{p-1}l_{p+1}}^{k_pk_{p+1}} = \sum_{m=1}^{d^2\chi^2}t_{m}[\varphi_m]^{k_pk_{p+1}}_{l_{p-1}l_{p+1}},
\end{align}
where $m$ indexes the one-hot tensors $\varphi_m$ and we use square braces to separate the sum index $m$ from the tensor indices. Replacing the two-site tensor $T$ with one of the one-hot tensors $\varphi_m$ results in a new matrix product state which we call a ``one-hot'' state, denoted $\ket{\varphi_m}$. Since $\ket{\phi}$ is in orthogonal form centred at $p$, we can use the orthogonality relations (Equations \ref{eq:orth_l}, \ref{eq:orth_r}) to trivially compute the overlap between two one-hot states as
\begin{align}
\braket{\varphi_m|\varphi_n} = \sum_{\substack{k_pk_{p+1}\\l_{p-1}l_{p+1}}} [\varphi_m]^{k_pk_{p+1}}_{l_{p-1}l_{p+1}}[\varphi^\dag_n]^{k_pk_{p+1}}_{l_{p-1}l_{p+1}} = \delta_{mn}.
\end{align}
Furthermore, by the same orthogonality relations, we have that
\begin{align}
\braket{\varphi_m|\phi} = t_m, \quad \braket{\phi|\phi} = \sum_{m}|t_m|^2=1.
\end{align}
Therefore the one-hot states must form a complete and orthonormal basis for all of the states that can be obtained by optimizing the two-site tensor,
\begin{align}
\ket{\phi} = \sum_{m=1}^{d^2\chi^2}t_m\ket{\varphi_m}, \quad \braket{\varphi_m|\varphi_n}=\delta_{mn}, \quad \sum_{m=1}^{d^2\chi^2}|t_m|^2=1.
\end{align}
Then by computing the $O(d^4\chi^4)$ matrix elements $H_{mn}=\braket{\varphi_m|\hat{H}|\varphi_n}$ and diagonalizing
\begin{align}
\textbf{H} = \textbf{C}\textbf{E}\textbf{C}^\dag
\end{align}
we find the optimal coefficients from the first column of the solution matrix, i.e. $t_m=C_{m1}$. Exact diagonalization has an asymptotic cost of $O(\chi^6)$, which is a factor of $\chi^3$ more expensive than standard variants of the two-site algorithm using iterative solvers ~\cite{Holtz_2012,Wouters_2014}. Provided that the bond dimension is kept small this remains only a modest additional cost.

Now we shall see how to generalize the two-site optimization algorithm to the multi-reference case. Given a set of MPS basis states $\{\ket{\phi_j}\}$, we may calculate the optimal coefficients $\{c_j\}$ to approximate the ground state of the Hamiltonian by solving the generalized eigenvalue problem
\begin{align}
    \textbf{HC}=\textbf{SCE},
\end{align}
where the matrix elements of $\textbf{H}$ and $\textbf{S}$ are given by 
\begin{align}
    H_{ij} &= \braket{\phi_i|\hat{H}|\phi_j}, \\
    S_{ij} &= \braket{\phi_i|\phi_j}.
\end{align}
The matrix $\textbf{E}$ is a diagonal matrix of eigenvalues $E_1,...,E_M$.
The optimal coefficients for the ground state estimate $E_1$ are then given by the first column of $\textbf{C}$, that is, $c_j=C_{j1}$. We can immediately see the connection to the two-site decomposition. By expanding a chosen reference state $\ket{\phi_j}=\sum_m t^{[j]}_m\ket{\varphi^{[j]}_m}$ and computing the additional matrix elements
\begin{align}
    H'_{jm,jn} = \braket{\varphi^{[j]}_m|\hat{H}|\varphi^{[j]}_n}, &\quad S'_{jm,jn} = \delta_{mn},\\
    H'_{jm,i} = \braket{\varphi^{[j]}_m|\hat{H}|\phi_i}, &\quad S'_{jm,i} = \braket{\varphi^{[j]}_m|\phi_i},
\end{align}
we can lower the ground state estimate $E'_1$ by solving the generalized eigenvalue problem on the expanded subspace matrices $\textbf{H}'$ and $\textbf{S}'$ (see Figure 8 in the main text). This can be extended in a straightforward manner to simultaneously expand any subset of the matrix product states at the same or different sites. To optimize all states simultaneously we compute the matrix elements between all of the expanded one-hot states,
\begin{align}
H'_{im,jn} = \braket{\varphi^{[i]}_m|\hat{H}|\varphi^{[j]}_n}, &\quad S'_{im,jn} = \braket{\varphi^{[i]}_m|\varphi^{[j]}_n}.
\end{align}
Having solved the generalized eigenvalue problem, we find the optimal coefficients for each two-site tensor from slicing and normalizing the first column of $\textbf{C}'$,
\begin{align}
t^{[j]}_m = \frac{C'_{jm,1}}{\sqrt{\sum_n|C'_{jn,1}|^2}}.
\end{align}
To optimize a single MPS this procedure requires $O(Md^2\chi^2)$ matrix element evaluations in the off-diagonal $(i\neq j)$ blocks per site decomposition, while to optimize all states simultaneously requires $O(M^2d^4\chi^4)$ off-diagonal matrix elements. Note that we need the QPU only for the $i<j$ blocks, since the $i=j$ blocks are classically tractable by tensor network contraction, and we may copy matrix elements into the adjoint blocks since $\textbf{H}'=\textbf{H}'^\dag$ and $\textbf{S}'=\textbf{S}'^\dag$. In practical calculations the $z$-spin and particle number symmetry are typically enforced by way of an internal block-sparse tensor structure ~\cite{Singh_2010,Singh_2011}, so the scalings in the number of matrix elements will have some prefactor less than one. To complete a full sweep over all $N-1$ bonds requires $O(NM^2d^4\chi^4)$ matrix elements per sweep. There is no requirement to decompose each MPS at the same bond, although we find this to be an effective strategy in small-scale calculations.

\subsection{Numerical stability}\label{app:A2}

As explained in Section \ref{sec:meth_reg}, the quantity of principal concern regarding numerical stability is the condition number of the expanded matrix $\textbf{S}'$. We can directly control this quantity by discarding one-hot states that have a high degree of linear dependence. We formalize the linear dependence condition by measuring the squared norm of the projection of each of the one-hot states onto the subspace of the previous states. We now derive an expression for the squared norm of the projection of $\ket{\varphi^{[j]}_n}$ onto the subspace $\{\ket{\varphi^{[i]}_m}\}_{im=1}^{jn-1}$ in terms of the elements of $\textbf{S}'$. Let $\textbf{S}^{\ddagger}$ denote the reduced overlap matrix consisting of the upper-left block of $\textbf{S}'$ over the rows and columns $im=1,...,jn-1$, and let $\vec{s}$ be the column vector of overlaps of the one-hot tensor $\ket{\varphi^{[j]}_n}$ with the states in the upper-left subspace, i.e. $s_{im}\equiv \braket{\varphi^{[i]}_m|\varphi^{[j]}_n}$. We assume that the upper-left subspace is not rank deficient, so it is spanned by a set of $jn-1$ orthonormal vectors. We may then construct an orthonormal basis $\{\ket{\beta_x}\}_{x=1}^{jn-1}$ that spans the same subspace by symmetric orthogonalization. Let $\textbf{B}\equiv[\textbf{S}^{\ddagger}]^{-\frac{1}{2}}$, then
\begin{align}
\ket{\beta_x} \equiv \sum_{im=1}^{jn-1} B_{x,im} \ket{\varphi^{[i]}_{m}},
\end{align}
and 
\begin{align}
\hat{P}^\ddagger = \sum_{x=1}^{jn-1}\ket{\beta_x}\bra{\beta_x}
\end{align}
is the projector onto the upper-left subspace. Then the squared norm of the projected state is given by
\begin{align}
    \|\hat{P}^\ddagger\ket{\varphi^{[j]}_{n}}\|^2 &= \sum_{x}\braket{\varphi^{[j]}_{n}|\beta_x}\braket{\beta_x|\varphi^{[j]}_{n}} \\
    &= \sum_{im}\sum_{i'm'}\sum_{x}B^*_{im,x}B_{x,i'm'} \braket{\varphi^{[j]}_{n}|\varphi^{i}_{m}}\braket{\varphi^{[i']}_{m'}|\varphi^{[j]}_{n}}\\
    &= \sum_{im}\sum_{i'm'}s^*_{im}[\textbf{B}^\dag\textbf{B}]_{im,i'm'}s_{i'm'} \\
    &= \vec{s}^{\,\dag}[\textbf{S}^{\ddagger}]^{-1}\vec{s},
\end{align}
where we have used that $\textbf{B}^\dag=\textbf{B}\implies \textbf{B}^\dag\textbf{B} = \textbf{B}^2=[\textbf{S}^\ddagger]^{-1}$. We discard the state $\ket{\varphi^{[j]}_n}$ if one minus the squared norm is less than some set tolerance, e.g. $10^{-3}$. As a secondary check we also discard the state if the condition number of the new reduced overlap matrix including $\ket{\varphi^{[j]}_n}$ is above a certain threshold. Since we have discarded the linearly dependent states in the upper left block the matrix inverse should remain be well-behaved (in practice we use the Moore-Penrose pseudoinverse with singular value tolerance $\sqrt{\varepsilon}\|\textbf{S}^{\ddagger}\|$ where $\varepsilon$ is the machine epsilon and $\|\cdot\|$ is the operator 2-norm).

We can further control the solution of the generalized eigenvalue problem by thresholding the singular values of $\textbf{S}'$ ~\cite{epperly_2021}. We described two common thresholding strategies in Section \ref{sec:meth_reg}, known as projection and inversion, which we re-state here for completeness. For both, we begin by computing
\begin{align}
    \textbf{S}' = \textbf{U}\boldsymbol\Lambda\textbf{U}^\dag.
\end{align}
We filter the eigenvalues to keep only those greater than a chosen singular value tolerance, i.e. $\lambda>\epsilon$. Then by the projection method, we take the rectangular matrix of the remaining eigenvectors $\textbf{U}_{\epsilon}$ and solve
\begin{align}
    [\textbf{U}_{\epsilon}^\dag\textbf{H}'\textbf{U}_{\epsilon}]\textbf{C}'= [\textbf{U}_{\epsilon}^\dag\textbf{S}'\textbf{U}_{\epsilon}]\textbf{C}'\textbf{E}'.
\end{align}
Then the columns of $\textbf{U}_{\epsilon}^\dag\textbf{C}'$ provide the approximate desired coefficients.
By the inversion method, we remove the eigenvalues $\lambda\leq\epsilon$ from $\boldsymbol\Lambda$, and then solve
\begin{align}
    \textbf{U}\boldsymbol\Lambda_\epsilon^{-1}\textbf{U}^\dag\textbf{H}' = \textbf{C}'\textbf{E}' \textbf{C}'^\dag.
\end{align}
The coefficients are provided by the renormalized columns $\vec{c}_i/\sqrt{\vec{c}_i^{\,\dag}\textbf{S}'\vec{c}_i}$.
Following ~\cite{epperly_2021} we can calculate the eigenvalue condition number of the ground state by
\begin{align}
    \kappa_1 = \frac{\|\vec{c}_1\|^2}{\sqrt{E_1^2+1}}.
\end{align}
Provided this number is not too large, then the ground state may still be relatively insensitive to small perturbations even when $\epsilon\ll1$, although this does not address the issue of the ordering of the subspace eigenstates. Using the measures outlined above we find that in practice the optimizer explores the space surrounding the ground state in a highly restricted manner, such that the more ill-conditioned regions of the non-orthogonal state space are largely avoided. This is backed up empirically by our finding that we can reliably converge the energy estimate with Gaussian noise in the overlap matrix elements on the order of $10^{-5}$.

\subsection{Orbital permutations and rotations}\label{app:A3}

As in the two-site DMRG algorithm, the optimized two-site tensor is decomposed back into single-site tensors by SVD and the bond dimension is truncated from $d\chi$ to $\chi$ singular values. Prior to the SVD we apply a local orbital permutation or rotation to the two-site tensor in order to reduce the truncation error. In the case of spin-orbital sites $d=2$, the FSWAP gate acting on neighboring orbitals has the form
\begin{align}
    \hat{f} = \begin{pmatrix}
        1 & 0 & 0 & 0 \\
        0 & 0 & 1 & 0 \\
        0 & 1 & 0 & 0 \\
        0 & 0 & 0 & -1
    \end{pmatrix}.
    \label{eq:fswap}
\end{align}
Note that this differs from a generic SWAP gate only in the phase on the last element, which encodes fermionic antisymmetry under the exchange of two electrons, i.e., $\hat{f}\ket{11}=-\ket{11}$. A Givens rotation with angle $\theta$ has the form
\begin{align}
    \hat{g}(\theta) = \begin{pmatrix}
        1 & 0 & 0 & 0 \\
        0 & \cos(\theta) & -\sin(\theta) & 0 \\
        0 & \sin(\theta) & \cos(\theta) & 0 \\
        0 & 0 & 0 & 1
    \end{pmatrix}.
\end{align}
We can also use a phased version of the FSWAP gate which corresponds exactly to a $\theta=\pi/2$ Givens rotation,
\begin{align}
(\hat{\mathds{1}}\otimes\hat{Z})\hat{f} = \hat{g}(\frac{\pi}{2}).
\end{align}
In the case of either $\hat{f}$ or $\hat{g}(\theta)$ this matrix is reshaped to a $2\times 2\times 2\times 2$ tensor, which we denote here as $F$ or $G_\theta$. In the case of orbital permutations, we contract
\begin{align}
\tilde{T}_{l_{p-1}l_{p+1}}^{k_pk_{p+1}} = \sum_{k'_pk'_{p+1}}T_{l_{p-1}l_{p+1}}^{k'_pk'_{p+1}}F^{k_pk_{p+1}}_{k'_pk'_{p+1}},
\end{align}
and perform SVDs on both the original and swapped two-site tensors
\begin{align}
T=U\Sigma V, \quad \tilde{T}=\tilde{U}\tilde{\Sigma}\tilde{V}.
\end{align}
We accept the FSWAP insertion if
\begin{align}
\sum_{l=1}^{\chi}\tilde{\sigma}^2_l > \sum_{l=1}^{\chi}\sigma^2_l.
\end{align}
For orbital rotations we contract
\begin{align}
[T_\theta]_{l_{p-1}l_{p+1}}^{k_pk_{p+1}} = \sum_{k'_pk'_{p+1}}T_{l_{p-1}l_{p+1}}^{k'_pk'_{p+1}}[G_\theta]^{k_pk_{p+1}}_{k'_pk'_{p+1}}
\end{align}
and perform the SVD
\begin{align}
T_\theta = U_\theta\Sigma_\theta V_\theta.
\end{align}
We then carry out a univariate black-box optimization on the angle $\theta$ to obtain
\begin{align}
\theta_\text{opt} = \text{arg}\min_\theta \left[1 -\sum_{l=1}^{\chi}[\sigma_\theta]^2_l\right].
\end{align}
The numerical bottleneck for this optimization will be the multiple SVD calls with cost $O(\chi^3)$. Provided that $\chi$ is kept small then this remains a low-cost classical co-processing step.
We can extend to permutations and rotations over spatial orbital sites (without spin-mixing) by defining FSWAP and Givens operations on the $d=4$ sites in terms of the above operations for $d=2$ to which we add the subscripts $\hat{f}_2$ and $\hat{g}_2(\theta)$:
\begin{align}
\hat{f}_4 &= (\hat{\mathds{1}}_2\otimes\hat{f}_2\otimes\hat{\mathds{1}}_2)(\hat{f}_2\otimes\hat{f}_2)(\hat{\mathds{1}}_2\otimes\hat{f}_2\otimes\hat{\mathds{1}}_2), \\
\hat{g}_4(\theta) &= (\hat{\mathds{1}}_2\otimes\hat{f}_2\otimes\hat{\mathds{1}}_2)(\hat{g}_2(\theta)\otimes\hat{g}_2(\theta))(\hat{\mathds{1}}_2\otimes\hat{f}_2\otimes\hat{\mathds{1}}_2),
\end{align}
where $\hat{\mathds{1}}_2$ is the single-qubit identity. Then we reshape to a $4\times4\times4\times4$ tensor and apply the same steps as before. In practical calculations we find that interleaving orbital permutation sweeps with orbital rotation sweeps is highly effective at breaking the optimizer out of local minima. This corresponds to alternating between rearranging the orbitals to different site positions via nearest neighbor swapping, and mixing of the orbitals through nearest neighbor quantum superpositions.
\subsection{Mitigation of the energy penalty}\label{app:A4}

Even after the orbital rotations the truncation error can significantly and nontrivially affect the energy estimate of the ansatz. We mitigate this truncation effect without growing the bond dimension by performing a sequence of alternating single-site decompositions at sites $p$ and $p+1$ following each two-site decomposition. We can do this with no additional QPU calls by contracting the single-site one-hot states with the two-site one-hot states, obtaining isometries that can be applied to the blocks of $\textbf{H}'$ and $\textbf{S}'$. 

First, we apply the new orbital rotation updates to the $\textbf{H}'$ and $\textbf{S}'$ matrices. Recall that the one-hot decomposition represents an orthonormal basis for all of the matrix product states that could be obtained by local transformations of the parameters at sites $p$ and $p+1$. Because a local orbital rotation implements such a local parameter transformation, it can be represented by a $d^2\chi^2\times d^2\chi^2$
\setcounter{algorithm}{0}
\renewcommand{\theHalgorithm}{\thesection\arabic{algorithm}}
\begin{algorithm}[H]
\caption{ \raggedright Generalized sweep algorithm}\label{alg:gen_sweep}
\begin{algorithmic}[1]
\State $p\gets 1$
\While{$p<N$}
\State{$E_\text{old}\gets E_1$}
\item[]
\For{$j$ in jset}
\State orthogonalize $\ket{\phi_j}$ to site $p$
\State contract two-site tensor $T^{[j]}$ over sites $p,p+1$
\State decompose $T^{[j]}$ to obtain $\{\ket{\varphi_m^{[j]}}\}$
\State classically contract $\braket{\varphi_m^{[j]}|\hat{H}|\varphi_n^{[j]}}$
\For{$i<j$}
\If{$i$ in jset}
\State call QPU for $\braket{\varphi_m^{[i]}|\hat{H}|\varphi_n^{[j]}}$, $\braket{\varphi_m^{[i]}|\varphi_n^{[j]}}$
\Else
\State call QPU for $\braket{\phi_i|\hat{H}|\varphi_n^{[j]}}$, $\braket{\phi_i|\varphi_n^{[j]}}$
\EndIf
\EndFor
\EndFor
\item[]
\State discard linearly dependent columns of $(\mathbf{H}',\mathbf{S}')$
\State solve $\mathbf{H}'\mathbf{C}'=\mathbf{S}'\mathbf{C}'\mathbf{E}'$
\item[]
\For{$j$ in jset}
\State update $T^{[j]}$ parameters from $\mathbf{C}'$
\If{rotation type `FSWAP'}
\State compute $\xi$ and $\tilde{\xi}$
\If{$\tilde{\xi}<\xi$}
\State $T^{[j]}\gets T^{[j]}F$
\State $\hat{G}_{ij}\gets \hat{f}_{p,p+1}\hat{G}_{ij}$
\EndIf
\ElsIf{rotation type `Givens'}
\State compute $\theta_\text{opt}=\text{arg}\min_{\theta}[\xi(\theta)]$
\State $T^{[j]}\gets T^{[j]}G_{\theta\text{opt}}$
\State $\hat{G}_{ij}\gets \hat{g}_{p,p+1}(-\theta_\text{opt})\hat{G}_{ij}$
\EndIf
\State truncate $T^{[j]}=U^{[j]}\Sigma^{[j]}V^{[j]}$ to $\chi$ singular values
\EndFor
\item[]
\State apply rotation matrices to $(\textbf{H}',\textbf{S}')$
\item[]
\State $\text{rep}\gets1$
\While{$\text{rep}\leq\text{nreps}$}
\For{$q\in\{p,p+1\}$}
\State apply single-site $q$ isometries to $(\textbf{H}',\textbf{S}')$
\State discard linearly dependent columns of $(\mathbf{H}',\mathbf{S}')$
\State solve $\mathbf{H}'\mathbf{C}'=\mathbf{S}'\mathbf{C}'\mathbf{E}'$
\For{$j$ in jset}
\State update $T^{[j]}$ parameters from $\mathbf{C}'$
\EndFor
\EndFor
\State $\text{rep}\gets \text{rep}+1$
\EndWhile
\item[]
\State{$E_\text{new}\gets E_1'$}
\If{$E_\text{new} < E_\text{old} + E_\text{tol}$}
\State accept updates to $T^{[j]}$, $\hat{G}_{ij}$
\State apply new rotations to Hamiltonian coefficients
\Else
\State revert $T^{[j]}$, $\hat{G}_{ij}$ to original values
\EndIf
\item[]
\State $p\gets p+1$
\EndWhile
\end{algorithmic}
\end{algorithm}
\noindent transformation matrix $\textbf{G}$ in the basis of the one-hot states. To compute this matrix, it is sufficient to compute the overlaps between each of the one-hot tensors with the orbital rotation tensor, i.e., 
\begin{align}
[\textbf{G}]_{mn} = \sum_{\{k\},\{k'\},\{l\}} [\varphi_m]^{k_pk_{p+1}}_{l_{p-1}l_{p+1}}G^{k_pk_{p+1}}_{k'_pk'_{p+1}}[\varphi_n]^{k'_pk'_{p+1}}_{l_{p-1}l_{p+1}}
\end{align}
(see Fig. 15 (a) in the main text). Let $\textbf{H}^{[ij]}$, $\textbf{S}^{[ij]}$ denote the blocks of the $\textbf{H}'$ and $\textbf{S}'$ expanded subspace matrices corresponding to a pair of reference states $i$ and $j$, and let $\textbf{G}^{[i]}$ denote the two-site orbital rotation matrix for reference state $i$. Then we update the blocks of the expanded matrices according to
\begin{align}
\textbf{H}^{[ij]} &\rightarrow \textbf{G}^{[i]\dag}\textbf{H}^{[ij]}\textbf{G}^{[j]}, \\
\textbf{S}^{[ij]} &\rightarrow \textbf{G}^{[i]\dag}\textbf{S}^{[ij]}\textbf{G}^{[j]}.
\end{align}
Thus the two-site subspace expansion matrices can be updated to include the new orbital rotations with no additional QPU calls required.

Now consider the SVD of the two-site tensor $T=U\Sigma V$, which is short-hand for Equation \ref{eq:svd_longform}. The truncation to $\chi$ singular values results in an undesirable increase in the expected energy. The optimal parameters for the two-site tensor are typically better than the previous parameters, however they may not be optimal for a pair of single-site tensors with a fixed singular value threshold. We find it effective to use the two-site parameters after truncation as the starting point for a sequence of single-site one-hot expansions, which explicitly take into account the fixed bond dimension cutoff. We contract the single-site tensor $T^{(1)}\equiv U\Sigma$ and expand it into $d\chi^2$ one-hot tensors $\{\varphi^{(1)}_m\}$. This defines a new expansion of the two-site tensor as
\begin{align}
T^{k_pk_{p+1}}_{l_{p-1}l_{p+1}}=\sum_{m}t^{(1)}_m\sum_{l_p}[\varphi^{(1)}_m]^{k_p}_{l_{p-1}l_p}V^{k_{p+1}}_{l_pl_{p+1}}
\end{align}
(see Fig. 15 (b) in the main text). Unlike the two-site one-hot expansion, this single-site one-hot expansion respects the fixed bond dimension cutoff, as any two-site tensor that satisfies this expansion also has a bond dimension $\leq\chi$ between sites $p$ and $p+1$. Contracting all of the single-site one-hot tensors $\varphi^{(1)}_mV$ with each the original two-site one-hot tensors $\varphi^{(2)}_n$ results in a rectangular $d\chi^2 \times d^2\chi^2$ transformation matrix $\mathbf{T}$, where the coefficients are obtained by
\begin{align}
[\mathbf{T}]_{mn} = \sum_{\{k\},\{l\}} [\varphi^{(1)}_m]^{k_p}_{l_{p-1}l_p}V^{k_{p+1}}_{l_pl_{p+1}}[\varphi^{(2)}_n]^{k_pk_{p+1}}_{l_{p-1}l_{p+1}}.
\end{align}
Let $\textbf{T}^{[i]}$ refer to the isometry thus obtained for reference state $i$. Multiplying the appropriate blocks of $\textbf{H}'$ and $\textbf{S}'$ yields the $Md\chi^2\times Md\chi^2$ matrices of the single-site decomposition over all the states, i.e.,
\begin{align}
\textbf{H}^{[ij]} &\rightarrow \textbf{T}^{[i]}\textbf{H}^{[ij]}\textbf{T}^{[j]\dag}, \\
\textbf{S}^{[ij]} &\rightarrow \textbf{T}^{[i]\dag}\textbf{S}^{[ij]}\textbf{T}^{[j]\dag}.
\end{align}
\begin{algorithm}[H]
\caption{ \raggedright Iterative subspace construction}\label{alg:sub_exp}
\begin{algorithmic}[1]
\State $M\gets 1$
\State compute RHF molecular orbitals
\State choose initial orbital ordering (random or heuristic)
\State initialize $\ket{\phi_1}$ with random parameters
\State optimize $\ket{\phi_1}$ with classical DMRG
\item[]
\While{$M<M_\text{max}$}
\State $M\gets M+1$
\State initialize $\ket{\phi_M}$ with random parameters
\For{$i<M$}
\State $\hat{G}_{iM}\gets\hat{G}_{i,M-1}$
\EndFor
\item[]
\State $s\gets1$
\While{$s\leq\text{ns1}$}
\State run S. Alg. \ref{alg:gen_sweep} with `FSWAP' and jset=$\{M\}$
\State $s\gets s+1$
\EndWhile
\item[]
\State $s\gets1$
\While{$s\leq\text{ns2}$}
\State run S. Alg. \ref{alg:gen_sweep} with `FSWAP' and jset=$\{1,...,M\}$
\State run S. Alg. \ref{alg:gen_sweep} with `Givens' and jset=$\{1,...,M\}$
\State $s\gets s+1$
\EndWhile
\item[]
\EndWhile
\end{algorithmic}
\end{algorithm}
\noindent Then by solving the generalized eigenvalue problem with these reduced $\textbf{H}'$ and $\textbf{S}'$ matrices we obtain the optimal coefficients for the single-site tensors. This procedure can be repeated as many times as desired, alternating between single-site expansions at sites $p$ and $p+1$. We find that in practice a handful of repetitions is sufficient to obtain substantially improved energy estimates and far more reliable convergence of the generalized sweep algorithm. After the single-site decompositions we impose a further condition that the parameter update is rejected if the new ground state energy estimate is higher than the previous estimate by a fixed tolerance on the order of $\sim 1$ mHa. We find that in combination these measures can effectively mitigate any remaining convergence issues due to truncation. This final step completes the generalized sweep algorithm, summarized with high-level pseudocode in Algorithm \ref{alg:gen_sweep}.
\subsection{Iterative subspace construction}\label{app:A5}
In practical calculations it is found to be an effective strategy to increase $M$ in stages, starting with a single matrix product state optimized with classical DMRG. At each stage a new MPS is added to the subspace, in the same orbital basis as the last MPS and with random parameters, and optimized individually for a number of sweeps with FSWAPs applied. Following this all the MPSs are optimized simultaneously with sweeps alternating between FSWAPs and Givens rotations. These steps are summarized in Algorithm \ref{alg:sub_exp}. We do not claim this iterative subspace construction to be optimal, although, as we have demonstrated, this procedure reliably converges to chemical accuracy in numerical experiments on small systems.

\section{C\MakeLowercase{ost estimates for ground state energy estimation of octahedral H$_6$ in the STO-3G basis}}\label{app:B}

\subsection{TNQE}

We optimize the ansatz via the procedure detailed in Algorithm \ref{alg:sub_exp} with $M_\text{max}=4$, performing four sweeps over each newly added MPS followed by 12 sweeps over all MPSs alternating between FSWAPs and Givens rotations (six of each). The TNQE algorithm requires the evaluation of $O(Md^2\chi^2)$ matrix elements per site decomposition when optimizing a single state, or $O(M^2d^4\chi^4)$ matrix elements when optimizing all states at once. As noted in the main text we have some reduction in the number of one-hot basis states for each site decomposition due to the block-sparse structure of the quantum number conserving MPS. We count up the number of matrix elements in the off-diagonal blocks giving a total number of QPU calls $n_\text{Q}=5.6\times 10^5$.
We may perform all the matrix element computations at each site decomposition in parallel, so the number of parallel sets of QPU calls is simply the number of sweeps multiplied by the number of bonds per sweep for a total of 240.
We include a small Gaussian noise term drawn from $\mathcal{N}(0,\delta^2)$ to simulate shot noise in the matrix elements. Empirically we find that we can reliably converge to chemical accuracy with noise levels less than or equal to $\delta=10^{-4}$ in the Hamiltonian matrix elements and $\delta=10^{-5}$ in the overlap matrix elements. We therefore require $n_\text{S}\approx 10^{10}$ shots per overlap matrix element to resolve the matrix elements to this level of precision, and, following Equation 25 in the main text, $n_\text{S}\approx 2.2\times10^{11}$ shots per Hamiltonian matrix element. We have made the approximation $\sum_\alpha |h_\alpha|\approx \sum_{pq}|h_{pq}|+\sum_{pqrs}|h_{pqrs}| = 46.95$ Hartree, with integral coefficients expressed in the initial RHF orbital basis. The total number of overlap and Hamiltonian matrix elements are each $n_\text{Q}/2$, which gives a total number of shots $6.44\times 10^{16}$.

We count the number of CNOT gates using Equation 13 in the main text with qubit count $N=12$. We find that we can prepare all of the one-hot matrix product states with very high fidelities (errors in the seventh decimal place) with disentangler depth $D=6$, for a total of $1.2\times 10^3$ CNOT gates per quantum circuit. In some cases we need to apply a small number of re-optimization loops over the disentanglers using a QR factorization of a black-box optimized matrix, somewhat similar to the strategies advocated in ~\cite{dov2022approximate,Rudolph_2024}. We use the layer depth formulae of Equations 14 and 15 in the main text to compute a layer depth of $3.1\times 10^3$, which as explained in the main text could be significantly reduced at the expense of six additional qubits and ten additional CNOT gates to $6.6\times 10^2$. We note that it is substantially more expensive in terms of disentangler depth $D$ to prepare the full MPS $\ket{\phi_i}$ than its basis vectors under the one-hot decomposition, which have reduced entanglement across many of the bonds. In practice we never need to prepare the full MPS as we can always compute the matrix elements using a one-hot decomposition with appropriate one-hot basis vector coefficients. Further research is needed to obtain a more detailed understanding of how the layer depth $D$ depends on the bond dimension $\chi$ in practice.

\subsection{UCCSD}

The UCCSD ansatz has the form
\begin{align}
\ket{\psi_\text{UCCSD}} = \exp(\hat{T}-\hat{T}^\dag)\ket{\psi_\text{HF}},
\end{align}
where $\ket{\psi_\text{HF}}$ is the Hartree-Fock Slater determinant and $\hat{T}=\hat{T}_1+\hat{T}_2$ is the sum of singles and doubles cluster operators,
\begin{align}
\hat{T}_1 = \sum_{\substack{p\in\text{vir}\\ q\in\text{occ}}}\theta_{pq}\hat{a}^\dag_p\hat{a}_q, \\ 
\hat{T}_2 = \sum_{\substack{p>q\in\text{vir}\\ r>s\in\text{occ}}}\theta_{pqrs}\hat{a}^\dag_p\hat{a}^\dag_q\hat{a}_r\hat{a}_s.
\end{align}
Since the Hamiltonian does not connect sectors of different $\hat{S}_z$ symmetry we use the same parameter for all of the different spin terms, and we find that this restriction has a negligible effect on the energy estimate. For the octahedral H$_6$ system, with $N=6$ spatial orbitals and $\eta/2=3$ occupied orbitals, we have a total of 9 singles parameters and 36 doubles parameters for 45 total parameters. We optimize the ansatz using the L-BFGS algorithm which calculates numerical gradients at each step, followed by a line search requiring additional QPU calls. We terminate the optimizer after 100 iterations, counting a total of $n_\text{Q}=2.5\times 10^4$ QPU calls. We may parallelize the energy evaluations for the numerical gradients, but not those for the line search. For 45 parameters this implies 90 energy evaluations per step to compute the gradients which can be done simultaneously, so subtracting from the total evaluations we arrive at $1.6\times 10^4$ sequential QPU calls with parallelization.

We add a small Gaussian noise term drawn from $\mathcal{N}(0,\delta^2)$ to simulate shot noise in the energy evaluations. Empirically we find that the optimizer fails to converge to the noiseless result at noise levels of $\delta=10^{-7}$, with energy errors above 10 mHa, and we must go to noise levels of $\delta=10^{-8}$ to achieve agreement with the noiseless result to within 3 mHa after 100 iterations. By Equation 25 in the main text we require a number of shots $n_\text{S}\approx 2.2\times 10^{19}$ to resolve each energy evaluation to this level of precision. This gives a total number of shots for the computation of $n_\text{Q}\cdot n_\text{S} \approx 5.4 \times 10^{23}$.

We use a CNOT-efficient encoding scheme for the fermionic excitation operators ~\cite{Yordanov_2020}. We count a total of $3.0\times 10^3$ CNOT gates for our system of $N=6$ spatial orbitals and $\eta=6$ electrons, with a layer depth (counting both CNOT gates and single-qubit rotations, using the equations and figures provided in ~\cite{Yordanov_2020}) of $3.9\times 10^3$ layers.

\section{D\MakeLowercase{iscussion of exponential quantum advantage and dequantization}}\label{app:C}

Here we discuss attempts at efficient classical computation, or `dequantization', of the off-diagonal matrix element contractions. We argue that while dequantization is possible in the restricted case of orbital permutations, there is no known classical algorithm enabling dequantization in the case of arbitrary orbital rotations. In the following analysis we choose $d=2$ corresponding to qubit sites for simplicity. 

Consider first the restricted case of orbital permutations. The application of an FSWAP gate between neighboring sites of an MPS can grow the bond dimension by up to a factor of 2. There are provable lower bounds of $O(N)$ for the depth of a sorting network on a one-dimensional graph ~\cite{banerjee_sorting_2019,Knuth98}, which means that the bond dimensions will in general grow to $O(2^N)$ under arbitrary permutations of the orbitals. However, since the FSWAP gate is contained within the Clifford subgroup, we know that we can simulate the FSWAP network circuit in polynomial time on a classical computer by the Gottesman-Knill theorem ~\cite{Gottesman_1996,Aaronson_2004}. If the input state is a computational basis vector (or `bitstring state') then it is obvious how to compute the output from the FSWAP network in linear time by simply rearranging the bits and updating the phase after each FSWAP. Since there exists an efficient perfect sampling algorithm for bitstring distributions defined by matrix product states ~\cite{Stoudenmire_2010,Ferris_2012}, we can in principle compute the overlap between two matrix product states by the random sampling algorithm suggested in Section D3 of the supplementary material of ~\cite{Huggins2022}, which we summarize here. Suppose that we have two matrix product states $\ket{\phi_i}$ and $\ket{\phi_j}$, and an efficient sampling algorithm for bitstring states $\ket{x}$ drawn with probabilities equal to $|\braket{x|\phi_i}|^2$. We may insert a resolution of the identity by summing over all bitstring states,
\begin{align}
    \braket{\phi_i|\phi_j} &= \sum_{x}\braket{\phi_i|x}\braket{x|\phi_j} \\
    &= \sum_{x}\frac{|\braket{\phi_i|x}|^2\braket{x|\phi_j}}{\braket{x|\phi_i}}.
    \label{eq:dequantize}
\end{align}
It has been shown in ~\cite{Tang_2019} that the quantity $\frac{\braket{x|\phi_j}}{\braket{x|\phi_i}}$ for randomly sampled states $\ket{x}$ drawn with probability $|\braket{x|\phi_i}|^2$ has mean equal to $\braket{\phi_i|\phi_j}$ and constant variance, and hence we may compute the overlap in polynomial time up to an additive error by randomly sampling bitstrings to reconstruct the sum in Equation \ref{eq:dequantize}, provided that we can efficiently compute the overlap $\braket{x|\phi_j}$. Since we can efficiently evaluate the transformation of a bitstring acted on by the FSWAP network, we can efficiently compute this overlap in the restricted case of orbital permutations.

In the case of more general orbital rotations, for which the corresponding quantum circuits include non-Clifford gates, there is no way to efficiently transform a bitstring from one representation into the other without an exponentially large memory/time cost. This is because the action of a Givens rotation between two qubits maps the output to a superposition of bitstrings with probabilities determined by the rotation angle $\theta$. For instance, the two-qubit bitstring $\ket{10}$ maps to 
\begin{align}
\hat{g}(\theta)\ket{10} = \cos(\theta)\ket{10} + \sin(\theta)\ket{01}.
\end{align}
Hence the sequence of Givens rotations described in \cite{Kivlichan_2018} to implement an arbitrary basis rotation on a bitstring with $\eta$ ones and $N-\eta$ zeros in the first basis will result in a superposition over ${N\choose \eta}$ bitstring states in the second basis. This prevents us from efficiently evaluating the overlaps $\braket{x|\phi_j}$. 

One might consider instead sampling from both matrix product states in their respective bases, and then computing the overlaps between the sampled bitstrings using techniques from classical quantum chemistry to evaluate overlaps between Slater determinants in rotated single-particle bases ~\cite{Sundstrom_2014,Thom_2009,Amos_1961}. However, it is easily proven by counter example that this approach cannot work. Consider the insertion of a double resolution of the identity into the overlap $\braket{\phi_i|\phi_j}$ over bitstrings $\ket{x}$ in the basis of $\ket{\phi_i}$ and bitstrings $\ket{y}$ in the basis of $\ket{\phi_j}$:
\begin{align}
    \braket{\phi_i|\phi_j} &= \sum_{x,y}\braket{\phi_i|x}\braket{x|y}\braket{y|\phi_j} \\
    &= \sum_{x,y}\frac{|\braket{\phi_i|x}|^2\braket{x|y}|\braket{y|\phi_j}|^2}{\braket{x|\phi_i}\braket{\phi_j|y}}.
\end{align}
Let us take the limiting case of identical states $\ket{\phi_i}=\ket{\phi_j}$ expressed in the same single-particle basis. Since the states are identical then their overlap should evaluate to 1. Starting from the all-zero state, it is easy to prepare an MPS of bond dimension $\chi=1$ that is an equally weighted superposition over all bitstring states by applying a Hadamard transformation on each qubit. Then $\ket{x}$ and $\ket{y}$ will be independently sampled from the uniform distribution over all bitstring states. It is then clear that for any polynomially large number of samples the probability of obtaining a nonzero overlap of $\ket{\phi_i}$ with itself will be $O(\frac{\text{poly}(N)}{\exp(N)})$. Thus it appears that dequantization attempts relying on independently sampling from both matrix product states cannot be successful.

\section{G\MakeLowercase{eneration of maximal entanglement via orbital rotations}}\label{app:D}

Here our objective is to find an orbital rotation that maps an unentangled bitstring state to a state with maximal Schmidt rank across the central partition. Since orbital rotations are particle number conserving, an input bistring with $\eta$ particles (ones) will be mapped to a superposition over the $\eta$-particle bitstrings. We will accomplish the stated goal using the starting state $\ket{\vec{k}_0} = \ket{1}^{\otimes N_A}\otimes\ket{0}^{\otimes N_B}$, where $N_A=N_B=N/2$, by rotating into an equally weighted superposition of states
\begin{align}
\hat{G}\ket{\vec{k}_0} = \frac{1}{2^{N/4}}\sum_{l=1}^{2^{N/2}}\sigma_l\ket{\vec{k}^A_l}\otimes\ket{\vec{k}^B_l},
\label{eq:rot_state}
\end{align}
where $\sigma_l\in\{\pm1\}$, and the set of states $\{\ket{\vec{k}^A_l}\}$ run over all the $2^{N/2}$ unique bitstring states defined on subsystem $A$, with each state $\ket{\vec{k}^B_l}$ obtained as the bit-flipped mirror image of $\ket{\vec{k}^A_l}$. For instance,
\begin{align}
    \ket{\vec{k}^A_l} = \ket{10110111} \implies \ket{\vec{k}^B_l} = \ket{00010010}.
\end{align}
Thus the superposition is entirely contained within the $N/2$-particle subspace, and each left-right pair is a unique pair of left- and right- bistrings, so we have that
\begin{align}
    \braket{\vec{k}^{A}_l|\vec{k}^{A}_{l'}} = \braket{\vec{k}^{B}_l|\vec{k}^{B}_{l'}} = \delta_{ll'}.
\end{align}
Consequently, Equation \ref{eq:rot_state} denotes a Schmidt decomposition of $\hat{G}\ket{\vec{k}_0}$ over the $A/B$ partition, with Schmidt rank equal to $2^{N/2}$ and maximal von Neumann entropy across the partition. We construct the transformation $\hat{G}$ using a sequence of $\pi/2$ and $\pi/4$ pairwise orbital rotations, where the $\pi/4$ rotation gate is given by
\begin{align}
\hat{g}(\frac{\pi}{4}) = \begin{pmatrix}
1 & 0 & 0 & 0 \\
0 & \frac{1}{\sqrt{2}} & \frac{-1}{\sqrt{2}} & 0 \\
0 & \frac{1}{\sqrt{2}} & \frac{1}{\sqrt{2}} & 0 \\
0 & 0 & 0 & 1
\end{pmatrix},
\end{align}
and we choose a phase convention for the $\pi/2$ gate so that it is equal to the FSWAP gate given in Equation \ref{eq:fswap}. The $\pi/2$ gate acts `deterministically' on a bitstring state, in that each input bitstring is mapped to a single output bitstring, and this gate will be used to move particles between neighboring positions. The $\pi/4$ gate acts `non-deterministically', mapping a unique bitstring to a superposition of `swapped' and `non-swapped' bitstrings, and will be used to generate entanglement across the $A/B$ partition. The procedure is as follows: starting from the bitstring $\ket{\vec{k}_0}$, the application of a $\pi/4$ gate on the qubits on either side of the partition boundary results in the state
\begin{align}
\frac{1}{\sqrt{2}}\Big(&\ket{1\cdots11}\otimes\ket{00\cdots0} \\ + &\ket{1\cdots10}\otimes\ket{10\cdots0}\Big).
\end{align}
A sequence of FSWAPs is then applied within the bulk of each partition to deterministically move the particle in the right partition to site $N$, and the `hole' in the left partition to site 1, resulting in the state
\begin{align}
\frac{1}{\sqrt{2}}\Big((-1)^{N/2-1}&\ket{11\cdots1}\otimes\ket{0\cdots00} \\ + &\ket{01\cdots1}\otimes\ket{0\cdots01}\Big).
\end{align}
Another $\pi/4$ gate is then applied at the boundary, resulting in a superposition over four unique bitstring state pairs, and the newly transferred particle and `hole' are moved to the `next available' sites, i.e. sites $N-1$ and $2$ respectively. Repeating this procedure $N/2$ times yields the desired superposition in Equation \ref{eq:rot_state}, and parallelizing these operations results in the `diamond-shaped' tensor network to construct $\hat{G}$ as depicted in Figure 6 in the main text.


\begin{thebibliography}{100}
\expandafter\ifx\csname url\endcsname\relax
  \def\url#1{\texttt{#1}}\fi
\expandafter\ifx\csname urlprefix\endcsname\relax\def\urlprefix{URL }\fi
\providecommand{\bibinfo}[2]{#2}
\providecommand{\eprint}[2][]{\url{#2}}

\section{References}

\bibitem{Hartree_1928}
\bibinfo{author}{Hartree, D.~R.}
\newblock \bibinfo{title}{The wave mechanics of an atom with a non-coulomb central field. part i. theory and methods}.
\newblock \emph{\bibinfo{journal}{Mathematical Proceedings of the Cambridge Philosophical Society}} \textbf{\bibinfo{volume}{24}}, \bibinfo{pages}{89–110} (\bibinfo{year}{1928}).

\bibitem{Penrose_1971}
\bibinfo{author}{Penrose, R.}
\newblock \bibinfo{title}{Applications of negative dimensional tensors}.
\newblock In \bibinfo{editor}{Welsh, D. J.~A.} (ed.) \emph{\bibinfo{booktitle}{Combinatorial Mathematics and its Applications}}, \bibinfo{pages}{221--244} (\bibinfo{publisher}{Academic Press}, \bibinfo{address}{New York}, \bibinfo{year}{1971}).

\bibitem{biamonte_tensor_2017}
\bibinfo{author}{Biamonte, J.} \& \bibinfo{author}{Bergholm, V.}
\newblock \bibinfo{title}{Tensor {Networks} in a {Nutshell}}.
\newblock Preprint at \url{http://arxiv.org/abs/1708.00006} (\bibinfo{year}{2017}).

\bibitem{Bridgeman_2017}
\bibinfo{author}{Bridgeman, J.~C.} \& \bibinfo{author}{Chubb, C.~T.}
\newblock \bibinfo{title}{Hand-waving and interpretive dance: an introductory course on tensor networks}.
\newblock \emph{\bibinfo{journal}{Journal of Physics A: Mathematical and Theoretical}} \textbf{\bibinfo{volume}{50}}, \bibinfo{pages}{223001} (\bibinfo{year}{2017}).

\bibitem{PhysRevLett.69.2863}
\bibinfo{author}{White, S.~R.}
\newblock \bibinfo{title}{Density matrix formulation for quantum renormalization groups}.
\newblock \emph{\bibinfo{journal}{Phys. Rev. Lett.}} \textbf{\bibinfo{volume}{69}}, \bibinfo{pages}{2863--2866} (\bibinfo{year}{1992}).

\bibitem{white_ab_1999}
\bibinfo{author}{White, S.~R.} \& \bibinfo{author}{Martin, R.~L.}
\newblock \bibinfo{title}{Ab {Initio} {Quantum} {Chemistry} using the {Density} {Matrix} {Renormalization} {Group}}.
\newblock \emph{\bibinfo{journal}{The Journal of Chemical Physics}} \textbf{\bibinfo{volume}{110}}, \bibinfo{pages}{4127--4130} (\bibinfo{year}{1999}).

\bibitem{chan_density_2011}
\bibinfo{author}{Chan, G. K.-L.} \& \bibinfo{author}{Sharma, S.}
\newblock \bibinfo{title}{The {Density} {Matrix} {Renormalization} {Group} in {Quantum} {Chemistry}}.
\newblock \emph{\bibinfo{journal}{Annual Review of Physical Chemistry}} \textbf{\bibinfo{volume}{62}}, \bibinfo{pages}{465--481} (\bibinfo{year}{2011}).

\bibitem{chan_matrix_2016}
\bibinfo{author}{Chan, G. K.-L.}, \bibinfo{author}{Keselman, A.}, \bibinfo{author}{Nakatani, N.}, \bibinfo{author}{Li, Z.} \& \bibinfo{author}{White, S.~R.}
\newblock \bibinfo{title}{Matrix product operators, matrix product states, and ab initio density matrix renormalization group algorithms}.
\newblock \emph{\bibinfo{journal}{The Journal of Chemical Physics}} \textbf{\bibinfo{volume}{145}}, \bibinfo{pages}{014102} (\bibinfo{year}{2016}).

\bibitem{baiardi_density_2020}
\bibinfo{author}{Baiardi, A.} \& \bibinfo{author}{Reiher, M.}
\newblock \bibinfo{title}{The density matrix renormalization group in chemistry and molecular physics: {Recent} developments and new challenges}.
\newblock \emph{\bibinfo{journal}{The Journal of Chemical Physics}} \textbf{\bibinfo{volume}{152}}, \bibinfo{pages}{040903} (\bibinfo{year}{2020}).

\bibitem{Eisert_2010}
\bibinfo{author}{Eisert, J.}, \bibinfo{author}{Cramer, M.} \& \bibinfo{author}{Plenio, M.~B.}
\newblock \bibinfo{title}{Colloquium: Area laws for the entanglement entropy}.
\newblock \emph{\bibinfo{journal}{Reviews of Modern Physics}} \textbf{\bibinfo{volume}{82}}, \bibinfo{pages}{277–306} (\bibinfo{year}{2010}).

\bibitem{Fannes1992}
\bibinfo{author}{Fannes, M.}, \bibinfo{author}{Nachtergaele, B.} \& \bibinfo{author}{Werner, R.~F.}
\newblock \bibinfo{title}{Finitely correlated states on quantum spin chains}.
\newblock \emph{\bibinfo{journal}{Communications in Mathematical Physics}} \textbf{\bibinfo{volume}{144}}, \bibinfo{pages}{443--490} (\bibinfo{year}{1992}).

\bibitem{Klumper_1993}
\bibinfo{author}{Klümper, A.}, \bibinfo{author}{Schadschneider, A.} \& \bibinfo{author}{Zittartz, J.}
\newblock \bibinfo{title}{Matrix product ground states for one-dimensional spin-1 quantum antiferromagnets}.
\newblock \emph{\bibinfo{journal}{Europhysics Letters (EPL)}} \textbf{\bibinfo{volume}{24}}, \bibinfo{pages}{293–297} (\bibinfo{year}{1993}).

\bibitem{sharma_spin-adapted_2012}
\bibinfo{author}{Sharma, S.} \& \bibinfo{author}{Chan, G. K.-L.}
\newblock \bibinfo{title}{Spin-adapted density matrix renormalization group algorithms for quantum chemistry}.
\newblock \emph{\bibinfo{journal}{The Journal of Chemical Physics}} \textbf{\bibinfo{volume}{136}}, \bibinfo{pages}{124121} (\bibinfo{year}{2012}).

\bibitem{sharma_low-energy_2014}
\bibinfo{author}{Sharma, S.}, \bibinfo{author}{Sivalingam, K.}, \bibinfo{author}{Neese, F.} \& \bibinfo{author}{Chan, G. K.-L.}
\newblock \bibinfo{title}{Low-energy spectrum of iron–sulfur clusters directly from many-particle quantum mechanics}.
\newblock \emph{\bibinfo{journal}{Nature Chemistry}} \textbf{\bibinfo{volume}{6}}, \bibinfo{pages}{927--933} (\bibinfo{year}{2014}).

\bibitem{Schuch_2008}
\bibinfo{author}{Schuch, N.}, \bibinfo{author}{Wolf, M.~M.}, \bibinfo{author}{Verstraete, F.} \& \bibinfo{author}{Cirac, J.~I.}
\newblock \bibinfo{title}{Entropy scaling and simulability by matrix product states}.
\newblock \emph{\bibinfo{journal}{Phys. Rev. Lett.}} \textbf{\bibinfo{volume}{100}}, \bibinfo{pages}{030504} (\bibinfo{year}{2008}).

\bibitem{verstraete2004renormalization}
\bibinfo{author}{Verstraete, F.} \& \bibinfo{author}{Cirac, J.~I.}
\newblock \bibinfo{title}{Renormalization algorithms for quantum-many body systems in two and higher dimensions}. 
\newblock Preprint at \url{https://arxiv.org/abs/cond-mat/0407066} (\bibinfo{year}{2004}).

\bibitem{Haferkamp_2020}
\bibinfo{author}{Haferkamp, J.}, \bibinfo{author}{Hangleiter, D.}, \bibinfo{author}{Eisert, J.} \& \bibinfo{author}{Gluza, M.}
\newblock \bibinfo{title}{Contracting projected entangled pair states is average-case hard}.
\newblock \emph{\bibinfo{journal}{Physical Review Research}} \textbf{\bibinfo{volume}{2}} (\bibinfo{year}{2020}).

\bibitem{Zalatel_2020}
\bibinfo{author}{Zaletel, M.~P.} \& \bibinfo{author}{Pollmann, F.}
\newblock \bibinfo{title}{Isometric tensor network states in two dimensions}.
\newblock \emph{\bibinfo{journal}{Phys. Rev. Lett.}} \textbf{\bibinfo{volume}{124}}, \bibinfo{pages}{037201} (\bibinfo{year}{2020}).

\bibitem{Feynman1986}
\bibinfo{author}{Feynman, R.~P.}
\newblock \bibinfo{title}{Quantum mechanical computers}.
\newblock \emph{\bibinfo{journal}{Foundations of Physics}} \textbf{\bibinfo{volume}{16}}, \bibinfo{pages}{507--531} (\bibinfo{year}{1986}).

\bibitem{Nielsen_Chuang_2010}
\bibinfo{author}{Nielsen, M.~A.} \& \bibinfo{author}{Chuang, I.~L.}
\newblock \emph{\bibinfo{title}{Quantum Computation and Quantum Information: 10th Anniversary Edition}} (\bibinfo{publisher}{Cambridge University Press}, \bibinfo{year}{2010}).

\bibitem{kitaev1995quantum}
\bibinfo{author}{Kitaev, A.~Y.}
\newblock \bibinfo{title}{Quantum measurements and the abelian stabilizer problem}.
\newblock Preprint at \url{https://arxiv.org/abs/quant-ph/9511026} (\bibinfo{year}{1995}).

\bibitem{Aspuru_Guzik_2005}
\bibinfo{author}{Aspuru-Guzik, A.}, \bibinfo{author}{Dutoi, A.~D.}, \bibinfo{author}{Love, P.~J.} \& \bibinfo{author}{Head-Gordon, M.}
\newblock \bibinfo{title}{Simulated quantum computation of molecular energies}.
\newblock \emph{\bibinfo{journal}{Science}} \textbf{\bibinfo{volume}{309}}, \bibinfo{pages}{1704–1707} (\bibinfo{year}{2005}).

\bibitem{Su_2021}
\bibinfo{author}{Su, Y.}, \bibinfo{author}{Berry, D.~W.}, \bibinfo{author}{Wiebe, N.}, \bibinfo{author}{Rubin, N.} \& \bibinfo{author}{Babbush, R.}
\newblock \bibinfo{title}{Fault-tolerant quantum simulations of chemistry in first quantization}.
\newblock \emph{\bibinfo{journal}{PRX Quantum}} \textbf{\bibinfo{volume}{2}} (\bibinfo{year}{2021}).

\bibitem{Ding_2023}
\bibinfo{author}{Ding, Z.} \& \bibinfo{author}{Lin, L.}
\newblock \bibinfo{title}{Even shorter quantum circuit for phase estimation on early fault-tolerant quantum computers with applications to ground-state energy estimation}.
\newblock \emph{\bibinfo{journal}{PRX Quantum}} \textbf{\bibinfo{volume}{4}} (\bibinfo{year}{2023}).

\bibitem{Peruzzo2014}
\bibinfo{author}{Peruzzo, A.} \emph{et~al.}
\newblock \bibinfo{title}{A variational eigenvalue solver on a photonic quantum processor}.
\newblock \emph{\bibinfo{journal}{Nature Communications}} \textbf{\bibinfo{volume}{5}}, \bibinfo{pages}{4213} (\bibinfo{year}{2014}).

\bibitem{Tilly_2022}
\bibinfo{author}{Tilly, J.} \emph{et~al.}
\newblock \bibinfo{title}{The variational quantum eigensolver: A review of methods and best practices}.
\newblock \emph{\bibinfo{journal}{Physics Reports}} \textbf{\bibinfo{volume}{986}}, \bibinfo{pages}{1–128} (\bibinfo{year}{2022}).

\bibitem{Grimsley2019}
\bibinfo{author}{Grimsley, H.~R.}, \bibinfo{author}{Economou, S.~E.}, \bibinfo{author}{Barnes, E.} \& \bibinfo{author}{Mayhall, N.~J.}
\newblock \bibinfo{title}{An adaptive variational algorithm for exact molecular simulations on a quantum computer}.
\newblock \emph{\bibinfo{journal}{Nature Communications}} \textbf{\bibinfo{volume}{10}}, \bibinfo{pages}{3007} (\bibinfo{year}{2019}).

\bibitem{Ryabinkin2018}
\bibinfo{author}{Ryabinkin, I.~G.}, \bibinfo{author}{Yen, T.-C.}, \bibinfo{author}{Genin, S.~N.} \& \bibinfo{author}{Izmaylov, A.~F.}
\newblock \bibinfo{title}{Qubit coupled cluster method: A systematic approach to quantum chemistry on a quantum computer}.
\newblock \emph{\bibinfo{journal}{Journal of Chemical Theory and Computation}} \textbf{\bibinfo{volume}{14}}, \bibinfo{pages}{6317--6326} (\bibinfo{year}{2018}).

\bibitem{Mizukami_2020}
\bibinfo{author}{Mizukami, W.} \emph{et~al.}
\newblock \bibinfo{title}{Orbital optimized unitary coupled cluster theory for quantum computer}.
\newblock \emph{\bibinfo{journal}{Phys. Rev. Res.}} \textbf{\bibinfo{volume}{2}}, \bibinfo{pages}{033421} (\bibinfo{year}{2020}).

\bibitem{McClean_2018}
\bibinfo{author}{McClean, J.~R.}, \bibinfo{author}{Boixo, S.}, \bibinfo{author}{Smelyanskiy, V.~N.}, \bibinfo{author}{Babbush, R.} \& \bibinfo{author}{Neven, H.}
\newblock \bibinfo{title}{Barren plateaus in quantum neural network training landscapes}.
\newblock \emph{\bibinfo{journal}{Nature Communications}} \textbf{\bibinfo{volume}{9}} (\bibinfo{year}{2018}).

\bibitem{PRXQuantum.4.030307}
\bibinfo{author}{Baek, U.} \emph{et~al.}
\newblock \bibinfo{title}{Say no to optimization: A nonorthogonal quantum eigensolver}.
\newblock \emph{\bibinfo{journal}{PRX Quantum}} \textbf{\bibinfo{volume}{4}}, \bibinfo{pages}{030307} (\bibinfo{year}{2023}).

\bibitem{motta_low_2021}
\bibinfo{author}{Motta, M.} \emph{et~al.}
\newblock \bibinfo{title}{Low rank representations for quantum simulation of electronic structure}.
\newblock \emph{\bibinfo{journal}{npj Quantum Information}} \textbf{\bibinfo{volume}{7}}, \bibinfo{pages}{1--7} (\bibinfo{year}{2021}).

\bibitem{bauer_quantum_2020}
\bibinfo{author}{Bauer, B.}, \bibinfo{author}{Bravyi, S.}, \bibinfo{author}{Motta, M.} \& \bibinfo{author}{Chan, G. K.-L.}
\newblock \bibinfo{title}{Quantum algorithms for quantum chemistry and quantum materials science}.
\newblock \emph{\bibinfo{journal}{Chemical Reviews}} \textbf{\bibinfo{volume}{120}}, \bibinfo{pages}{12685--12717} (\bibinfo{year}{2020}).

\bibitem{haghshenas_variational_2022}
\bibinfo{author}{Haghshenas, R.}, \bibinfo{author}{Gray, J.}, \bibinfo{author}{Potter, A.~C.} \& \bibinfo{author}{Chan, G. K.-L.}
\newblock \bibinfo{title}{Variational {Power} of {Quantum} {Circuit} {Tensor} {Networks}}.
\newblock \emph{\bibinfo{journal}{Physical Review X}} \textbf{\bibinfo{volume}{12}}, \bibinfo{pages}{011047} (\bibinfo{year}{2022}).

\bibitem{kim_robust_2017}
\bibinfo{author}{Kim, I.~H.} \& \bibinfo{author}{Swingle, B.}
\newblock \bibinfo{title}{Robust entanglement renormalization on a noisy quantum computer}. 
\newblock Preprint at \url{http://arxiv.org/abs/1711.07500} (\bibinfo{year}{2017}).

\bibitem{huggins_towards_2019}
\bibinfo{author}{Huggins, W.}, \bibinfo{author}{Patel, P.}, \bibinfo{author}{Whaley, K.~B.} \& \bibinfo{author}{Stoudenmire, E.~M.}
\newblock \bibinfo{title}{Towards {Quantum} {Machine} {Learning} with {Tensor} {Networks}}.
\newblock \emph{\bibinfo{journal}{Quantum Science and Technology}} \textbf{\bibinfo{volume}{4}}, \bibinfo{pages}{024001} (\bibinfo{year}{2019}).

\bibitem{borregaard_noise-robust_2021}
\bibinfo{author}{Borregaard, J.}, \bibinfo{author}{Christandl, M.} \& \bibinfo{author}{Stilck~França, D.}
\newblock \bibinfo{title}{Noise-robust exploration of many-body quantum states on near-term quantum devices}.
\newblock \emph{\bibinfo{journal}{npj Quantum Information}} \textbf{\bibinfo{volume}{7}}, \bibinfo{pages}{1--6} (\bibinfo{year}{2021}).

\bibitem{jamet2023anderson}
\bibinfo{author}{Jamet, F.} \emph{et~al.}
\newblock \bibinfo{title}{Anderson impurity solver integrating tensor network methods with quantum computing}.
\newblock Preprint at \url{https://arxiv.org/abs/2304.06587} (\bibinfo{year}{2023}).

\bibitem{Lubasch_2020}
\bibinfo{author}{Lubasch, M.}, \bibinfo{author}{Joo, J.}, \bibinfo{author}{Moinier, P.}, \bibinfo{author}{Kiffner, M.} \& \bibinfo{author}{Jaksch, D.}
\newblock \bibinfo{title}{Variational quantum algorithms for nonlinear problems}.
\newblock \emph{\bibinfo{journal}{Phys. Rev. A}} \textbf{\bibinfo{volume}{101}}, \bibinfo{pages}{010301} (\bibinfo{year}{2020}).

\bibitem{Keever_2023}
\bibinfo{author}{Mc~Keever, C.} \& \bibinfo{author}{Lubasch, M.}
\newblock \bibinfo{title}{Classically optimized hamiltonian simulation}.
\newblock \emph{\bibinfo{journal}{Phys. Rev. Res.}} \textbf{\bibinfo{volume}{5}}, \bibinfo{pages}{023146} (\bibinfo{year}{2023}).

\bibitem{Huang_2023}
\bibinfo{author}{Huang, J.} \emph{et~al.}
\newblock \bibinfo{title}{Tensor-network-assisted variational quantum algorithm}.
\newblock \emph{\bibinfo{journal}{Physical Review A}} \textbf{\bibinfo{volume}{108}} (\bibinfo{year}{2023}).

\bibitem{Watanabe_2024}
\bibinfo{author}{Watanabe, R.}, \bibinfo{author}{Fujii, K.} \& \bibinfo{author}{Ueda, H.}
\newblock \bibinfo{title}{Variational quantum eigensolver with embedded entanglement using a tensor-network ansatz}.
\newblock \emph{\bibinfo{journal}{Phys. Rev. Res.}} \textbf{\bibinfo{volume}{6}}, \bibinfo{pages}{023009} (\bibinfo{year}{2024}).

\bibitem{Miao_2023}
\bibinfo{author}{Miao, Q.} \& \bibinfo{author}{Barthel, T.}
\newblock \bibinfo{title}{Quantum-classical eigensolver using multiscale entanglement renormalization}.
\newblock \emph{\bibinfo{journal}{Phys. Rev. Res.}} \textbf{\bibinfo{volume}{5}}, \bibinfo{pages}{033141} (\bibinfo{year}{2023}).

\bibitem{miao2023convergence}
\bibinfo{author}{Miao, Q.} \& \bibinfo{author}{Barthel, T.}
\newblock \bibinfo{title}{Convergence and quantum advantage of trotterized mera for strongly-correlated systems}.
\newblock \emph{\bibinfo{journal}{Quantum}} \textbf{\bibinfo{volume}{9}}, \bibinfo{pages}{1631} (\bibinfo{year}{2025}).

\bibitem{McClean_2017}
\bibinfo{author}{McClean, J.~R.}, \bibinfo{author}{Kimchi-Schwartz, M.~E.}, \bibinfo{author}{Carter, J.} \& \bibinfo{author}{de~Jong, W.~A.}
\newblock \bibinfo{title}{Hybrid quantum-classical hierarchy for mitigation of decoherence and determination of excited states}.
\newblock \emph{\bibinfo{journal}{Phys. Rev. A}} \textbf{\bibinfo{volume}{95}}, \bibinfo{pages}{042308} (\bibinfo{year}{2017}).

\bibitem{Huggins_2020}
\bibinfo{author}{Huggins, W.~J.}, \bibinfo{author}{Lee, J.}, \bibinfo{author}{Baek, U.}, \bibinfo{author}{O’Gorman, B.} \& \bibinfo{author}{Whaley, K.~B.}
\newblock \bibinfo{title}{A non-orthogonal variational quantum eigensolver}.
\newblock \emph{\bibinfo{journal}{New Journal of Physics}} \textbf{\bibinfo{volume}{22}}, \bibinfo{pages}{073009} (\bibinfo{year}{2020}).

\bibitem{PhysRevA.105.022417}
\bibinfo{author}{Cortes, C.~L.} \& \bibinfo{author}{Gray, S.~K.}
\newblock \bibinfo{title}{Quantum krylov subspace algorithms for ground- and excited-state energy estimation}.
\newblock \emph{\bibinfo{journal}{Phys. Rev. A}} \textbf{\bibinfo{volume}{105}}, \bibinfo{pages}{022417} (\bibinfo{year}{2022}).

\bibitem{francis2022subspace}
\bibinfo{author}{Francis, A.}, \bibinfo{author}{Agrawal, A.~A.}, \bibinfo{author}{Howard, J.~H.}, \bibinfo{author}{Kökcü, E.} \& \bibinfo{author}{Kemper, A.~F.}
\newblock \bibinfo{title}{Subspace diagonalization on quantum computers using eigenvector continuation}.
\newblock Preprint at \url{https://arxiv.org/abs/2209.10571} (\bibinfo{year}{2022}).

\bibitem{motta2023subspace}
\bibinfo{author}{Motta, M.} \emph{et~al.}
\newblock \bibinfo{title}{Subspace methods for electronic structure simulations on quantum computers}.
\newblock \emph{\bibinfo{journal}{Electronic Structure}} \textbf{\bibinfo{volume}{6}}, \bibinfo{pages}{013001} (\bibinfo{year}{2024}).

\bibitem{Fan2023}
\bibinfo{author}{Fan, Y.}, \bibinfo{author}{Liu, J.}, \bibinfo{author}{Li, Z.} \& \bibinfo{author}{Yang, J.}
\newblock \bibinfo{title}{Quantum circuit matrix product state ansatz for large-scale simulations of molecules}.
\newblock \emph{\bibinfo{journal}{Journal of Chemical Theory and Computation}} \textbf{\bibinfo{volume}{19}}, \bibinfo{pages}{5407--5417} (\bibinfo{year}{2023}).

\bibitem{Meth_2022}
\bibinfo{author}{Meth, M.} \emph{et~al.}
\newblock \bibinfo{title}{Probing phases of quantum matter with an ion-trap tensor-network quantum eigensolver}.
\newblock \emph{\bibinfo{journal}{Phys. Rev. X}} \textbf{\bibinfo{volume}{12}}, \bibinfo{pages}{041035} (\bibinfo{year}{2022}).

\bibitem{Rudolph2023}
\bibinfo{author}{Rudolph, M.~S.} \emph{et~al.}
\newblock \bibinfo{title}{Synergistic pretraining of parametrized quantum circuits via tensor networks}.
\newblock \emph{\bibinfo{journal}{Nature Communications}} \textbf{\bibinfo{volume}{14}}, \bibinfo{pages}{8367} (\bibinfo{year}{2023}).

\bibitem{khan2023preoptimizing}
\bibinfo{author}{Khan, A.}, \bibinfo{author}{Clark, B.~K.} \& \bibinfo{author}{Tubman, N.~M.}
\newblock \bibinfo{title}{Pre-optimizing variational quantum eigensolvers with tensor networks}. 
\newblock Preprint at \url{https://arxiv.org/abs/2310.12965} (\bibinfo{year}{2023}).

\bibitem{miao2024isometric}
\bibinfo{author}{Miao, Q.} \& \bibinfo{author}{Barthel, T.}
\newblock \bibinfo{title}{Isometric tensor network optimization for extensive hamiltonians is free of barren plateaus}.
\newblock \emph{\bibinfo{journal}{Physical Review A}} \textbf{\bibinfo{volume}{109}} (\bibinfo{year}{2024}).

\bibitem{barthel2023absence}
\bibinfo{author}{Barthel, T.} \& \bibinfo{author}{Miao, Q.}
\newblock \bibinfo{title}{Absence of barren plateaus and scaling of gradients in the energy optimization of isometric tensor network states}.
\newblock Preprint at \url{https://arxiv.org/abs/2304.00161} (\bibinfo{year}{2023}).

\bibitem{CerveroMartin2023barrenplateausin}
\bibinfo{author}{Cervero~Mart{\'{i}}n, E.}, \bibinfo{author}{Plekhanov, K.} \& \bibinfo{author}{Lubasch, M.}
\newblock \bibinfo{title}{Barren plateaus in quantum tensor network optimization}.
\newblock \emph{\bibinfo{journal}{{Quantum}}} \textbf{\bibinfo{volume}{7}}, \bibinfo{pages}{974} (\bibinfo{year}{2023}).

\bibitem{Helgaker_2000}
\bibinfo{author}{Helgaker, T.}, \bibinfo{author}{Jørgensen, P.} \& \bibinfo{author}{Olsen, J.}
\newblock \emph{\bibinfo{title}{Second Quantization}}, chap.~\bibinfo{chapter}{1}, \bibinfo{pages}{1--33} (\bibinfo{publisher}{John Wiley \& Sons, Ltd}, \bibinfo{year}{2000}).

\bibitem{doi:10.1137/090752286}
\bibinfo{author}{Oseledets, I.~V.}
\newblock \bibinfo{title}{Tensor-train decomposition}.
\newblock \emph{\bibinfo{journal}{SIAM Journal on Scientific Computing}} \textbf{\bibinfo{volume}{33}}, \bibinfo{pages}{2295--2317} (\bibinfo{year}{2011}).

\bibitem{Vidal_2003}
\bibinfo{author}{Vidal, G.}
\newblock \bibinfo{title}{Efficient classical simulation of slightly entangled quantum computations}.
\newblock \emph{\bibinfo{journal}{Physical Review Letters}} \textbf{\bibinfo{volume}{91}} (\bibinfo{year}{2003}).

\bibitem{Schuch_2007}
\bibinfo{author}{Schuch, N.}, \bibinfo{author}{Wolf, M.~M.}, \bibinfo{author}{Verstraete, F.} \& \bibinfo{author}{Cirac, J.~I.}
\newblock \bibinfo{title}{Computational complexity of projected entangled pair states}.
\newblock \emph{\bibinfo{journal}{Phys. Rev. Lett.}} \textbf{\bibinfo{volume}{98}}, \bibinfo{pages}{140506} (\bibinfo{year}{2007}).

\bibitem{Arad_2010}
\bibinfo{author}{Arad, I.} \& \bibinfo{author}{Landau, Z.}
\newblock \bibinfo{title}{Quantum computation and the evaluation of tensor networks}.
\newblock \emph{\bibinfo{journal}{SIAM Journal on Computing}} \textbf{\bibinfo{volume}{39}}, \bibinfo{pages}{3089--3121} (\bibinfo{year}{2010}).

\bibitem{Jordan1928}
\bibinfo{author}{Jordan, P.} \& \bibinfo{author}{Wigner, E.}
\newblock \bibinfo{title}{{\"U}ber das paulische {\"a}quivalenzverbot}.
\newblock \emph{\bibinfo{journal}{Zeitschrift f{\"u}r Physik}} \textbf{\bibinfo{volume}{47}}, \bibinfo{pages}{631--651} (\bibinfo{year}{1928}).

\bibitem{Baek_2022}
\bibinfo{author}{Baek, U.} \emph{et~al.}
\newblock \bibinfo{title}{Say no to optimization: A non-orthogonal quantum eigensolver} (\bibinfo{year}{2022}).

\bibitem{Sundstrom_2014}
\bibinfo{author}{Sundstrom, E.~J.} \& \bibinfo{author}{Head-Gordon, M.}
\newblock \bibinfo{title}{{Non-orthogonal configuration interaction for the calculation of multielectron excited states}}.
\newblock \emph{\bibinfo{journal}{The Journal of Chemical Physics}} \textbf{\bibinfo{volume}{140}}, \bibinfo{pages}{114103} (\bibinfo{year}{2014}).

\bibitem{Thom_2009}
\bibinfo{author}{Thom, A. J.~W.} \& \bibinfo{author}{Head-Gordon, M.}
\newblock \bibinfo{title}{{Hartree–Fock solutions as a quasidiabatic basis for nonorthogonal configuration interaction}}.
\newblock \emph{\bibinfo{journal}{The Journal of Chemical Physics}} \textbf{\bibinfo{volume}{131}}, \bibinfo{pages}{124113} (\bibinfo{year}{2009}).

\bibitem{Krumnow_2016}
\bibinfo{author}{Krumnow, C.}, \bibinfo{author}{Veis, L.}, \bibinfo{author}{Legeza, O.} \& \bibinfo{author}{Eisert, J.}
\newblock \bibinfo{title}{Fermionic orbital optimization in tensor network states}.
\newblock \emph{\bibinfo{journal}{Phys. Rev. Lett.}} \textbf{\bibinfo{volume}{117}}, \bibinfo{pages}{210402} (\bibinfo{year}{2016}).

\bibitem{Zgid_2008}
\bibinfo{author}{Zgid, D.} \& \bibinfo{author}{Nooijen, M.}
\newblock \bibinfo{title}{{The density matrix renormalization group self-consistent field method: Orbital optimization with the density matrix renormalization group method in the active space}}.
\newblock \emph{\bibinfo{journal}{The Journal of Chemical Physics}} \textbf{\bibinfo{volume}{128}}, \bibinfo{pages}{144116} (\bibinfo{year}{2008}).

\bibitem{Ghosh_2008}
\bibinfo{author}{Ghosh, D.}, \bibinfo{author}{Hachmann, J.}, \bibinfo{author}{Yanai, T.} \& \bibinfo{author}{Chan, G. K.-L.}
\newblock \bibinfo{title}{Orbital optimization in the density matrix renormalization group, with applications to polyenes and $\beta$-carotene}.
\newblock \emph{\bibinfo{journal}{The Journal of Chemical Physics}} \textbf{\bibinfo{volume}{128}} (\bibinfo{year}{2008}).

\bibitem{Verstraete_2009}
\bibinfo{author}{Verstraete, F.}, \bibinfo{author}{Cirac, J.~I.} \& \bibinfo{author}{Latorre, J.~I.}
\newblock \bibinfo{title}{Quantum circuits for strongly correlated quantum systems}.
\newblock \emph{\bibinfo{journal}{Phys. Rev. A}} \textbf{\bibinfo{volume}{79}}, \bibinfo{pages}{032316} (\bibinfo{year}{2009}).

\bibitem{banerjee_sorting_2019}
\bibinfo{author}{Banerjee, I.}, \bibinfo{author}{Richards, D.} \& \bibinfo{author}{Shinkar, I.}
\newblock \bibinfo{title}{Sorting {Networks} on {Restricted} {Topologies}}.
\newblock In \bibinfo{editor}{Catania, B.}, \bibinfo{editor}{Královič, R.}, \bibinfo{editor}{Nawrocki, J.} \& \bibinfo{editor}{Pighizzini, G.} (eds.) \emph{\bibinfo{booktitle}{{SOFSEM} 2019: {Theory} and {Practice} of {Computer} {Science}}}, \bibinfo{pages}{54--66} (\bibinfo{publisher}{Springer International Publishing}, \bibinfo{address}{Cham}, \bibinfo{year}{2019}).

\bibitem{Knuth98}
\bibinfo{author}{Knuth, D.~E.}
\newblock \emph{\bibinfo{title}{The Art of Computer Programming}}, vol. \bibinfo{volume}{3: Sorting and Searching} (\bibinfo{publisher}{Addison-Wesley}, \bibinfo{year}{1998}).

\bibitem{10.1145/320831.320833}
\bibinfo{author}{Friend, E.~H.}
\newblock \bibinfo{title}{Sorting on electronic computer systems}.
\newblock \emph{\bibinfo{journal}{J. ACM}} \textbf{\bibinfo{volume}{3}}, \bibinfo{pages}{134–168} (\bibinfo{year}{1956}).

\bibitem{Qian_2024}
\bibinfo{author}{Qian, X.}, \bibinfo{author}{Huang, J.} \& \bibinfo{author}{Qin, M.}
\newblock \bibinfo{title}{Augmenting density matrix renormalization group with clifford circuits}.
\newblock \emph{\bibinfo{journal}{Phys. Rev. Lett.}} \textbf{\bibinfo{volume}{133}}, \bibinfo{pages}{190402} (\bibinfo{year}{2024}).

\bibitem{lami2024quantumstatedesignsclifford}
\bibinfo{author}{Lami, G.}, \bibinfo{author}{Haug, T.} \& \bibinfo{author}{Nardis, J.~D.}
\newblock \bibinfo{title}{Quantum state designs with clifford enhanced matrix product states}. 
\newblock Preprint at \url{https://arxiv.org/abs/2404.18751} (\bibinfo{year}{2024}).

\bibitem{Kivlichan_2018}
\bibinfo{author}{Kivlichan, I.~D.} \emph{et~al.}
\newblock \bibinfo{title}{Quantum simulation of electronic structure with linear depth and connectivity}.
\newblock \emph{\bibinfo{journal}{Phys. Rev. Lett.}} \textbf{\bibinfo{volume}{120}}, \bibinfo{pages}{110501} (\bibinfo{year}{2018}).

\bibitem{THOULESS1960225}
\bibinfo{author}{Thouless, D.}
\newblock \bibinfo{title}{Stability conditions and nuclear rotations in the hartree-fock theory}.
\newblock \emph{\bibinfo{journal}{Nuclear Physics}} \textbf{\bibinfo{volume}{21}}, \bibinfo{pages}{225--232} (\bibinfo{year}{1960}).

\bibitem{Li_2022}
\bibinfo{author}{Li, W.}, \bibinfo{author}{Ren, J.}, \bibinfo{author}{Yang, H.} \& \bibinfo{author}{Shuai, Z.}
\newblock \bibinfo{title}{On the fly swapping algorithm for ordering of degrees of freedom in density matrix renormalization group}.
\newblock \emph{\bibinfo{journal}{Journal of Physics: Condensed Matter}} \textbf{\bibinfo{volume}{34}}, \bibinfo{pages}{254003} (\bibinfo{year}{2022}).

\bibitem{Schon_2005}
\bibinfo{author}{Sch\"on, C.}, \bibinfo{author}{Solano, E.}, \bibinfo{author}{Verstraete, F.}, \bibinfo{author}{Cirac, J.~I.} \& \bibinfo{author}{Wolf, M.~M.}
\newblock \bibinfo{title}{Sequential generation of entangled multiqubit states}.
\newblock \emph{\bibinfo{journal}{Phys. Rev. Lett.}} \textbf{\bibinfo{volume}{95}}, \bibinfo{pages}{110503} (\bibinfo{year}{2005}).

\bibitem{fomichev2024initial}
\bibinfo{author}{Fomichev, S.} \emph{et~al.}
\newblock \bibinfo{title}{Initial state preparation for quantum chemistry on quantum computers}.
\newblock \emph{\bibinfo{journal}{PRX Quantum}} \textbf{\bibinfo{volume}{5}}, \bibinfo{pages}{040339} (\bibinfo{year}{2024}).

\bibitem{ran_encoding_2020}
\bibinfo{author}{Ran, S.-J.}
\newblock \bibinfo{title}{Encoding of matrix product states into quantum circuits of one- and two-qubit gates}.
\newblock \emph{\bibinfo{journal}{Physical Review A}} \textbf{\bibinfo{volume}{101}}, \bibinfo{pages}{032310} (\bibinfo{year}{2020}).

\bibitem{Malz_2024}
\bibinfo{author}{Malz, D.}, \bibinfo{author}{Styliaris, G.}, \bibinfo{author}{Wei, Z.-Y.} \& \bibinfo{author}{Cirac, J.~I.}
\newblock \bibinfo{title}{Preparation of matrix product states with log-depth quantum circuits}.
\newblock \emph{\bibinfo{journal}{Physical Review Letters}} \textbf{\bibinfo{volume}{132}} (\bibinfo{year}{2024}).

\bibitem{Shende_2006}
\bibinfo{author}{Shende, V.}, \bibinfo{author}{Bullock, S.} \& \bibinfo{author}{Markov, I.}
\newblock \bibinfo{title}{Synthesis of quantum-logic circuits}.
\newblock \emph{\bibinfo{journal}{IEEE Transactions on Computer-Aided Design of Integrated Circuits and Systems}} \textbf{\bibinfo{volume}{25}}, \bibinfo{pages}{1000–1010} (\bibinfo{year}{2006}).

\bibitem{low2018trading}
\bibinfo{author}{Low, G.~H.}, \bibinfo{author}{Kliuchnikov, V.} \& \bibinfo{author}{Schaeffer, L.}
\newblock \bibinfo{title}{Trading t gates for dirty qubits in state preparation and unitary synthesis}.
\newblock \emph{\bibinfo{journal}{Quantum}} \textbf{\bibinfo{volume}{8}}, \bibinfo{pages}{1375} (\bibinfo{year}{2024}).

\bibitem{dov2022approximate}
\bibinfo{author}{Ben-Dov, M.}, \bibinfo{author}{Shnaiderov, D.}, \bibinfo{author}{Makmal, A.} \& \bibinfo{author}{Dalla~Torre, E.~G.}
\newblock \bibinfo{title}{Approximate encoding of quantum states using shallow circuits}.
\newblock \emph{\bibinfo{journal}{npj Quantum Information}} \textbf{\bibinfo{volume}{10}} (\bibinfo{year}{2024}).

\bibitem{Rudolph_2024}
\bibinfo{author}{Rudolph, M.~S.}, \bibinfo{author}{Chen, J.}, \bibinfo{author}{Miller, J.}, \bibinfo{author}{Acharya, A.} \& \bibinfo{author}{Perdomo-Ortiz, A.}
\newblock \bibinfo{title}{Decomposition of matrix product states into shallow quantum circuits}.
\newblock \emph{\bibinfo{journal}{Quantum Science and Technology}} \textbf{\bibinfo{volume}{9}}, \bibinfo{pages}{015012} (\bibinfo{year}{2023}).

\bibitem{harris_digital_2010}
\bibinfo{author}{Harris, D.} \& \bibinfo{author}{Harris, S.}
\newblock \emph{\bibinfo{title}{Digital {Design} and {Computer} {Architecture}}} (\bibinfo{publisher}{Morgan Kaufmann}, \bibinfo{year}{2010}).

\bibitem{Singh_2010}
\bibinfo{author}{Singh, S.}, \bibinfo{author}{Pfeifer, R. N.~C.} \& \bibinfo{author}{Vidal, G.}
\newblock \bibinfo{title}{Tensor network decompositions in the presence of a global symmetry}.
\newblock \emph{\bibinfo{journal}{Phys. Rev. A}} \textbf{\bibinfo{volume}{82}}, \bibinfo{pages}{050301} (\bibinfo{year}{2010}).

\bibitem{Singh_2011}
\bibinfo{author}{Singh, S.}, \bibinfo{author}{Pfeifer, R. N.~C.} \& \bibinfo{author}{Vidal, G.}
\newblock \bibinfo{title}{Tensor network states and algorithms in the presence of a global u(1) symmetry}.
\newblock \emph{\bibinfo{journal}{Phys. Rev. B}} \textbf{\bibinfo{volume}{83}}, \bibinfo{pages}{115125} (\bibinfo{year}{2011}).

\bibitem{epperly_2021}
\bibinfo{author}{Epperly, E.~N.}, \bibinfo{author}{Lin, L.} \& \bibinfo{author}{Nakatsukasa, Y.}
\newblock \bibinfo{title}{A theory of quantum subspace diagonalization}.
\newblock \emph{\bibinfo{journal}{SIAM Journal on Matrix Analysis and Applications}} \textbf{\bibinfo{volume}{43}}, \bibinfo{pages}{1263-1290}
(\bibinfo{year}{2022}).

\bibitem{li_spin-projected_2017}
\bibinfo{author}{Li, Z.} \& \bibinfo{author}{Chan, G. K.-L.}
\newblock \bibinfo{title}{Spin-{Projected} {Matrix} {Product} {States}: {Versatile} {Tool} for {Strongly} {Correlated} {Systems}}.
\newblock \emph{\bibinfo{journal}{Journal of Chemical Theory and Computation}} \textbf{\bibinfo{volume}{13}}, \bibinfo{pages}{2681--2695} (\bibinfo{year}{2017}).

\bibitem{Yordanov_2020}
\bibinfo{author}{Yordanov, Y.~S.}, \bibinfo{author}{Arvidsson-Shukur, D. R.~M.} \& \bibinfo{author}{Barnes, C. H.~W.}
\newblock \bibinfo{title}{Efficient quantum circuits for quantum computational chemistry}.
\newblock \emph{\bibinfo{journal}{Phys. Rev. A}} \textbf{\bibinfo{volume}{102}}, \bibinfo{pages}{062612} (\bibinfo{year}{2020}).

\bibitem{Imada_1998}
\bibinfo{author}{Imada, M.}, \bibinfo{author}{Fujimori, A.} \& \bibinfo{author}{Tokura, Y.}
\newblock \bibinfo{title}{Metal-insulator transitions}.
\newblock \emph{\bibinfo{journal}{Rev. Mod. Phys.}} \textbf{\bibinfo{volume}{70}}, \bibinfo{pages}{1039--1263} (\bibinfo{year}{1998}).

\bibitem{askerka_o2-evolving_2017}
\bibinfo{author}{Askerka, M.}, \bibinfo{author}{Brudvig, G.~W.} \& \bibinfo{author}{Batista, V.~S.}
\newblock \bibinfo{title}{The {O2}-{Evolving} {Complex} of {Photosystem} {II}: {Recent} {Insights} from {Quantum} {Mechanics}/{Molecular} {Mechanics} ({QM}/{MM}), {Extended} {X}-ray {Absorption} {Fine} {Structure} ({EXAFS}), and {Femtosecond} {X}-ray {Crystallography} {Data}}.
\newblock \emph{\bibinfo{journal}{Accounts of Chemical Research}} \textbf{\bibinfo{volume}{50}}, \bibinfo{pages}{41--48} (\bibinfo{year}{2017}).

\bibitem{Rissler_2006}
\bibinfo{author}{Rissler, J.}, \bibinfo{author}{Noack, R.~M.} \& \bibinfo{author}{White, S.~R.}
\newblock \bibinfo{title}{Measuring orbital interaction using quantum information theory}.
\newblock \emph{\bibinfo{journal}{Chemical Physics}} \textbf{\bibinfo{volume}{323}}, \bibinfo{pages}{519–531} (\bibinfo{year}{2006}).

\bibitem{Wecker_2015}
\bibinfo{author}{Wecker, D.}, \bibinfo{author}{Hastings, M.~B.} \& \bibinfo{author}{Troyer, M.}
\newblock \bibinfo{title}{Progress towards practical quantum variational algorithms}.
\newblock \emph{\bibinfo{journal}{Phys. Rev. A}} \textbf{\bibinfo{volume}{92}}, \bibinfo{pages}{042303} (\bibinfo{year}{2015}).

\bibitem{Huang_2020}
\bibinfo{author}{Huang, H.-Y.}, \bibinfo{author}{Kueng, R.} \& \bibinfo{author}{Preskill, J.}
\newblock \bibinfo{title}{Predicting many properties of a quantum system from very few measurements}.
\newblock \emph{\bibinfo{journal}{Nature Physics}} \textbf{\bibinfo{volume}{16}}, \bibinfo{pages}{1050–1057} (\bibinfo{year}{2020}).

\bibitem{Brassard_2002}
\bibinfo{author}{Brassard, G.}, \bibinfo{author}{Høyer, P.}, \bibinfo{author}{Mosca, M.} \& \bibinfo{author}{Tapp, A.}
\newblock \bibinfo{title}{Quantum amplitude amplification and estimation} (\bibinfo{year}{2002}).

\bibitem{Begusic_2024}
\bibinfo{author}{Begušić, T.}, \bibinfo{author}{Gray, J.} \& \bibinfo{author}{Chan, G. K.-L.}
\newblock \bibinfo{title}{Fast and converged classical simulations of evidence for the utility of quantum computing before fault tolerance}.
\newblock \emph{\bibinfo{journal}{Science Advances}} \textbf{\bibinfo{volume}{10}}, \bibinfo{pages}{eadk4321} (\bibinfo{year}{2024}).

\bibitem{shin2023dequantizing}
\bibinfo{author}{Shin, S.}, \bibinfo{author}{Teo, Y.~S.} \& \bibinfo{author}{Jeong, H.}
\newblock \bibinfo{title}{Dequantizing quantum machine learning models using tensor networks}.
\newblock \emph{\bibinfo{journal}{Phys. Rev. Res.}} \textbf{\bibinfo{volume}{6}}, \bibinfo{pages}{023218} (\bibinfo{year}{2024}).

\bibitem{lee2023evaluating}
\bibinfo{author}{Lee, S.} \emph{et~al.}
\newblock \bibinfo{title}{Evaluating the evidence for exponential quantum advantage in ground-state quantum chemistry}.
\newblock \emph{\bibinfo{journal}{Nature Communications}} \textbf{\bibinfo{volume}{14}}, \bibinfo{pages}{1952} (\bibinfo{year}{2023}).

\bibitem{itensor}
\bibinfo{author}{Fishman, M.}, \bibinfo{author}{White, S.~R.} \& \bibinfo{author}{Stoudenmire, E.~M.}
\newblock \bibinfo{title}{{The ITensor Software Library for Tensor Network Calculations}}.
\newblock \emph{\bibinfo{journal}{SciPost Phys. Codebases}} \bibinfo{pages}{4} (\bibinfo{year}{2022}).

\bibitem{Sun_2020}
\bibinfo{author}{Sun, Q.} \emph{et~al.}
\newblock \bibinfo{title}{{Recent developments in the PySCF program package}}.
\newblock \emph{\bibinfo{journal}{The Journal of Chemical Physics}} \textbf{\bibinfo{volume}{153}}, \bibinfo{pages}{024109} (\bibinfo{year}{2020}).

\bibitem{Sun_2018}
\bibinfo{author}{Sun, Q.} \emph{et~al.}
\newblock \bibinfo{title}{Pyscf: the python-based simulations of chemistry framework}.
\newblock \emph{\bibinfo{journal}{WIREs Computational Molecular Science}} \textbf{\bibinfo{volume}{8}}, \bibinfo{pages}{e1340} (\bibinfo{year}{2018}).

\bibitem{Sun_2015}
\bibinfo{author}{Sun, Q.}
\newblock \bibinfo{title}{Libcint: An efficient general integral library for gaussian basis functions}.
\newblock \emph{\bibinfo{journal}{Journal of Computational Chemistry}} \textbf{\bibinfo{volume}{36}}, \bibinfo{pages}{1664--1671} (\bibinfo{year}{2015}).

\bibitem{ali_ordering_2021}
\bibinfo{author}{Ali, M.}
\newblock \bibinfo{title}{On the {Ordering} of {Sites} in the {Density} {Matrix} {Renormalization} {Group} using {Quantum} {Mutual} {Information}}.
\newblock Preprint at \url{http://arxiv.org/abs/2103.01111} (\bibinfo{year}{2021}).

\bibitem{Liu1989}
\bibinfo{author}{Liu, D.~C.} \& \bibinfo{author}{Nocedal, J.}
\newblock \bibinfo{title}{On the limited memory bfgs method for large scale optimization}.
\newblock \emph{\bibinfo{journal}{Mathematical Programming}} \textbf{\bibinfo{volume}{45}}, \bibinfo{pages}{503--528} (\bibinfo{year}{1989}).

\bibitem{hirsbrunner2023mp2}
\bibinfo{author}{Hirsbrunner, M.~R.}, \bibinfo{author}{Chamaki, D.}, \bibinfo{author}{Mullinax, J.~W.} \& \bibinfo{author}{Tubman, N.~M.}
\newblock \bibinfo{title}{Beyond mp2 initialization for unitary coupled cluster quantum circuits}.
\newblock \emph{\bibinfo{journal}{Quantum}} \textbf{\bibinfo{volume}{8}}, \bibinfo{pages}{1538} (\bibinfo{year}{2024}).

\bibitem{Izmaylov_2021}
\bibinfo{author}{Izmaylov, A.~F.}, \bibinfo{author}{Lang, R.~A.} \& \bibinfo{author}{Yen, T.-C.}
\newblock \bibinfo{title}{Analytic gradients in variational quantum algorithms: Algebraic extensions of the parameter-shift rule to general unitary transformations}.
\newblock \emph{\bibinfo{journal}{Phys. Rev. A}} \textbf{\bibinfo{volume}{104}}, \bibinfo{pages}{062443} (\bibinfo{year}{2021}).

\bibitem{Cleve_1969}
\bibinfo{author}{Cleve, R.}, \bibinfo{author}{Ekert, A.}, \bibinfo{author}{Macchiavello, C.} \& \bibinfo{author}{Mosca, M.}
\newblock \bibinfo{title}{Quantum algorithms revisited}.
\newblock \emph{\bibinfo{journal}{Proceedings of the Royal Society of London. Series A: Mathematical, Physical and Engineering Sciences}} \textbf{\bibinfo{volume}{454}}, \bibinfo{pages}{339--354} (\bibinfo{year}{1998}).

\bibitem{Polla_2023}
\bibinfo{author}{Polla, S.}, \bibinfo{author}{Anselmetti, G.-L.~R.} \& \bibinfo{author}{O’Brien, T.~E.}
\newblock \bibinfo{title}{Optimizing the information extracted by a single qubit measurement}.
\newblock \emph{\bibinfo{journal}{Physical Review A}} \textbf{\bibinfo{volume}{108}} (\bibinfo{year}{2023}).

\bibitem{Huggins_2021}
\bibinfo{author}{Huggins, W.~J.} \emph{et~al.}
\newblock \bibinfo{title}{Efficient and noise resilient measurements for quantum chemistry on near-term quantum computers}.
\newblock \emph{\bibinfo{journal}{npj Quantum Information}} \textbf{\bibinfo{volume}{7}} (\bibinfo{year}{2021}).

\bibitem{inoue2023optimal}
\bibinfo{author}{Inoue, W.} \emph{et~al.}
\newblock \bibinfo{title}{Almost optimal measurement scheduling of molecular hamiltonian via finite projective plane}.
\newblock \emph{\bibinfo{journal}{Phys. Rev. Res.}} \textbf{\bibinfo{volume}{6}}, \bibinfo{pages}{013096} (\bibinfo{year}{2024}).

\bibitem{wei2012decomposition}
\bibinfo{author}{Wei, H.-R.} \& \bibinfo{author}{Di, Y.-M.}
\newblock \bibinfo{title}{Decomposition of orthogonal matrix and synthesis of two-qubit and three-qubit orthogonal gates}.
\newblock \emph{\bibinfo{journal}{Quantum Info. Comput.}} \textbf{\bibinfo{volume}{12}}, \bibinfo{pages}{262–270} (\bibinfo{year}{2012}).

\bibitem{zickert2021hands}
\bibinfo{author}{Zickert, F.}
\newblock \emph{\bibinfo{title}{Hands-On Quantum Machine Learning With Python: Volume 1: Get Started}} (\bibinfo{publisher}{Amazon Digital Services LLC - KDP Print US}, \bibinfo{year}{2021}).

\bibitem{Shende_2009}
\bibinfo{author}{Shende, V.~V.} \& \bibinfo{author}{Markov, I.~L.}
\newblock \bibinfo{title}{On the cnot-cost of toffoli gates}.
\newblock \emph{\bibinfo{journal}{Quantum Info. Comput.}} \textbf{\bibinfo{volume}{9}}, \bibinfo{pages}{461–486} (\bibinfo{year}{2009}).

\end{thebibliography}

\begin{thebibliography}{10}
\expandafter\ifx\csname url\endcsname\relax
  \def\url#1{\texttt{#1}}\fi
\expandafter\ifx\csname urlprefix\endcsname\relax\def\urlprefix{URL }\fi
\providecommand{\bibinfo}[2]{#2}
\providecommand{\eprint}[2][]{\url{#2}}

\bibitem{Holtz_2012}
\bibinfo{author}{Holtz, S.}, \bibinfo{author}{Rohwedder, T.} \& \bibinfo{author}{Schneider, R.}
\newblock \bibinfo{title}{The alternating linear scheme for tensor optimization in the tensor train format}.
\newblock \emph{\bibinfo{journal}{SIAM Journal on Scientific Computing}} \textbf{\bibinfo{volume}{34}}, \bibinfo{pages}{A683--A713} (\bibinfo{year}{2012}).

\bibitem{Wouters_2014}
\bibinfo{author}{Wouters, S.} \& \bibinfo{author}{Van~Neck, D.}
\newblock \bibinfo{title}{The density matrix renormalization group for ab initio quantum chemistry}.
\newblock \emph{\bibinfo{journal}{The European Physical Journal D}} \textbf{\bibinfo{volume}{68}} (\bibinfo{year}{2014}).

\bibitem{Singh_2010}
\bibinfo{author}{Singh, S.}, \bibinfo{author}{Pfeifer, R. N.~C.} \& \bibinfo{author}{Vidal, G.}
\newblock \bibinfo{title}{Tensor network decompositions in the presence of a global symmetry}.
\newblock \emph{\bibinfo{journal}{Phys. Rev. A}} \textbf{\bibinfo{volume}{82}}, \bibinfo{pages}{050301} (\bibinfo{year}{2010}).

\bibitem{Singh_2011}
\bibinfo{author}{Singh, S.}, \bibinfo{author}{Pfeifer, R. N.~C.} \& \bibinfo{author}{Vidal, G.}
\newblock \bibinfo{title}{Tensor network states and algorithms in the presence of a global u(1) symmetry}.
\newblock \emph{\bibinfo{journal}{Phys. Rev. B}} \textbf{\bibinfo{volume}{83}}, \bibinfo{pages}{115125} (\bibinfo{year}{2011}).

\bibitem{epperly_2021}
\bibinfo{author}{Epperly, E.~N.}, \bibinfo{author}{Lin, L.} \& \bibinfo{author}{Nakatsukasa, Y.}
\newblock \bibinfo{title}{A theory of quantum subspace diagonalization}.
\newblock \emph{\bibinfo{journal}{SIAM Journal on Matrix Analysis and Applications}} \textbf{\bibinfo{volume}{43}}, \bibinfo{pages}{1263-1290}
(\bibinfo{year}{2022}).

\bibitem{dov2022approximate}
\bibinfo{author}{Ben-Dov, M.}, \bibinfo{author}{Shnaiderov, D.}, \bibinfo{author}{Makmal, A.} \& \bibinfo{author}{Dalla~Torre, E.~G.}
\newblock \bibinfo{title}{Approximate encoding of quantum states using shallow circuits}.
\newblock \emph{\bibinfo{journal}{npj Quantum Information}} \textbf{\bibinfo{volume}{10}} (\bibinfo{year}{2024}).

\bibitem{Rudolph_2024}
\bibinfo{author}{Rudolph, M.~S.}, \bibinfo{author}{Chen, J.}, \bibinfo{author}{Miller, J.}, \bibinfo{author}{Acharya, A.} \& \bibinfo{author}{Perdomo-Ortiz, A.}
\newblock \bibinfo{title}{Decomposition of matrix product states into shallow quantum circuits}.
\newblock \emph{\bibinfo{journal}{Quantum Science and Technology}} \textbf{\bibinfo{volume}{9}}, \bibinfo{pages}{015012} (\bibinfo{year}{2023}).

\bibitem{Yordanov_2020}
\bibinfo{author}{Yordanov, Y.~S.}, \bibinfo{author}{Arvidsson-Shukur, D. R.~M.} \& \bibinfo{author}{Barnes, C. H.~W.}
\newblock \bibinfo{title}{Efficient quantum circuits for quantum computational chemistry}.
\newblock \emph{\bibinfo{journal}{Phys. Rev. A}} \textbf{\bibinfo{volume}{102}}, \bibinfo{pages}{062612} (\bibinfo{year}{2020}).

\bibitem{banerjee_sorting_2019}
\bibinfo{author}{Banerjee, I.}, \bibinfo{author}{Richards, D.} \& \bibinfo{author}{Shinkar, I.}
\newblock \bibinfo{title}{Sorting {Networks} on {Restricted} {Topologies}}.
\newblock In \bibinfo{editor}{Catania, B.}, \bibinfo{editor}{Královič, R.}, \bibinfo{editor}{Nawrocki, J.} \& \bibinfo{editor}{Pighizzini, G.} (eds.) \emph{\bibinfo{booktitle}{{SOFSEM} 2019: {Theory} and {Practice} of {Computer} {Science}}}, \bibinfo{pages}{54--66} (\bibinfo{publisher}{Springer International Publishing}, \bibinfo{address}{Cham}, \bibinfo{year}{2019}).

\bibitem{Knuth98}
\bibinfo{author}{Knuth, D.~E.}
\newblock \emph{\bibinfo{title}{The Art of Computer Programming}}, vol. \bibinfo{volume}{3: Sorting and Searching} (\bibinfo{publisher}{Addison-Wesley}, \bibinfo{year}{1998}).

\bibitem{Gottesman_1996}
\bibinfo{author}{Gottesman, D.}
\newblock \bibinfo{title}{Class of quantum error-correcting codes saturating the quantum hamming bound}.
\newblock \emph{\bibinfo{journal}{Phys. Rev. A}} \textbf{\bibinfo{volume}{54}}, \bibinfo{pages}{1862--1868} (\bibinfo{year}{1996}).

\bibitem{Aaronson_2004}
\bibinfo{author}{Aaronson, S.} \& \bibinfo{author}{Gottesman, D.}
\newblock \bibinfo{title}{Improved simulation of stabilizer circuits}.
\newblock \emph{\bibinfo{journal}{Physical Review A}} \textbf{\bibinfo{volume}{70}} (\bibinfo{year}{2004}).

\bibitem{Stoudenmire_2010}
\bibinfo{author}{Stoudenmire, E.~M.} \& \bibinfo{author}{White, S.~R.}
\newblock \bibinfo{title}{Minimally entangled typical thermal state algorithms}.
\newblock \emph{\bibinfo{journal}{New Journal of Physics}} \textbf{\bibinfo{volume}{12}}, \bibinfo{pages}{055026} (\bibinfo{year}{2010}).

\bibitem{Ferris_2012}
\bibinfo{author}{Ferris, A.~J.} \& \bibinfo{author}{Vidal, G.}
\newblock \bibinfo{title}{Perfect sampling with unitary tensor networks}.
\newblock \emph{\bibinfo{journal}{Phys. Rev. B}} \textbf{\bibinfo{volume}{85}}, \bibinfo{pages}{165146} (\bibinfo{year}{2012}).

\bibitem{Huggins2022}
\bibinfo{author}{Huggins, W.~J.} \emph{et~al.}
\newblock \bibinfo{title}{Unbiasing fermionic quantum monte carlo with a quantum computer}.
\newblock \emph{\bibinfo{journal}{Nature}} \textbf{\bibinfo{volume}{603}}, \bibinfo{pages}{416--420} (\bibinfo{year}{2022}).

\bibitem{Tang_2019}
\bibinfo{author}{Tang, E.}
\newblock \bibinfo{title}{A quantum-inspired classical algorithm for recommendation systems}.
\newblock In \emph{\bibinfo{booktitle}{Proceedings of the 51st Annual ACM SIGACT Symposium on Theory of Computing}}, STOC 2019, \bibinfo{pages}{217–228} (\bibinfo{publisher}{Association for Computing Machinery}, \bibinfo{address}{New York, NY, USA}, \bibinfo{year}{2019}).

\bibitem{Kivlichan_2018}
\bibinfo{author}{Kivlichan, I.~D.} \emph{et~al.}
\newblock \bibinfo{title}{Quantum simulation of electronic structure with linear depth and connectivity}.
\newblock \emph{\bibinfo{journal}{Phys. Rev. Lett.}} \textbf{\bibinfo{volume}{120}}, \bibinfo{pages}{110501} (\bibinfo{year}{2018}).

\bibitem{Sundstrom_2014}
\bibinfo{author}{Sundstrom, E.~J.} \& \bibinfo{author}{Head-Gordon, M.}
\newblock \bibinfo{title}{{Non-orthogonal configuration interaction for the calculation of multielectron excited states}}.
\newblock \emph{\bibinfo{journal}{The Journal of Chemical Physics}} \textbf{\bibinfo{volume}{140}}, \bibinfo{pages}{114103} (\bibinfo{year}{2014}).

\bibitem{Thom_2009}
\bibinfo{author}{Thom, A. J.~W.} \& \bibinfo{author}{Head-Gordon, M.}
\newblock \bibinfo{title}{{Hartree–Fock solutions as a quasidiabatic basis for nonorthogonal configuration interaction}}.
\newblock \emph{\bibinfo{journal}{The Journal of Chemical Physics}} \textbf{\bibinfo{volume}{131}}, \bibinfo{pages}{124113} (\bibinfo{year}{2009}).

\bibitem{Amos_1961}
\bibinfo{author}{Amos, A.~T.}, \bibinfo{author}{Hall, G.~G.} \& \bibinfo{author}{Jones, H.}
\newblock \bibinfo{title}{Single determinant wave functions}.
\newblock \emph{\bibinfo{journal}{Proceedings of the Royal Society of London. Series A. Mathematical and Physical Sciences}} \textbf{\bibinfo{volume}{263}}, \bibinfo{pages}{483--493} (\bibinfo{year}{1961}).

\end{thebibliography}
\end{document}